\documentclass[11pt,a4paper]{article}
\usepackage[margin=21mm]{geometry}
\usepackage{amsmath,amssymb,mathtools,bm,mathrsfs}
\usepackage{booktabs,tabularx,array,longtable}
\usepackage{microtype}
\usepackage{hyperref}
\usepackage{xcolor}
\usepackage{enumitem}
\usepackage{tikz}
\usetikzlibrary{arrows.meta,decorations.markings,decorations.pathmorphing,decorations.pathreplacing}
\hypersetup{colorlinks=true,linkcolor=blue!45!black,citecolor=blue!45!black,urlcolor=blue!45!black}
\newcommand{\ii}{\mathrm{i}}
\newcommand{\dd}{\mathrm{d}}
\newcommand{\ee}{\mathrm{e}}
\newcommand{\Tr}{\operatorname{Tr}}
\newcommand{\Res}{\operatorname{Res}}
\newcommand{\cS}{\mathfrak S}

\newcommand{\cZ}{\mathcal Z}
\newcommand{\cW}{\mathcal W}
\newcommand{\cO}{\mathcal O}
\newcommand{\cI}{\mathcal I}
\newcommand{\cR}{\mathcal R}
\newcommand{\bB}{\mathbf B}

\title{Multi-Instantons and Exact-WKB Stokes Jumps in Multi-Well Quantum Mechanics}
\author{%
  Alireza Behtash\textsuperscript{1}\quad and\quad Syo Kamata\textsuperscript{2}\\[8pt]
  \small\textsuperscript{1}\itshape Stacks, 6841 Middleboro Dr, Raleigh NC 27612, USA\\[2pt]
  \small\textsuperscript{2}\itshape Department of Physics, The University of Tokyo, 7-3-1 Hongo, Bunkyo-ku, Tokyo 113-0033, Japan\\[6pt]
  \small\ttfamily ali@stacksai.com,\quad skamata11phys@gmail.com
}
\date{}

\begin{document}
\maketitle
\vspace{-2.0em}

\begin{abstract}
We study the relation between the exact-WKB Delabaere--Dillinger--Pham Stokes jumps and the multi-instanton expansion of a one-dimensional potential with $N$ equal-depth minima. For a single barrier this relation is the Bogomolny--Zinn-Justin cancellation between the instanton--anti-instanton ambiguity and the Borel ambiguity of perturbation theory, and we extend it to every consecutive barrier interval and to the integration contours themselves. In the dilute large-$T$ regime at fixed nonperturbative grade, we derive the quasi-zero mode integral over the instanton separations from the Euclidean path integral and reduce it to Gamma and Hankel periods. With the local Weber level function of each well and one normalization per well fixed by the local Weber connection matrix, these periods reproduce the two lateral connected exact-WKB coefficients. The two lateral contours of each instanton--anti-instanton separation differ by exactly one Hankel cycle, and this linear Picard--Lefschetz relation reproduces the Stokes jump of the connected coefficient. A numerical solution of the Schr\"odinger equation for a symmetric triple well confirms the resulting two-barrier level splitting.
At higher fixed grade, the map from ordered event sequences to normalized scalar coefficient products is not one-to-one. At leading dilute order the scalar weight of a closed event sequence depends only on the numbers of barrier crossings, $\mathcal I\bar{\mathcal I}$ pairs, and $\mathcal I\mathcal I$/$\bar{\mathcal I}\bar{\mathcal I}$ pairs. We construct two inequivalent grade-six triple-well event sequences with the same product of local Weber amplitudes, which embed in every potential with $N\ge3$ wells. Our results hold within the regular dilute, one-loop, local Weber approximation.
\end{abstract}

\tableofcontents
\section{Introduction}
\label{sec:intro}

One-dimensional multi-well quantum mechanics provides two complementary descriptions of tunneling.  Exact-WKB encodes the nonperturbative spectrum in Voros symbols and their Delabaere--Dillinger--Pham (DDP) Stokes automorphism \cite{DDP1997,SueishiExactWKBI,SueishiExactWKBII}, while the path integral expresses the same physics as a sum over instanton events, integrated over their quasi-zero modes and grouped into bion sectors \cite{ZinnJustinJentschuraI,DunneTripleWell2020}.  Picard--Lefschetz theory gives a natural language for the complexified integration cycles that enter the latter description \cite{Witten2010,TanizakiKoike,BehtashInfinity}.

Both descriptions are shadows of a single geometric structure.  The WKB differential is a quadratic differential on the spectral curve, and its trajectory structure fixes which cycles carry Voros symbols and where those symbols jump \cite{GMN}.  Deformations of that trajectory structure are governed by wall-crossing formulas \cite{KSWallCrossing}, and under a mutation of the Stokes graph the Voros symbols obey the mutation rule of a cluster algebra, which is the form the DDP formula takes in this language \cite{IwakiNakanishi}.  The same data is read in symplectic geometry as a stability condition on a category generated by vanishing cycles \cite{BridgelandSmith,HKK,SeidelVC,SeidelDef}, and holomorphic Floer theory places the path integral, viewed as an exponential integral, in the same framework \cite{KS-HFT}.  In this picture a Stokes jump acts linearly on integration cycles, by the Picard--Lefschetz formula, which adds integer multiples of the cycles that are exchanged across the Stokes direction \cite{Witten2010,KS-HFT}.  The lateral ambiguity of a resummed quantity is then the period of the cycles picked up in that crossing.  Cluster algebras have the same structure.  A mutation changes the basis of cycles by an integer linear map and acts nonlinearly only on the exponentiated periods, which for the Schr\"odinger equation are the Voros symbols.

The modern form of the subject rests on two developments.  Instantons gave a semiclassical account of tunneling in the path integral \cite{Coleman1985}, and the study of large orders of perturbation theory showed that the perturbative series diverges factorially at a rate set by the same tunneling action \cite{BenderWu1969,BenderWu1973}.  High orders of that series can now be generated directly for a wide class of potentials \cite{SulejmanpasicUnsalBenderWu}.  Making the tunneling contributions themselves well defined turned out to be the harder problem.  An instanton and an anti-instanton attract each other, so the integral over their separation is dominated by the region in which the pair is no longer dilute.  Bogomolny and Zinn-Justin showed that this integral becomes well defined after continuation in the coupling, and that the imaginary part generated by the continuation cancels the Borel ambiguity of the perturbative series \cite{Bogomolny1980,ZinnJustin1981,ZinnJustin1983}.  The construction was later recast as generalized Bohr--Sommerfeld quantization conditions for the double well and the periodic potential \cite{ZinnJustinJentschuraI,JentschuraZinnJustinII}, and related derivations were obtained from Borel-summed and uniform WKB \cite{Silverstone1985,Alvarez2004}.  That cancellation was the first concrete evidence that the perturbative and the nonperturbative sectors of a theory are not independent.

Resurgence provides the general framework for such cancellations \cite{Ecalle1981,Costin2008}.  A transseries treats the two as a single object, and the Stokes automorphism relates the ambiguity of one sector to the discontinuity of another.  Accounts written for a field theory audience are given in Refs.~\cite{AnicetoReview,Dorigoni2019,MarinoBook2015,ZinnJustinBook}.  The same structure appears well beyond quantum mechanics.  Bion and renormalon effects govern the nonperturbative structure of the semiclassical expansion of gauge theories and sigma models compactified on a small circle \cite{ArgyresUnsal,DunneUnsalCPN,FujimoriMultiBion}, and a resurgent reading of perturbation theory has been used to locate new saddles and the Lefschetz thimbles attached to them \cite{ChermanDorigoniUnsal,BasarDunneUnsal}.  Quantum mechanics remains the setting in which each step can be controlled, so it continues to serve as the laboratory for the general problem.

Exact-WKB grew out of the complex WKB method and gives the spectral description in this correspondence an exact formulation \cite{DDP1997,Voros1983,KawaiTakei2005}.  The Voros symbols attached to cycles of the classical spectral curve are Borel resummed, the quantization condition is written in terms of them, and the DDP formula states how they jump across a Stokes line.  Uniform WKB supplied the first systematic dictionary between this description and the multi-instanton expansion \cite{DunneUnsalUniform}.  The quantization conditions for a multi-well potential that we take as spectral input below were derived in that framework \cite{SueishiExactWKBI,SueishiExactWKBII}, and the method extends to supersymmetric, deformed, and $\mathcal{PT}$-symmetric potentials \cite{KamataExactWKB,KamataPT}.  In such settings the nonperturbative content can survive even where the naive instanton contribution cancels \cite{KozcazCheshire}.  Exact-WKB also links the spectral problem to integrable structures and to topological string data \cite{GMN,ItoMarinoShu,GrassiHatsudaMarino}.

The counterpart of these developments is the complexified path integral \cite{Witten2010,TanizakiKoike}.  Complex and exact saddles of quantum mechanical path integrals have been constructed explicitly, together with the hidden topological angles that they carry \cite{BehtashHTA,BehtashComplex,BehtashPL,DunneUnsalZero}.  The Morse-theoretic content of those constructions is made explicit in the homological formulation of supersymmetric quantum mechanics \cite{BehtashHomological}.  A bion is not an ordinary saddle of the bare Euclidean action.  Its quasi-zero mode integral (QZI) is instead controlled by a critical point that runs off to infinite separation as the coupling goes to zero \cite{BehtashInfinity}.  We compare that configuration, and the lateral continuation of the integral attached to it, with the DDP jump.

Much of the foundational work concerns the double well, the periodic potential, or a single barrier. Multi-well bions and spectral path integrals have also been studied \cite{DunneTripleWell2020,TureUnsal2025}, and they admit a question that the double well cannot pose. Once three wells are present, a closed sequence of instanton and anti-instanton events has a nontrivial order in which the wells are visited, whereas in the double well a closed event sequence simply alternates between the two wells and is fixed by its length. We therefore ask, for a general one-dimensional potential with $N$ equal-depth wells, which parts of an ordered event sequence are retained by the ordinary scalar spectral data.

The oriented-event transfer determinant has the algebraic structure of a directed-bond evolution matrix on a quantum graph. Distinct periodic codes with equal scalar products and their grouping into classes are familiar there, and the triple-well excursion words have the two-letter combinatorics of the 2-star model of Ref.~\cite{SchanzSmilansky1999}. The present comparison concerns the Weber/DDP amplitudes and their reduced QZI periods. Ordering degeneracy itself is not claimed as a new general graph-theoretic phenomenon.

The local relation between exact-WKB Stokes jumps and QZIs is well known, and our question is more specific than that relation. We work at fixed finite nonperturbative grade\footnote{In the physics literature this is more often called a nonperturbative sector, as in the two-instanton sector, and the two words refer to the same thing here. We say grade because the barrier weights $B_e$ multiply, each carrying the action of one bion, so the nonperturbative expansion carries a grading by total bion number in the algebraic sense. Several results below, such as the invariance of the bion grade under the Airy-to-Weber reduction of Sec.~\ref{subsec:airyWeberDDP}, are properties of that grading rather than of a choice of sector.} in the dilute large-$T$ regime, in which the inverse-Laplace saddle has already entered the Weber regime while a finite-grade expansion has not yet broken down because of the exponentially small level splitting. By the inverse-Laplace saddle we mean the saddle point of the Bromwich integral \eqref{eq:Bromwich} in the complex energy plane, which reconstructs the partition function from the spectral determinant. It is a complex energy rather than a field configuration, and by the Legendre relation of Sec.~\ref{subsec:finiteT_WKB_period} it is the energy of the periodic orbit of period $T$.

The order of limits is essential.  We first fix the nonperturbative grade and only then take $T$ large, so all results hold coefficient by coefficient in the nonperturbative expansion, and we never take the strict zero-temperature limit obtained by truncating the nonperturbative series and then sending $T\to\infty$.  In this ordering the retained contributions are described by a dilute gas of widely separated events, with Gaussian (one-loop) fluctuations supplying the leading perturbative corrections about each event.  As $T$ grows the inverse-Laplace saddle is driven toward the minima of the wells, so the two simple turning points of a well coalesce into a double turning point, and the appropriate local uniform-WKB model is the Weber equation rather than the generic Airy model of an isolated simple turning point.  The wells themselves remain nondegenerate, and the classical saddle configuration does not become degenerate as $E\to V_{\min}$, since the degeneracy comes solely from the coalescence of the pair of turning points.  The path-space reduction is likewise local rather than global.  It is restricted to a regular dilute neighborhood in which the collective coordinates and the normal one-loop sector remain nondegenerate.

There are two parts to the analysis. First, for any consecutive barrier interval, the large-$T$ QZI reduces to Gamma/Hankel periods that reproduce the two lateral exact-WKB coefficients, and the Picard--Lefschetz relation of the contours at the two $\cI\bar\cI$ pairs reproduces the DDP jump between them. We derive the same QZI directly from a regular neighborhood constructed by the normal exponential map around the multi-instanton configuration in the path integral, in which the normal directions remain nondegenerate and the Stokes jump is carried by the quasi-zero modes. Second, at higher fixed grade we assemble the local Weber amplitudes into a transfer matrix acting on oriented tunneling events, and the scalar weight of each closed event sequence depends only on the counts $(\mathbf k,\mathbf R,\mathbf X)$ of barrier crossings, $\cI\bar\cI$ pairs, and $\cI\cI$/$\bar\cI\bar\cI$ pairs. Notably, the map from ordered event sequences to normalized scalar coefficient products is not one-to-one. An explicit grade-six degeneracy already occurs in the triple-well case and, since every $N$-well potential with $N\ge3$ contains three consecutive wells, embeds in all of them. For the more general coefficient-level projection, physical-cycle selection and its compatibility with the Stokes jump are explicit assumptions.

The double and triple wells serve only as special cases of the general construction.  The double well checks the recovery of the one-instanton splitting from the bion grading and the QZI/DDP relation at an arbitrary fixed grade.  The triple well checks the first connected two-barrier sector and gives the simplest explicit ordering degeneracy.  The general results are formulated for an arbitrary $N$-well potential.

For clarity, the assumptions and the regime of validity of the analysis are summarized below.
\begin{itemize}
\item \textbf{Multi-well setting.} We consider a one-dimensional confining polynomial multi-well potential with equal-depth, nondegenerate minima and nearest-neighbor tunneling between adjacent wells. The semiclassical limit \(\hbar\to0\) is taken with the relevant instanton actions kept finite. This setting is fixed by Eqs.~\eqref{eq:Hamiltonian} and \eqref{eq:minima} of Sec.~\ref{subsec:hamiltonian}.
\item \textbf{Fixed-grade large-$T$ regime.} The analysis is performed at fixed finite nonperturbative grade \(K\), in the parametrically large window $K\log(1/|\hbar|)\lesssim T\ll \Delta_{\rm np}^{-1}$,
where the constituent instantons are well separated while a fixed-grade expansion remains valid. Thus the results do not refer to the strict \(T\to\infty\) limit at fixed \(\hbar\), which would require an all-grade resummation. The two bounds of this window are derived in Sec.~\ref{subsec:regime} as Eq.~\eqref{eq:largeTPhysicsWindow} and compared in Sec.~\ref{sec:fixedGradeLargeT}.
\item \textbf{Dilute one-loop/local Weber approximation.} The path integral is reduced to the dilute quasi-moduli sector with large separations, retaining the leading tail interactions and one-loop fluctuations. The corresponding spectral problem is treated in the local Weber regime near the double turning points.  Finite-separation corrections, the regions where instanton events merge, and higher-loop corrections are outside the present approximation. The reduction to the QZI is carried out in Sec.~\ref{sec:fullPI}, and the local Weber description of the spectral problem is obtained in Sec.~\ref{subsec:largeTWeber}.
\item \textbf{Regularity and spectral input.} The reduction is restricted to regular complexified path-space patches in which the collective-coordinate Gram matrix and the reduced normal Hessian remain nondegenerate, as discussed in Sec.~\ref{sec:fullPI}. The DDP transformation law and the lateral quantization functions are taken as input from exact-WKB, as explained in Sec.~\ref{subsec:questions}, while the absolute QZI normalization is fixed once per well by the off-diagonal entry of the local Weber connection matrix in Sec.~\ref{subsec:weberMatching}. Attractive QZI cycles are defined by the standard Bogomolny--Zinn-Justin (BZJ) lateral continuation of Sec.~\ref{subsec:gammaPeriods}.
\end{itemize}

It is useful to state at the outset which results are derived and which rest on input.  Exact-WKB supplies the lateral quantization functions $Q_N^\pm$, the DDP action on the Voros symbols including its nonlinear factors $(1+B_e)^{\pm1}$, and the compatibility of the Airy-to-Weber reduction with that action.  The local Weber equation supplies the level function $\mathsf n_v$ in the exponents of the QZI periods and one normalization per well, imposed on the repulsive separations.  The path integral, on a regular dilute patch, supplies the reduction to the QZI and its Gamma/Hankel periods.  Given these inputs, the attractive periods, the unit period of the Hankel cycle, and hence the lateral entries of the local Weber connection matrix are derived, as are the integer contour relation $\widetilde{\mathsf c}_-=\mathsf c_+-H$ at each $\cI\bar\cI$ pair and the four-term change of the connected coefficient that follows from it.  For a single barrier this change equals the first-order DDP change of $\log P_N$, so the contour relation is the QZI form of the familiar cancellation between the $\cI\bar\cI$ ambiguity and the Borel ambiguity of perturbation theory.  The nonlinear form of the full DDP action is not derived from the QZI, and at higher grade the selection of the physical cycle for arbitrary event sequences, together with its compatibility with the Stokes jump, is assumed.  The two-barrier coefficient of the triple well is also tested against the numerically computed spectrum in App.~\ref{app:twNumerics}.

The scope of the results is limited in four respects, which we state once here rather than at each use.  First, every path-space statement is local.  It concerns a regular dilute neighborhood of the multi-instanton configuration, and we neither construct the integration cycle of the full complexified path space, nor fix a global basis of Lefschetz thimbles, nor compute a global path-space intersection pairing, so the integer contour relations are those of the reduced QZI.  Second, the absolute normalization of the elementary one-barrier block is fixed by the local Weber connection matrix rather than computed from the path integral.  It is checked independently only for the symmetric double well, and its path-integral derivation for barriers between wells of unequal frequency remains open.  Third, the classical inverse-Laplace saddle motivates the Weber energy range but is not shown to dominate the Bromwich integral uniformly, and all asymptotic statements hold coefficient by coefficient at fixed grade, without uniform control of the error constants over the joint large-$T$, small-$\hbar$ window.  Fourth, the statements about event sequences concern formal products of local Weber amplitudes.  They are not reconstruction or impossibility theorems for unrestricted spectral data, and the relations among the $B_e$ and $u_v^\pm$ in a specific potential can identify further event sequences.

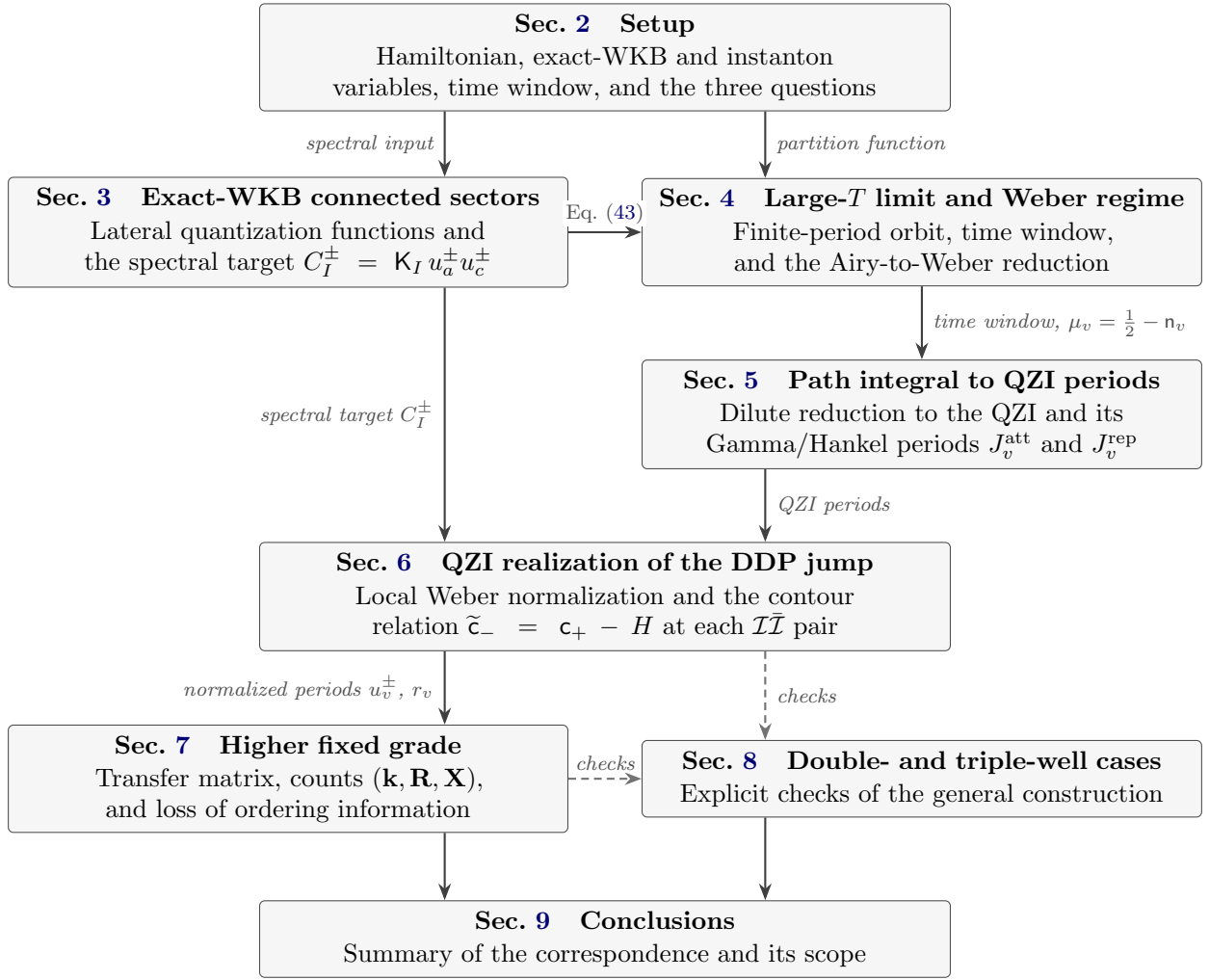
\begin{figure}[!b]
\centering
\begin{tikzpicture}[
  >=Stealth,
  box/.style={draw=black!70,rounded corners=2pt,fill=gray!7,align=center,
    text width=#1,inner sep=4pt,font=\small,
    execute at begin node={\hyphenpenalty=10000\exhyphenpenalty=10000}},
  box/.default=7.4cm,
  flow/.style={->,thick,black!75},
  check/.style={->,thick,black!55,densely dashed},
  lab/.style={font=\scriptsize\itshape,black!70,fill=white,inner sep=1pt},
]
\node[box=9.2cm] (s1) at (0,0)
  {\textbf{Sec.~\ref{sec:setup}\quad Setup}\\[1pt]
   Hamiltonian, exact-WKB and instanton variables, time window, and the three questions};
\node[box] (s2) at (-4.35,-2.4)
  {\textbf{Sec.~\ref{sec:spectralTarget}\quad Exact-WKB connected sectors}\\[1pt]
   Lateral quantization functions and the spectral target $C_I^\pm=\mathsf K_I\,u_a^\pm u_c^\pm$};
\node[box] (s3) at (4.35,-2.4)
  {\textbf{Sec.~\ref{sec:largeTWeber}\quad Large-$T$ limit and Weber regime}\\[1pt]
   Finite-period orbit, time window, and the Airy-to-Weber reduction};
\node[box] (s4) at (4.35,-4.9)
  {\textbf{Sec.~\ref{sec:fullPI}\quad Path integral to QZI periods}\\[1pt]
   Dilute reduction to the QZI and its Gamma/Hankel periods $J_v^{\rm att}$ and $J_v^{\rm rep}$};
\node[box=9.2cm] (s5) at (0,-7.4)
  {\textbf{Sec.~\ref{sec:QZIStokes}\quad QZI realization of the DDP jump}\\[1pt]
   Local Weber normalization and the contour relation $\widetilde{\mathsf c}_-=\mathsf c_+-H$ at each $\cI\bar\cI$ pair};
\node[box] (s6) at (-4.35,-9.9)
  {\textbf{Sec.~\ref{sec:higher_order_projection}\quad Higher fixed grade}\\[1pt]
   Transfer matrix, counts $(\mathbf k,\mathbf R,\mathbf X)$, and loss of ordering information};
\node[box] (s7) at (4.35,-9.9)
  {\textbf{Sec.~\ref{sec:physical_cases}\quad Double- and triple-well cases}\\[1pt]
   Explicit checks of the general construction};
\node[box=9.2cm] (s8) at (0,-12.1)
  {\textbf{Sec.~\ref{sec:conclusions}\quad Conclusions}\\[1pt]
   Summary of the correspondence and its scope};
\coordinate (L) at (-2.2,0);
\coordinate (R) at (2.2,0);
\draw[flow] (s1.south -| L) -- node[lab,left=3pt]{spectral input} (s2.north -| L);
\draw[flow] (s1.south -| R) -- node[lab,right=3pt]{partition function} (s3.north -| R);
\draw[flow] (s2.east) -- node[lab,above=2pt,font=\scriptsize]{Eq.~\eqref{eq:Bromwich}} (s2.east -| s3.west);
\draw[flow] (s3.south) -- node[lab,right=3pt]{time window, $\mu_v=\tfrac12-\mathsf n_v$} (s4.north);
\draw[flow] (s2.south -| L) -- node[lab,left=3pt]{spectral target $C_I^\pm$} (s5.north -| L);
\draw[flow] (s4.south -| R) -- node[lab,right=3pt]{QZI periods} (s5.north -| R);
\draw[flow] (s5.south -| L) -- node[lab,left=3pt]{normalized periods $u_v^\pm$, $r_v$} (s6.north -| L);
\draw[check] (s5.south -| R) -- node[lab,right=3pt]{checks} (s7.north -| R);
\draw[check] (s6.east) -- node[lab,above=2pt]{checks} (s6.east -| s7.west);
\draw[flow] (s6.south -| L) -- (s8.north -| L);
\draw[flow] (s7.south -| R) -- (s8.north -| R);
\end{tikzpicture}
\caption{Logical structure of the paper.  Solid arrows indicate that the results of one section are used as input in another, and dashed arrows indicate where the double- and triple-well cases of Sec.~\ref{sec:physical_cases} check the general construction of Secs.~\ref{sec:QZIStokes} and \ref{sec:higher_order_projection}.}
\label{fig:section_flow}
\end{figure}

Section~\ref{sec:setup} fixes the Hamiltonian, the exact-WKB and instanton variables, and the time window, and states the three questions addressed in the paper together with the exact-WKB input. Section~\ref{sec:spectralTarget} then derives the connected exact-WKB coefficient $C_I^\pm$ of an arbitrary consecutive barrier interval. Section~\ref{sec:largeTWeber} shows that the large-$T$ inverse-Laplace saddle is the finite-period Euclidean orbit and that it enters the Weber regime. In Sec.~\ref{sec:fullPI} we construct the corresponding QZI directly from a regular dilute neighborhood of the path integral, reduce it to Gamma/Hankel periods, and identify the spectral pole that fixes their Mellin exponents. Section~\ref{sec:QZIStokes} fixes the local Weber normalization, shows that the two lateral QZI periods reproduce the two exact-WKB coefficients, and derives their DDP jump from the Picard--Lefschetz relation at the two $\cI\bar\cI$ pairs. Section~\ref{sec:higher_order_projection} assembles the same local blocks into a transfer matrix at higher fixed grade and determines which information about ordered event sequences survives the specified scalar coefficient projection. The double- and triple-well cases of Sec.~\ref{sec:physical_cases} check the general construction in the smallest cases, and Sec.~\ref{sec:conclusions} summarizes the results and their scope. The logical relations among the sections are shown in Fig.~\ref{fig:section_flow}. Appendix~\ref{app:notation} collects the notation, and Appendix~\ref{app:QZI} rederives the QZI in the standard language of the instanton calculus.
\section{Setup}
\label{sec:setup}

We construct the quantum mechanical problem, the exact-WKB and instanton variables in which it is analyzed, and the time window in which the analysis applies, and then state the three questions addressed in the rest of the paper.  The main symbols are collected in Table~\ref{tab:notation} of Appendix~\ref{app:notation}.

\subsection{Hamiltonian, wells, and the partition function}
\label{subsec:hamiltonian}

We consider a one-dimensional Hamiltonian
\begin{equation}
 \widehat H=-\frac{\hbar^2}{2}\frac{\dd^2}{\dd x^2}+V(x),
 \label{eq:Hamiltonian}
\end{equation}
where $V(x)$ is a real confining polynomial, analytic in the complex domain needed below, with $N$ adjacent nondegenerate minima
\begin{equation}
 x_1<x_2<\cdots<x_N,
 \qquad
 V(x_v)=V_{\min},
 \qquad
 V''(x_v)=\omega_v^2>0.
 \label{eq:minima}
\end{equation}
The equal-depth assumption gives all wells a common low-energy scaling.  Only adjacent minima in this sequence are separated by an elementary barrier, so every tunneling event considered below connects nearest neighbors and no direct transition between distant wells is retained.  The semiclassical limit is $\hbar\to0$ taken with $V$ held fixed, so the barrier actions stay finite while the tunneling weights they control become exponentially small.

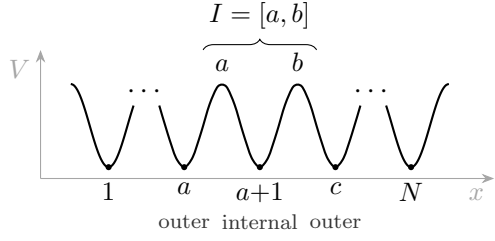
\begin{figure}[t]
\centering
\begin{tikzpicture}[>=Stealth]
\begin{scope}
  \draw[->,gray!70] (-0.9,-0.15) -- (5.1,-0.15) node[below left] {\small$x$};
  \draw[->,gray!70] (-0.9,-0.15) -- (-0.9,1.55) node[below left] {\small$V$};
  \draw[thick,domain=-0.5:0.33,smooth,variable=\x] plot ({\x},{0.55*(1-cos(360*\x))});
  \draw[thick,domain=0.67:3.33,smooth,variable=\x] plot ({\x},{0.55*(1-cos(360*\x))});
  \draw[thick,domain=3.67:4.5,smooth,variable=\x] plot ({\x},{0.55*(1-cos(360*\x))});
  \node at (0.5,1.0) {$\cdots$};
  \node at (3.5,1.0) {$\cdots$};
  \foreach \x/\l in {0/1,1/a,2/{a{+}1},3/c,4/N} {\fill (\x,0) circle (0.04); \node[below=2pt,font=\small] at (\x,0) {$\l$};}
  \node[font=\scriptsize,gray!40!black] at (1,-0.72) {outer};
  \node[font=\scriptsize,gray!40!black] at (2,-0.72) {internal};
  \node[font=\scriptsize,gray!40!black] at (3,-0.72) {outer};
  \node[above=1pt,font=\small] at (1.5,1.1) {$a$};
  \node[above=1pt,font=\small] at (2.5,1.1) {$b$};
  \draw[decorate,decoration={brace,amplitude=4pt}] (1.25,1.55) -- node[above=4pt,font=\small] {$I=[a,b]$} (2.75,1.55);
\end{scope}
\end{tikzpicture}
\caption{Labeling of wells and barriers.  Wells are labeled $v=1,\ldots,N$ and barriers $e=1,\ldots,N-1$, with barrier $e$ between wells $e$ and $e+1$.  A consecutive barrier interval $I=[a,b]$ of Eq.~\eqref{eq:interval}, shown here with two barriers, has outer wells $a$ and $c=b+1$ and internal wells $a+1,\ldots,c-1$.}
\label{fig:wells}
\end{figure}

The wells of this sequence are labeled $v=1,\ldots,N$ from left to right and the barriers $e=1,\ldots,N-1$, with barrier $e$ separating wells $e$ and $e+1$, so that well $v$ is bounded by barrier $v-1$ on the left and barrier $v$ on the right.  Since the wells lie in a row, every nonempty connected set of barriers is a consecutive interval, and we write
\begin{equation}
 I=[a,b],
 \qquad c=b+1,
 \label{eq:interval}
\end{equation}
for the barriers $a,a+1,\ldots,b$.  The wells touched by $I$ are $a$ through $c$, so $a$ and $c$ are the two outer wells of the interval and $a+1,\ldots,c-1$ are the internal ones.  The index $a$ labels both the leftmost barrier of $I$ and the left outer well.  Fig.~\ref{fig:wells} shows this labeling.

We write the partition function as
\begin{equation}
 Z(T,\hbar)=\Tr\exp\left(-\frac{T\widehat H}{\hbar}\right),
 \qquad T=\beta\hbar.
 \label{eq:thermaltrace}
\end{equation}
Here $\beta$ is the inverse temperature in the conventional thermal notation, while $T$ is the Euclidean time of the path integral.  The corresponding Euclidean action is
\begin{equation}
 S_E[x]=\int_0^T \dd t\left[\frac12\dot x^2+V(x)\right],
 \qquad x(0)=x(T).
 \label{eq:EuclideanAction}
\end{equation}
Thus
\begin{equation}
 Z(T,\hbar)=\int_{x(0)=x(T)}\mathcal D x\;\ee^{-S_E[x]/\hbar}.
 \label{eq:pathintegral}
\end{equation}
We often subtract the common classical potential minimum $V_{\min}$, which only removes the overall factor $\exp(-TV_{\min}/\hbar)$.

\subsection{Exact-WKB and instanton variables}
\label{subsec:variables}

In exact-WKB, the classical spectral curve is
\begin{equation}
 p^2=2\bigl(V(x)-E\bigr).
 \label{eq:spectralcurve}
\end{equation}
For each selected well we use a local perturbative cycle, or $A$-cycle, $\gamma_v$, and for each barrier $e$ between $x_e$ and $x_{e+1}$ a nonperturbative cycle, or $B$-cycle, $\delta_e$, which stems from the standard exact-WKB nomenclature \cite{SueishiExactWKBI}. Their Borel-resummed Voros symbols are denoted by
\begin{equation}
 A_v=X_{\gamma_v},
 \qquad
 B_e=X_{\delta_e}.
 \label{eq:VorosSymbols}
\end{equation}
Both are functions of the spectral energy and of $\hbar$.  In a convention-independent leading-order notation, a Voros symbol behaves as
\begin{equation}
 X_\Gamma(E,\hbar)
 \sim
 \exp\left[\frac{1}{\hbar}\oint_\Gamma p(x,E)\,\dd x+\cdots\right],
 \label{eq:VorosLeading}
\end{equation}
where the dots denote the higher WKB corrections.  The perturbative cycle $\gamma_v$ probes the local oscillatory motion in one well and encodes the perturbative energy levels through the local Bohr--Sommerfeld or exact-WKB quantization condition, whereas $\delta_e$ crosses the classically forbidden barrier and therefore carries the tunneling effects. We orient the nonperturbative cycles so that $B_e$ is exponentially small in the semiclassical regime under consideration.  Since the partition function is a sum over periodic paths, a single instanton cannot close on itself, so $B_e$ is naturally attached to a closed crossing of barrier $e$, once in each direction.  Let $S_{I,e}$ be the action of one oriented instanton crossing barrier $e$. A closed event sequence must cross the same barrier equally often in the two directions, so one instanton and one anti-instanton across $e$ together contribute the classical weight
\begin{equation}
 \exp\left[-\frac{S_{I,e}}{\hbar}\right]
 \exp\left[-\frac{S_{I,e}}{\hbar}\right]
 =\exp\left[-\frac{2S_{I,e}}{\hbar}\right].
 \label{eq:openToClosedAction}
\end{equation}
Exact-WKB gives a prescription for getting a resummed version of $B_e$, including its fluctuation/WKB dressing.  Thus, at leading exponential order,
\begin{equation}
 B_e\sim \ee^{-2S_{I,e}/\hbar},
 \qquad
 \text{one simple tunneling event}\sim B_e^{1/2}.
 \label{eq:BfromInstantonPair}
\end{equation}
We call $B_e$ the {\it barrier weight} of $e$.  At leading order it equals the classical bion weight $\ee^{-2S_{I,e}/\hbar}$, but it is not the full amplitude of a bion.  That amplitude also contains the QZI over the instanton/anti-instanton separation, which carries the lateral choice, whereas $B_e$ is left invariant by the DDP Stokes automorphism.  We therefore reserve the word {\it bion} for the instanton--anti-instanton configuration of the path integral and for the grading it induces.  Over a consecutive barrier interval $I=[a,b]$, defined in Eq.~\eqref{eq:interval}, the minimal closed bion trajectory crosses every barrier once in each direction.  The factor $\prod_{e\in I}B_e$ therefore appears both in the exact-WKB coefficient of the interval and in its path-integral contribution.  Note that the absolute one-loop normalization of the bion solution is fixed at the elementary one-barrier level. A superscript $\pm$ always denotes the two lateral Borel resummations at $\arg\hbar=0_+$ and $0_-$. 

In the Euclidean path integral, an elementary tunneling event is represented by an instanton profile $q_e(t)$, a solution joining adjacent minima, whose weight is the $B_e^{1/2}$ of Eq.~\eqref{eq:BfromInstantonPair}.  A dilute multi-event configuration consists of a sequence of instantons and anti-instantons centered at times $t_r$ and separated by long time intervals
\begin{equation}
 \tau_r=t_{r+1}-t_r.
 \label{eq:separations}
\end{equation}
We fix the notation for times once and for all.  Euclidean time is denoted by $t$, while $\tau$, with or without a subscript and in bold for the collection, always denotes a separation or a duration rather than a time coordinate.  Where a local time measured from the start of a single separation is needed, it is called $\sigma$.  The separation $\tau_r$ spans the visit to the well $v_r$ that lies between the events at $t_r$ and $t_{r+1}$, so the index $r$ labels a separation and a visit at the same time.  The $t_r$, or equivalently the separations $\tau_r$, are collective coordinates.  At infinite separation they are exact zero modes, whereas at large but finite separation they become quasi-zero modes.  The space they parametrize is the quasi-moduli space, and the finite-dimensional integral over it is the QZI.

\subsection{Fixed-grade time window}
\label{subsec:regime}

The analysis applies in a time window whose two ends we now estimate. At fixed bion grade $K$, a closed configuration has $2K$ separations. The tail interaction in separation $r$ becomes of order one at
\begin{equation}
 \tau_{r,\mathrm{int}}=\frac{1}{\omega_r}\log\frac{1}{|\hbar|}+O(1).
\end{equation}
This is an interaction scale, rather than a universal fixed-$T$ saddle position. After the Laplace transform of Sec.~\ref{subsec:laplaceFactorization}, the exponent of a single separation is
\begin{equation}
 \mathcal W_r(\tau_r;s)=(s+\varrho_r)\tau_r+\lambda_r\Lambda_r\ee^{-\omega_r\tau_r},
 \qquad \omega_r\mu_r=s+\varrho_r.
\end{equation}
Here $s$ is the Laplace variable conjugate to $T$, $\varrho_r$ is the relative zero-point energy of the visited well, and $\lambda_r=\pm1$ and $\Lambda_r>0$ are the sign and magnitude of the tail interaction, all defined in Sec.~\ref{sec:fullPI}. A finite stationary point obeys $\Lambda_r\ee^{-\omega_r\tau_r^*}=\lambda_r\mu_r$. The logarithmic estimate for $\Re\tau_r^*$ follows only when $|\mu_r|$ is bounded above and away from zero as $\hbar\to0$, with the logarithmic sheet fixed. At fixed $T$, the multiplier $s$ is determined by $\sum_r\tau_r=T$ and need not satisfy that condition. In particular, the double-well bion has $\tau_1^*=\tau_2^*=T/2$, with $|\mu|=|\Lambda|\ee^{-\omega T/2}$, which tends to zero if $T$ grows faster than $(2/\omega)\log(1/|\hbar|)$. Nor does the stationary equation by itself identify the occupied BZJ contour.

To accommodate a dilute region in which all $2K$ tail interactions are bounded, the total time must be at least the sum of these interaction scales. At fixed frequencies this gives the scale estimate
\begin{equation}
 T\gtrsim T_{\rm QZI}(K,\hbar)\sim K\log \frac{1}{|\hbar|}.
\end{equation}
 A margin above this scale is required for uniformly small interactions. The estimate only describes the available dilute region and does not assert that every constrained saddle or every part of the original real integration domain lies there.

On the other hand, let \(\Delta_{\rm np}\) denote the relevant exponentially small displacement or splitting of the scaled levels $\epsilon_n=(E_n-V_{\min})/\hbar$. A physical energy splitting is $\hbar\Delta_{\rm np}$. One-instanton splitting requires degenerate neighboring perturbative levels, and equal classical well depths alone do not imply that degeneracy. Euclidean evolution becomes sensitive to this splitting through the dimensionless combination
$T\Delta_{\rm np}$, and a necessary smallness condition for the fixed-grade truncation is $T\Delta_{\rm np}\ll1$, or equivalently
\begin{equation}  T\ll T_*\sim\Delta_{\rm np}^{-1}. 
\end{equation} 
Once \(T\) reaches \(T_*\), factors involving the level splitting, such as \(\ee^{-T\Delta_{\rm np}}\), can no longer be approximated using finitely many powers of \(\Delta_{\rm np}\). Contributions from arbitrarily high nonperturbative grades become comparable, so the full transseries or multi-instanton expansion must be resummed before taking \(T\to\infty\). Putting together the lower and upper bounds, the time scale must satisfy
\begin{equation}
 T_{\rm QZI}(K,\hbar)\sim K\log\frac1{|\hbar|}\ \lesssim\ T\ \ll\ T_*\sim \Delta_{\rm np}^{-1},
 \label{eq:largeTPhysicsWindow}
\end{equation}
so $T$ must be large enough to fit all the widely separated events at the chosen grade, but not so large that infinitely many nonperturbative grades must be resummed.

\subsection{Three questions and spectral input}
\label{subsec:questions}

In terms of the Voros symbols and the separations introduced above, the paper addresses three questions.
\begin{enumerate}[leftmargin=2.0em]
\item[(1)] What coefficient does exact-WKB predict at a given barrier interval, and what is its lateral DDP jump? Choose a consecutive interval of barriers, say from well \(a\) to well \(c\). Exact-WKB expands the spectral determinant or quantization condition in the small barrier weights \(B_e\). What is the coefficient multiplying the product of barrier weights associated with that interval and how does that coefficient change when the WKB series is Borel-resummed just above versus just below the Stokes ray? If the interval contribution is
$$ C_I^\pm(E,\hbar)\prod_{e\in I}B_e, $$
the question asks for both \(C_I^+\) and \(C_I^-\), or equivalently the DDP discontinuity
$$ \operatorname{Disc}_0 C_I = C_I^- - C_I^+. $$
Here, {\it lateral} refers to upper and lower Borel resummations, and {\it DDP jump} refers to the DDP transformation relating them.
\item[(2)] If the same multi-instanton sector is evaluated by QZI, does the integral give the same coefficient and the same lateral jump? The same physical contribution can be calculated from the Euclidean path integral by considering a sequence of instantons and anti-instantons and integrating over their separations \(\tau_r\), which is the QZI. The question is whether the QZI reproduces both the exact-WKB coefficient and its DDP jump, the latter through the jump of the QZI contour. Matching only the jump would show that the two descriptions have the same Stokes topology. Matching the complete coefficient additionally requires the normalizations, determinant factors, instanton amplitudes, and multiplicities to agree, in which case exact-WKB and the multi-instanton calculation are genuinely the same nonperturbative calculation.
\item[(3)] At higher order, how much of the ordered event sequence survives in the spectral scalar? At higher bion grade, many different event sequences are possible, where by an {\it event sequence} we mean the ordered list of instanton and anti-instanton events of a configuration, closed on the circle $t\sim t+T$ of the path integral \eqref{eq:pathintegral}. For example, two sequences may cross the same barriers the same number of times but do so in different chronological orders. As configurations in path space these ordered sequences,
$$ w=(e_1,e_2,\ldots,e_m), $$
are distinct, but the final spectral determinant is a scalar function of \(E\) and \(\hbar\). The question is whether this scalar retains enough information to reconstruct the full ordered sequence \(w\), or whether different event sequences become indistinguishable after projection to spectral data.
 In the normalized local-amplitude expansion an event sequence is characterized by the aggregate counts
$$(\mathbf k,\mathbf R,\mathbf X), $$
where
\(\mathbf k\) shows how often each barrier is crossed in each direction,
\(\mathbf R\) counts the \(\cI\bar\cI\) pairs at each well, where the trajectory turns back, and
\(\mathbf X\) counts the \(\cI\cI\) and \(\bar\cI\bar\cI\) pairs, where it passes through the well. Since different ordered event sequences can have identical \((\mathbf k,\mathbf R,\mathbf X)\), one expects that they can produce the same spectral scalar, in which case the map from ordered event sequences to scalar spectral coefficients is not one-to-one.
\end{enumerate}

The paper does not start from the Schr\"odinger equation and reconstruct the complete exact-WKB connection problem.  Instead, it takes three established exact-WKB results as input.  First, we need the DDP transformation of Voros symbols, which specifies how the Borel-resummed Voros symbols change when the resummation direction crosses a Stokes ray.  It supplies the exact-WKB Stokes-jump rule that the QZI calculation is expected to reproduce.  Second, we use the reduced quantization functions \(Q_N^\pm\).  These are the upper and lower lateral quantization functions for the \(N\)-well system, after the relevant common or purely perturbative factors have been removed, and expanding them in the small barrier weights \(B_e\) reveals the spectral coefficient associated with each barrier interval.  Both inputs are recalled in Sec.~\ref{subsec:lateralQC}.  Third, we use the local Weber equation of each well.  Near a double turning point, the local Schr\"odinger problem is described by the Weber equation, with the harmonic level number $\epsilon/\omega_v$ replaced by the local Weber level function $\mathsf n_v(\epsilon,\hbar)$.  This function enters the exponents of the QZI periods, which goes beyond the one-loop path-integral reduction, so we state it as a separate input.  The standard connection formula of the Weber equation then fixes one normalization per well, imposed on the repulsive separations alone, and this fixes the conversion between the exact-WKB barrier weight and the elementary instanton or QZI contribution.

\section{Exact-WKB connected sectors and the spectral target}
\label{sec:spectralTarget}

We first isolate the connected spectral coefficient that the multi-instanton calculation must reproduce, keeping the DDP transformation itself as exact-WKB input, and then relate the quantization function to the partition function computed by the path integral \eqref{eq:pathintegral}.

\subsection{Lateral quantization conditions for a polynomial potential}
\label{subsec:lateralQC}

We take $V$ to be a polynomial with $N$ equal-depth minima, with wells and barriers labeled as in Fig.~\ref{fig:wells}.  There is no barrier beyond the two outermost wells, and we encode this by setting
\begin{equation}
 B_0=B_N=0.
 \label{eq:endConvention}
\end{equation}
As before, $A_v$ and $B_e$ are the Borel-resummed Voros symbols of the perturbative and nonperturbative cycles of Eq.~\eqref{eq:VorosSymbols}.  Thus $A_v$ carries the local Bohr--Sommerfeld data of well $v$, while $B_e$ is the exponentially small barrier weight of $e$ introduced after Eq.~\eqref{eq:BfromInstantonPair}.

Let $\cS_0^{\rm DDP}$ denote the DDP Stokes automorphism across the Stokes ray $\arg\hbar=0$, defined by $\mathcal S_{0+}=\mathcal S_{0-}\circ\cS_0^{\rm DDP}$, where $\mathcal S_{0\pm}$ are the lateral Borel resummations along $\arg\hbar=0_\pm$.  It acts on the two families of symbols as \cite{DDP1997,SueishiExactWKBI}
\begin{equation}
 \cS_0^{\rm DDP}[A_v]
 =A_v\bigl[(1+B_{v-1})(1+B_v)\bigr]^{(-1)^v},
 \qquad
 \cS_0^{\rm DDP}[B_e]=B_e,
 \label{eq:DDPaction}
\end{equation}
so every barrier weight is fixed exactly, and each perturbative symbol is dressed only by the two barriers that bound its own well.  The exponent $(-1)^v$ is the intersection pairing of the $A$-cycle of well $v$ with the two adjacent $B$-cycles, and it alternates because consecutive wells are traversed in opposite senses.  That alternation is the source of every parity\footnote{Here and below, the parity of a well means whether its label $v$ is even or odd, that is, the sign $(-1)^v$.  It is unrelated to any spatial reflection symmetry of the potential, such as $x\to-x$, and to the parity of an eigenstate.  The potential is not assumed to be reflection symmetric, and reflection parity enters only for the symmetric double well of Sec.~\ref{subsec:doubleWell}, where it is named explicitly.} condition that follows.

The two lateral quantization functions are sums over the subsets $S$ of the barrier set $\{1,\ldots,N-1\}$,
\begin{align}
 Q_N^+
 &=\sum_{S}
 \left(\prod_{e\in S}B_e\right)
 \left(\prod_{v\in\partial S,\;v\;{\rm odd}}A_v\right)
 \left(\prod_{v\notin\partial S}(1+A_v)\right),
 \label{eq:Qplus}
 \\
 Q_N^-
 &=\sum_{S}
 \left(\prod_{e\in S}B_e\right)
 \left(\prod_{v\in\partial S,\;v\;{\rm even}}A_v\right)
 \left(\prod_{v\notin\partial S}(1+A_v)\right).
 \label{eq:QZInus}
\end{align}
Here $\partial S$ is the set of wells incident on at least one barrier of $S$.  Each term is read as follows.  The barriers of $S$ are active and supply the weight $\prod_{e\in S}B_e$.  A well untouched by $S$ contributes its full perturbative factor $1+A_v$.  A well touched by $S$ contributes $A_v$ when its parity matches the lateral choice and contributes $1$ otherwise.  The upper and lower functions differ only in which parity is selected, and that single interchange is the whole of the lateral ambiguity at this order.  These are the connection-matrix quantization conditions of the exact-WKB description derived in Refs.~\cite{SueishiExactWKBI,SueishiExactWKBII} and studied further in Ref.~\cite{KamataPT}.

For the double well, $N=2$, there is one barrier and two wells, and
\begin{equation}
 Q_2^+=(1+A_1)(1+A_2)+B_1A_1,
 \qquad
 Q_2^-=(1+A_1)(1+A_2)+B_1A_2.
 \label{eq:Q2example}
\end{equation}
The purely $ A$-cyclic parts give the common perturbative term, and the single active barrier attaches $A_1$ in the upper function and $A_2$ in the lower one, making up the nonperturbative parts.  The triple well is written out in the same way in Eq.~\eqref{eq:Z3example} below.

The purely perturbative part is
\begin{equation}
 P_N:=\prod_{v=1}^N P_v,
 \qquad
 P_v:=1+A_v,
 \label{eq:PN}
\end{equation}
which we divide out,
\begin{equation}
 \cZ_N^\pm:=\frac{Q_N^\pm}{P_N}.
 \label{eq:normalizedZ}
\end{equation}
Two distinct reductions are therefore in play.  Already $Q_N^\pm$ is a reduced representative of anything spectral, in the sense that the common factors carrying no zeros have been dropped from the full spectral determinant, and $\cZ_N^\pm$ removes in addition the purely perturbative factor $P_N$.  After both steps every term of $\cZ_N^\pm$ beyond the leading $1$ carries at least one barrier weight, so the expansion begins at bion grade one.\footnote{The division by $P_N$ is adapted to the nonperturbative expansion and is not used for the purely perturbative part, since the zeros of $P_N$ are the perturbative levels and become poles of $\cZ_N^\pm$.}  The logarithm $\log\cZ_N^\pm$ then isolates the connected nonperturbative contributions in the usual way.

Only the zeros of a quantization function are spectral data.  Let $D(E,\hbar)$ be any function whose zeros in the spectral domain under consideration are the exact levels.  If
\begin{equation}
 D(E,\hbar)\longrightarrow M(E,\hbar)D(E,\hbar),
 \qquad M\neq0
 \label{eq:multiplicativeNormalization}
\end{equation}
with $M$ holomorphic and nowhere vanishing in the spectral domain, then
\begin{equation}
 -\partial_E\log(MD)
 =-\partial_E\log D-\partial_E\log M.
 \label{eq:normalizationResolvent}
\end{equation}
The second term has no poles there, so neither the exact levels nor the local resolvent residues are changed.  The formal decomposition of $\log D$ into perturbative and bion-graded pieces, however, does depend on the representative.  For that reason we fix the reduced representative $Q_N^\pm$ before introducing the connected bion expansion.  This choice fixes the transseries sectors and leaves the spectrum untouched.

It remains to check that a fixed finite grade is a meaningful truncation under DDP.  Grade counts barrier weights, so let $\mathfrak m_B^K$ denote those series in $B_1,\ldots,B_{N-1}$ in which every term carries at least $K$ barrier weights, with coefficients built from the $A_v$ and from inverses of the $P_v$.  A term of $\mathfrak m_B^K$ therefore has bion grade at least $K$.  Since Eq.~\eqref{eq:DDPaction} fixes each barrier weight exactly and dresses each $A_v$ only by factors $1+B_e$, the perturbative symbols satisfy
\begin{equation}
 \cS_0^{\rm DDP}[A_v]
 =A_v\bigl[1+O(\mathfrak m_B)\bigr].
\end{equation}
An $A$-dependent coefficient multiplying a grade-$K$ monomial in the barrier weights may therefore generate terms of higher grade under DDP, but its grade-$K$ component is unchanged.  Equivalently,
\begin{equation}
 (\cS_0^{\rm DDP}-1)\,\mathfrak m_B^K
 \subseteq \mathfrak m_B^{K+1}.
 \label{eq:generalBionTriangularity}
\end{equation}
The induced DDP action on the associated quotient $\mathfrak m_B^K/\mathfrak m_B^{K+1}$ is the identity. The two explicit lateral coefficient functions can nevertheless differ, because lower-grade terms, including $P_N$, contribute once the DDP action \eqref{eq:DDPaction} is applied to the Voros symbols, as explained in Sec.~\ref{subsec:nonlinearDDP}. The Stokes comparison can accordingly be carried out coefficient by coefficient at fixed total bion grade. This makes the fixed finite nonperturbative grade of Sec.~\ref{subsec:regime} a well-defined truncation.

\subsection{Local perturbative oscillations and the connected interval coefficient}
\label{subsec:intervalCoefficient}

In each term of $Q_N^\pm$ a well touched by the active barriers contributes $A_v$ or $1$ according to its parity, while an untouched well contributes its full factor $P_v$.  Dividing by $P_N$ as in Eq.~\eqref{eq:normalizedZ} therefore attaches to every touched well the local perturbative oscillations
\begin{equation}
 u_v^+=
 \begin{cases}
 A_v/P_v,&v\text{ odd},\\[1mm]
 1/P_v,&v\text{ even},
 \end{cases}
 \qquad
 u_v^-=
 \begin{cases}
 1/P_v,&v\text{ odd},\\[1mm]
 A_v/P_v,&v\text{ even}.
 \end{cases}
 \label{eq:uDef}
\end{equation}
  Since $P_v=1+A_v$, the two weights at a given well sum to $u_v^++u_v^-=(A_v+1)/P_v=1$, so that for either parity of $v$
\begin{equation}
 u_v^-=1-u_v^+.
 \label{eq:uComplement}
\end{equation}
This complementary relation compares the two coefficient expressions in a common formal $(A,B)$ chart. Comparing the physical lateral evaluations also requires applying the DDP action \eqref{eq:DDPaction} to the Voros symbols.

Every nonempty connected set of barriers is a consecutive interval $I=[a,b]$ of the form \eqref{eq:interval}, with outer wells $a$ and $c=b+1$ and internal wells $a+1,\ldots,c-1$.  After division by $P_N$, Eqs.~\eqref{eq:Qplus}--\eqref{eq:QZInus} imply that such an interval carries the normalized weight
\begin{equation}
 w_I^\pm=
 \left(\prod_{e\in I}B_e\right)
 \left(\prod_{v\in V(I)}u_v^\pm\right),
 \label{eq:intervalActivity}
\end{equation}
where $V(I)=\{a,\ldots,c\}$ is the set of wells touched by $I$.  Two intervals can appear in the same term only when they share no well, so $\cZ_N^\pm$ has the form of the grand partition function of a one-dimensional gas of mutually nonoverlapping intervals, not to be confused with the quantum partition function $Z$.

Below, $[X]f$ denotes the coefficient of the monomial $X$ in the expansion of $f$.  The first nontrivial case is the triple well, in which wells $1$, $2$, and $3$ are joined by barriers $1$ and $2$.  Its two single-barrier intervals and one two-barrier interval give
\begin{equation}
 \cZ_3^\pm
 =1+B_1u_1^\pm u_2^\pm
   +B_2u_2^\pm u_3^\pm
   +B_1B_2u_1^\pm u_2^\pm u_3^\pm.
 \label{eq:Z3example}
\end{equation}
Expanding the logarithm,
\begin{align}
 [B_1B_2]\log\cZ_3^\pm
 &=u_1^\pm u_2^\pm u_3^\pm
 -(u_1^\pm u_2^\pm)(u_2^\pm u_3^\pm)
 \nonumber\\
 &=u_1^\pm u_3^\pm
 u_2^\pm(1-u_2^\pm).
 \label{eq:tripleConnected}
\end{align}
The subtraction is exactly the removal of the disconnected product of the two single-barrier sectors.

For a general interval $I=[a,b]$ the same inclusion and exclusion can be carried out as a sum over the cuts made at its internal wells.  A choice of $r-1$ cuts splits $I$ into $r$ consecutive subintervals, and such a splitting enters $\log\cZ_N^\pm$ with the sign $(-1)^{r-1}$.  The subtraction in Eq.~\eqref{eq:tripleConnected} is the case $r=2$.  The sum over cut patterns then factorizes well by well.  An uncut internal well contributes $u_v^\pm$, whereas a cut well belongs to both adjacent subintervals and contributes $-(u_v^\pm)^2$.  Each internal well therefore contributes
\begin{equation}
 u_v^\pm-(u_v^\pm)^2
 =u_v^\pm(1-u_v^\pm)
 =u_v^+u_v^-.
 \label{eq:internalFactorDerivation}
\end{equation}
The two outer wells are never cut, so their factors remain $u_a^\pm$ and $u_c^\pm$.  Hence the coefficient of the monomial in which every barrier of $I$ appears exactly once is
\begin{align}
 C_I^\pm
 &:=[\prod_{e=a}^{b}B_e]\log\cZ_N^\pm
 \nonumber\\
 &=u_a^\pm u_c^\pm
 \prod_{v=a+1}^{c-1}u_v^\pm(1-u_v^\pm)
 =u_a^\pm u_c^\pm
 \prod_{v=a+1}^{c-1}u_v^+u_v^-.
 \label{eq:CIconnected}
\end{align}
Using Eq.~\eqref{eq:uDef},
\begin{equation}
 u_v^+u_v^-=\frac{A_v}{(1+A_v)^2}.
 \label{eq:uvproductA}
\end{equation}
Defining
\begin{equation}
 \mathsf K_I:=\prod_{v=a+1}^{c-1}\frac{A_v}{(1+A_v)^2}
 =\prod_{v=a+1}^{c-1}u_v^+u_v^-,
 \label{eq:KI}
\end{equation}
we get
\begin{equation}
C_I^\pm=\mathsf K_Iu_a^\pm u_c^\pm.
 \label{eq:spectralTarget}
\end{equation}
This is the exact-WKB coefficient that the QZI of the minimal event sequence of $I$ must reproduce.  Its lateral dependence sits entirely in the two outer-well factors $u_a^\pm$ and $u_c^\pm$, because the internal factor $\mathsf K_I$ is built from the products $u_v^+u_v^-$ and is therefore the same on both laterals.

\subsection{Logarithmic resolvent and the partition function}
\label{subsec:logResolvent}

The quantization condition of Sec.~\ref{subsec:lateralQC} is an equation for the energy levels, whereas the Euclidean path integral of Eq.~\eqref{eq:pathintegral} computes the partition function.  The bridge between them is the logarithmic resolvent.  In this subsection $D$ denotes either lateral representative $Q_N^\pm$, or any function differing from it by a multiplicative normalization in the sense of Eq.~\eqref{eq:multiplicativeNormalization}.  We suppress the label $\pm$ wherever a relation holds for both laterals.  If $D(E)$ has simple zeros at the exact levels $E_n$, then locally
\begin{equation}
 -\partial_E\log D(E)
 =\sum_n\frac{1}{E_n-E}+\text{terms analytic at the levels}.
 \label{eq:logResolvent}
\end{equation}
All $N$ wells share the $V_{\min}$ of Eq.~\eqref{eq:minima}, so it is convenient to measure energies from it in units of $\hbar$,
\begin{equation}
 E=V_{\min}+\hbar\epsilon,
 \qquad
 E_n=V_{\min}+\hbar\epsilon_n.
 \label{eq:scaledEnergy}
\end{equation}
The variable $\epsilon$ is the scaled energy.  In these variables, the partition function \eqref{eq:thermaltrace} loses exactly the potential minimum factor noted below Eq.~\eqref{eq:pathintegral}, and the shifted partition function
\begin{equation}
 \widetilde Z(T,\hbar)
 :=\ee^{TV_{\min}/\hbar}Z(T,\hbar)
 =\sum_n\ee^{-T\epsilon_n}
 \label{eq:shiftedTrace}
\end{equation}
is conjugate to $\epsilon$ in the same way that $Z$ is conjugate to $E$.
The scaled logarithmic resolvent then obeys the Laplace representation
\begin{equation}
 -\partial_\epsilon\log D(\epsilon)
 =\int_0^\infty\dd T\;\widetilde Z(T,\hbar)\ee^{T\epsilon},
 \label{eq:resolventLaplace}
\end{equation}
which converges for $\Re\epsilon<\epsilon_0$, with $\epsilon_0$ the lowest scaled level, since each term integrates to $1/(\epsilon_n-\epsilon)$. One qualification is needed. The level sum in Eq.~\eqref{eq:logResolvent} converges for confining polynomial potentials of degree four or higher, whose levels grow faster than linearly. It diverges logarithmically for the equally spaced harmonic reference. A common subtraction may be used for the comparison and is needed for that reference, but it is not a universal requirement on the actual resolvent trace. Any such subtraction amounts to dividing $D$ by a nowhere vanishing factor, so by Eq.~\eqref{eq:normalizationResolvent} it shifts neither the levels nor the residues. In the explicit calculations the reference is the harmonic-oscillator partition function $Z_{\rm SHO}$ of a single well. With that understood, an inverse Laplace transform or {\it Bromwich} integral reconstructs the partition function,
\begin{equation}
 \widetilde Z(T,\hbar)
 =-\frac{1}{2\pi\ii}
 \int_{C_{\rm B}}\dd\epsilon\;
 \ee^{-T\epsilon}\partial_\epsilon\log D(\epsilon,\hbar).
 \label{eq:Bromwich}
\end{equation}
The vertical contour $C_{\rm B}$ lies in the convergence strip $\Re\epsilon<\epsilon_0$, hence to the left of every spectral pole, and for $T>0$ the factor $\ee^{-T\epsilon}$ decays to the right, so the contour is closed that way.  Near a simple zero one has $\partial_\epsilon\log D=(\epsilon-\epsilon_n)^{-1}+\cdots$, so the residue of the integrand at $\epsilon_n$ is $\ee^{-T\epsilon_n}$.  Closing to the right encircles the poles clockwise and supplies a factor $-2\pi\ii$, which the prefactor cancels.  Each simple zero therefore contributes $+\ee^{-T\epsilon_n}$, and Eq.~\eqref{eq:Bromwich} returns Eq.~\eqref{eq:shiftedTrace}.

The logarithmic derivative exposes the spectral poles, while expanding $\log D$, or equivalently $\log\cZ_N^\pm$, in the small barrier weights separates the result into connected nonperturbative sectors, labeled by monomials in the barrier weights, before the inverse Laplace transform is taken.  The interval coefficient $C_I^\pm$ of Eq.~\eqref{eq:spectralTarget} is one such sector.  Since DDP preserves each bion grade by Eq.~\eqref{eq:generalBionTriangularity}, the comparison with the QZI can be made coefficient by coefficient in this connected expansion.

\section{Large-\texorpdfstring{$T$}{T} spectral limit and the Weber regime}
\label{sec:largeTWeber}

Before introducing the QZI, we use the classical Euclidean trajectory attached to the spectral inverse Laplace transform to motivate the energy range in which each well is described by the Weber equation. The classical argument is heuristic, and the rest of the paper uses only its conclusion, namely the range $\epsilon=O(1)$ of low-lying levels.  The argument applies to the general $N$-well potential and does not rely on the double- or triple-well examples.

\subsection{Finite-period Euclidean motion from the WKB period}
\label{subsec:finiteT_WKB_period}

Let $\Gamma$ be a homology class of the complexified spectral curve \eqref{eq:spectralcurve} that carries a regular periodic Euclidean classical trajectory.  Conservation of energy for the action \eqref{eq:EuclideanAction} gives the Euclidean first integral
\begin{equation}
 \frac12\dot x^2=V(x)-E
 \label{eq:EuclideanFirstIntegralPhysics}
\end{equation}
which introduces the spectral energy $E$ and identifies the Euclidean velocity along the orbit with the momentum $p$ of Eq.~\eqref{eq:spectralcurve}.  Define the leading WKB period
\begin{equation}
 \Pi_\Gamma(E)
 :=\oint_\Gamma \sqrt{2\bigl(V(x)-E\bigr)}\,\dd x.
 \label{eq:WKBPeriodPhysics}
\end{equation}
The integrand is the momentum $p$, so $\Pi_\Gamma$ is the tunneling action of the orbit.  We orient $\Gamma$ so that this action is positive on a barrier, which is the orientation of Sec.~\ref{subsec:variables} that makes the nonperturbative contributions exponentially small.  In that orientation the Voros symbol of Eq.~\eqref{eq:VorosLeading} reads
\begin{equation}
 X_\Gamma\sim\ee^{-\Pi_\Gamma(E)/\hbar}.
 \label{eq:VorosFromPeriod}
\end{equation}
Along the orbit $\dd t=\dd x/\sqrt{2(V-E)}$, so differentiating Eq.~\eqref{eq:WKBPeriodPhysics} under the integral gives $\partial_E\Pi_\Gamma=-\oint_\Gamma\dd t$, that is
\begin{equation}
 T_\Gamma(E)=-\partial_E\Pi_\Gamma(E),
 \label{eq:LegendrePeriodPhysics}
\end{equation}
where $T_\Gamma(E)$ is the period of the orbit in the class $\Gamma$.  The same first integral gives the on-shell Euclidean action.  Using $V=\tfrac12\dot x^2+E$ the Lagrangian becomes $\tfrac12\dot x^2+V=\dot x^2+E$, and $\int_0^T\dot x^2\,\dd t=\oint_\Gamma\dot x\,\dd x=\Pi_\Gamma(E)$, so that
\begin{equation}
 S_{E,\Gamma}(T)=\Pi_\Gamma(E)+ET,
 \label{eq:LegendreActionPhysics}
\end{equation}
where $E$ is fixed by Eq.~\eqref{eq:LegendrePeriodPhysics} and the subscript $E$ denotes Euclidean as in Eq.~\eqref{eq:EuclideanAction} rather than the energy.  Equations \eqref{eq:LegendrePeriodPhysics} and \eqref{eq:LegendreActionPhysics} are a Legendre pair exchanging $E$ and $T$.  In the Bromwich integral \eqref{eq:Bromwich} the explicit factor is $\ee^{-T\epsilon}$, while by Eq.~\eqref{eq:VorosFromPeriod} the Voros content of $\log D$ carries $-\Pi_\Gamma(E)/\hbar$, so the integrand goes as
\begin{equation}
 \exp\left[-\frac{1}{\hbar}\bigl(T(E-V_{\min})+\Pi_\Gamma(E)\bigr)\right]
 =\ee^{-S_{E,\Gamma}(T)/\hbar}
 \label{eq:BromwichOnShell}
\end{equation}
up to the vacuum energy factor already removed in Eq.~\eqref{eq:shiftedTrace}.  The spectral Laplace phase is therefore the on-shell Euclidean action itself, and its stationarity $T+\partial_E\Pi_\Gamma(E)=0$ is, by Eq.~\eqref{eq:LegendrePeriodPhysics}, the condition $T=T_\Gamma(E)$ that the orbit in the class $\Gamma$ have period $T$.  At leading semiclassical order the inverse-Laplace saddle is thus the spectral representation of that finite-period orbit, and we now solve this condition in the large-$T$ limit.

The logarithmic sheets that appear in the QZI also have a direct spectral interpretation.  Near a nondegenerate minimum the perturbative cycle $\gamma_v$ of Eq.~\eqref{eq:VorosSymbols} has period
\begin{equation}
 \Pi_{\gamma_v}(E)
 =\frac{2\pi\ii}{\omega_v}(E-V_{\min})+\cdots ,
 \label{eq:vacuumCycleNearMinimumPhysics}
\end{equation}
with $\omega_v$ the harmonic frequency of Eq.~\eqref{eq:minima}.  Let $\ell_v\in\mathbb Z$ count how many times the orbit winds around $\gamma_v$.  Adding such a winding to $\Gamma$ shifts the complex period by
\begin{equation}
 \Delta T_v=-\frac{2\pi\ii \ell_v}{\omega_v}.
 \label{eq:vacuumWindingShiftPhysics}
\end{equation}
In the Weber regime the different logarithmic sheets of the complex separation variables are therefore the image of ordinary integer winding around the perturbative $A$-cycles.  The winding number $\ell_v$ is attached to a well and should not be confused with the barrier multiplicity $k_e$, which counts how often barrier $e$ is crossed in each direction.

\subsection{Large-\texorpdfstring{$T$}{T} approach to the Weber regime}
\label{subsec:largeTWeber}

Fix a homology class $\Gamma$ as in Sec.~\ref{subsec:finiteT_WKB_period}.  Its period diverges logarithmically as the spectral saddle approaches a nondegenerate vacuum.  The reason is that Euclidean motion is governed by the inverted potential, so a minimum $x_v$ of $V$ is a quadratic maximum of $-V$.  By Eq.~\eqref{eq:EuclideanFirstIntegralPhysics} an orbit at energy $E$ turns around where $V=E$, at a distance $\sqrt{2\delta E}/\omega_v$ from that maximum, and it lingers near the turning point for a time of order $\omega_v^{-1}\log(1/\delta E)$.  Writing $\delta E=E-V_{\min}$, the large-period behavior therefore takes the generic form
\begin{equation}
 T_\Gamma(E)
 =\kappa_\Gamma\log\!\left(\frac{E_\Gamma}{\delta E}\right)
 +c_\Gamma+O(1),
 \qquad \delta E\to0^+,
 \label{eq:generalLogPeriodPhysics}
\end{equation}
in which the time the orbit spends near each well $v$ contributes $\omega_v^{-1}$ to the coefficient $\kappa_\Gamma>0$. This has the same logarithmic scale as the interaction threshold of Sec.~\ref{subsec:regime}, but solves a different stationarity problem. The constant $E_\Gamma$ is fixed by the global shape of the orbit and its only role is to make the logarithm dimensionless, and $c_\Gamma$ is the finite part of the large-period expansion. Solving $T=T_\Gamma(E)$ for $\delta E$ gives $\delta E=E_\Gamma\exp[-(T-c_\Gamma)/\kappa_\Gamma]$, that is
\begin{equation}
 E^{*}_\Gamma(T)-V_{\min}
 =C_\Gamma\exp\!\left[-\frac{T}{\kappa_\Gamma}\right]
 \bigl(1+ O(1)\bigr),
 \label{eq:generalSaddleApproachPhysics}
\end{equation}
with $C_\Gamma=E_\Gamma\exp(c_\Gamma/\kappa_\Gamma)$ up to the indicated $O(1)$ corrections.  Here $E^{*}_\Gamma(T)$ is the position of the inverse-Laplace saddle at period $T$, and it should not be confused with the constant scale $E_\Gamma$ appearing inside the logarithm.  The saddle therefore approaches the minimum point of the well exponentially in $T$. This classical estimate is reliable only while $E^*_\Gamma-V_{\min}\gg\hbar$. Once the offset reaches the quantum scale $\hbar\omega_v$, the saddle meets the low-lying levels, and for larger $T$ the Bromwich integral \eqref{eq:Bromwich} is controlled by the spectral poles at $\epsilon=O(1)$ rather than by an isolated saddle. The estimate therefore does not locate the dominant contribution at large $T$, but it shows that the relevant energies are driven into the range $\epsilon=O(1)$, which is all that is used below.

In the region near the potential minimum, the natural energy variable is already the scaled energy $\epsilon$ of Eq.~\eqref{eq:scaledEnergy}, and the range of interest is $\epsilon=O(1)$, that is $E-V_{\min}=O(\hbar)$.  It remains to choose the matching local coordinate around the well,
\begin{equation}
 x-x_v=\sqrt{\hbar}\,y,
 \label{eq:WeberScalingPhysics}
\end{equation}
whose power of $\hbar$ is fixed by requiring the harmonic energy $\tfrac12\omega_v^2(x-x_v)^2$ to be of order $\hbar$ as well.  Expanding $V$ about the minimum as $V=V_{\min}+\tfrac12\omega_v^2(x-x_v)^2+O\bigl((x-x_v)^3\bigr)$ and dividing the eigenvalue equation for the Hamiltonian~\eqref{eq:Hamiltonian} by $\hbar$, the kinetic term, the harmonic term, and the energy all become of the same order under Eqs.~\eqref{eq:scaledEnergy} and \eqref{eq:WeberScalingPhysics}, and the local equation turns into
\begin{equation}
 \left[-\frac12\partial_y^2
 +\frac12\omega_v^2 y^2-\epsilon
 +O(\sqrt{\hbar})\right]\psi(y)=0,
 \label{eq:LocalWeberEquationPhysics}
\end{equation}
where the $O(\sqrt{\hbar})$ error comes from the cubic term of the expansion.  The two local turning points sit at $y=\pm\sqrt{2\epsilon}/\omega_v$, so they are separated by $O(1)$ in the Weber variable and by $O(\sqrt{\hbar})$ in $x$.  They have therefore merged into a double turning point, and the uniform local description is the Weber (parabolic-cylinder) equation rather than the Airy equation appropriate to isolated simple turning points \cite{BehtashHomological}.

Equating the exponentially small energy offset $E^*_\Gamma-V_{\min}$ of Eq.~\eqref{eq:generalSaddleApproachPhysics} with the local quantum scale $\hbar\omega_v$ gives $T=\kappa_\Gamma\log\bigl[C_\Gamma/(\hbar\omega_v)\bigr]$, that is the parametric crossover
\begin{equation}
 T_{\rm AW,\Gamma}
 \sim
 \kappa_\Gamma
 \log\!\left(\frac{E_\Gamma}{\hbar\omega_v}\right),
 \label{eq:generalAiryWeberCrossoverPhysics}
\end{equation}
up to the finite $O(1)$ term $c_\Gamma$, where the subscript ${\rm AW}$ stands for Airy--Weber.  The large-$T$ inverse-Laplace problem therefore enters the Weber regime parametrically before the resummation scale $T_*$ of Eq.~\eqref{eq:largeTPhysicsWindow} is reached.

\subsection{Airy--Weber matching of the DDP action}
\label{subsec:airyWeberDDP}

Section~\ref{subsec:largeTWeber} placed the large-$T$ saddle in the Weber regime, so the appropriate local description of each well is the Weber one.  Let $\mathcal D_{\rm AW}$ denote the corresponding Airy-to-Weber reduction \cite{SueishiExactWKBII}.  It replaces the generic Airy-type description of an elementary well/barrier block, built from the connection formulas at its simple turning points, by the Weber-type description, in which the two turning points of the well are close together and the local equation is the parabolic-cylinder (Weber) equation \eqref{eq:LocalWeberEquationPhysics}.  Applied barrier by barrier, this reduction sends the barrier weight $B_e$ of Eq.~\eqref{eq:VorosSymbols} to
\begin{equation}
 \mathcal D_{\rm AW}[B_e]
 =\mathfrak q_e\,U_{W,e}(\epsilon,\hbar),
 \qquad U_{W,e}\neq0 .
 \label{eq:generalAWBarrierPhysics}
\end{equation}
Here the Weber regime is the range $\epsilon=O(1)$ identified in Sec.~\ref{subsec:largeTWeber}, on which each well is described by the Weber equation \eqref{eq:LocalWeberEquationPhysics}. By Eq.~\eqref{eq:generalSaddleApproachPhysics} the relevant energies enter that energy range once $T$ exceeds the crossover \eqref{eq:generalAiryWeberCrossoverPhysics}, so Eq.~\eqref{eq:generalAWBarrierPhysics} is the appropriate description over the whole upper part of the time window \eqref{eq:largeTPhysicsWindow}. At leading exponential order $\mathfrak q_e=\ee^{-2S_{I,e}/\hbar}$ is the classical bion weight of Eq.~\eqref{eq:BfromInstantonPair}, the leading exponential of the barrier weight $B_e$. The factor $U_{W,e}$ is the accompanying local Weber dressing, which is nonvanishing on that window and whose normalization does not affect the bion grading. Its absolute value matters for coefficient matching in Sec.~\ref{subsec:weberMatching}. Raising Eq.~\eqref{eq:generalAWBarrierPhysics} to the barrier multiplicities $k_e$, the number of crossings of barrier $e$ in each direction, gives
\begin{equation}
 \prod_e B_e^{k_e}
 \longmapsto
 \prod_e \mathfrak q_e^{k_e}U_{W,e}^{k_e},
 \label{eq:generalAWGradePhysics}
\end{equation}
so the Airy-to-Weber reduction preserves the bion grade exactly.

The same reduction also commutes with the DDP action.  We write $\cS_0^{\rm DDP,Ai}$ and $\cS_0^{\rm DDP,W}$ for the Stokes automorphism of Sec.~\ref{subsec:lateralQC} expressed in the Airy-type and Weber-type descriptions, respectively.  To leading order in the Weber-type description at fixed bion grade,
\begin{equation}
 \mathcal D_{\rm AW}\circ\cS_0^{\rm DDP,Ai}
 =\cS_0^{\rm DDP,W}\circ\mathcal D_{\rm AW}.
 \label{eq:generalAWDDPCovariancePhysics}
\end{equation}
Equation~\eqref{eq:generalAWDDPCovariancePhysics} states that taking the Stokes jump and then reducing to the Weber-type description agrees with reducing first and then taking the jump, so the DDP action of Eq.~\eqref{eq:DDPaction} may be read in either description.

\subsection{Fixed-grade large \texorpdfstring{$T$}{T} versus the zero-temperature limit}
\label{sec:fixedGradeLargeT}

Section~\ref{subsec:regime} bounded $T$ from below by the dilute-gas scale $T_{\rm QZI}(K,\hbar)$ and from above by the resummation scale $T_*$, and Sec.~\ref{subsec:largeTWeber} added the Airy--Weber crossover $T_{\rm AW,\Gamma}$. At fixed grade both lower scales are of order $\log(1/|\hbar|)$, with $T_{\rm QZI}$ growing with $K$ and $T_{\rm AW,\Gamma}$ with $\kappa_\Gamma$, so it suffices to work above the larger of the two. The upper scale is exponentially larger, so the window \eqref{eq:largeTPhysicsWindow} is exponentially wide.

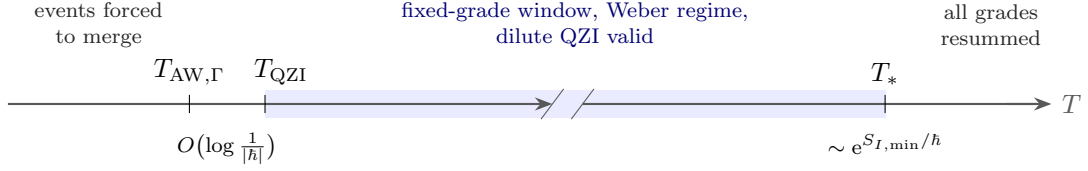
\begin{figure}[t]
\centering
\begin{tikzpicture}[>=Stealth]
  \fill[blue!8] (3.4,-0.18) rectangle (11.6,0.18);
  \draw[->,thick,gray!70!black] (0,0) -- (7.2,0);
  \draw[thick,gray!70!black] (7.6,0) -- (13.6,0);
  \draw[->,thick,gray!70!black] (13.2,0) -- (13.8,0) node[right,font=\small] {$T$};
  \draw[gray!70!black] (7.05,-0.2) -- (7.35,0.2);
  \draw[gray!70!black] (7.45,-0.2) -- (7.75,0.2);
  \draw (2.4,-0.12) -- (2.4,0.12) node[above,font=\small] {$T_{\rm AW,\Gamma}$};
  \draw (3.4,-0.12) -- (3.4,0.12) node[above,font=\small,xshift=6pt] {$T_{\rm QZI}$};
  \draw (11.6,-0.12) -- (11.6,0.12) node[above,font=\small] {$T_*$};
  \node[below,font=\scriptsize,align=center] at (2.9,-0.25) {$O\bigl(\log\frac1{|\hbar|}\bigr)$};
  \node[below,font=\scriptsize,align=center] at (11.6,-0.25) {$\sim\ee^{S_{I,\min}/\hbar}$};
  \node[font=\scriptsize,align=center,gray!30!black] at (1.2,1.05) {events forced\\ to merge};
  \node[font=\scriptsize,align=center,blue!45!black] at (7.5,1.05) {fixed-grade window, Weber regime,\\ dilute QZI valid};
  \node[font=\scriptsize,align=center,gray!30!black] at (13.0,1.05) {all grades\\ resummed};
\end{tikzpicture}
\caption{The scales of the period $T$ at fixed grade. The Airy--Weber crossover $T_{\rm AW,\Gamma}$ of Eq.~\eqref{eq:generalAiryWeberCrossoverPhysics} and the dilute-gas scale $T_{\rm QZI}$ of Sec.~\ref{subsec:regime} are both logarithmic in $1/\hbar$, in either order, while the resummation scale $T_*\sim\Delta_{\rm np}^{-1}$ of Eq.~\eqref{eq:splittingScalePhysics} is exponentially larger, as indicated by the break in the axis. The shaded region denotes the sufficient fixed-grade window \eqref{eq:fixedGradeBreakdownPhysics}. The upper scale drawn is the one-instanton estimate for degenerate neighboring levels, whereas nondegenerate cases can have a later breakdown time.}
\label{fig:timeWindow}
\end{figure}

For degenerate adjacent levels the scaled splitting is of one-instanton order. More generally, a conservative upper estimate on the size of the relevant nonperturbative scaled correction is, up to powers of $\hbar$,
\begin{equation}
 \Delta_{\rm np} \lesssim\ee^{-S_{I,\min}/\hbar},
 \qquad
 S_{I,\min}:=\min_e S_{I,e},
 \label{eq:splittingScalePhysics}
\end{equation}
with $S_{I,e}$ being the one-instanton actions of Sec.~\ref{subsec:variables}. The smallest barrier action supplies a conservative one-instanton scale. When neighboring perturbative levels are not degenerate, the leading spectral correction can start at bion or higher order, so the actual breakdown time can be later. It is therefore sufficient, at the level of exponential scales, to impose
\begin{equation}
 \max\bigl(T_{\rm QZI}, T_{\rm AW,\Gamma}\bigr)
 =O\!\left(\log\frac{1}{|\hbar|}\right)
 \ \lesssim\ T\ \ll\  \ee^{S_{I,\min}/\hbar},
 \label{eq:fixedGradeBreakdownPhysics}
\end{equation}
with the appropriate algebraic prefactors included when a quantitative upper bound is needed. 
A logarithm at one end and an exponential at the other leave a parametrically broad semiclassical window in which the Weber description of Sec.~\ref{subsec:airyWeberDDP} is already mandatory while the fixed-grade expansion is still uniform. All fixed-grade large-$T$ results below hold in this window, and the three scales are compared in Fig.~\ref{fig:timeWindow}. The strict zero-temperature limit sends $T\to\infty$ at fixed $\hbar$ and therefore leaves it, so the grades must be resummed before that limit is taken.

\section{From the Euclidean path integral to QZI periods}
\label{sec:fullPI}

Sections~\ref{sec:spectralTarget} and \ref{sec:largeTWeber} worked with the exact-WKB description of the spectrum.  We now construct the object to be compared with the connected coefficient $C_I^\pm$ of Eq.~\eqref{eq:spectralTarget}, namely the integral over the collective coordinates of the instanton events and its reduction to Gamma and Hankel periods. The dilute multi-event configuration is first embedded in the path integral, its normal directions are then integrated out on a regular patch, and the resulting QZI is finally factorized by a Laplace transform.  Each step has a standard counterpart in the multi-instanton calculus \cite{Bogomolny1980,ZinnJustin1981,ZinnJustin1983,ZinnJustinJentschuraI}.  Appendix~\ref{app:QZI} rederives the QZI in that conventional language, with collective coordinates, Gaussian integration, and an explicit tail interaction, works out the double-well bion as an example, and gives references for each step.

From here on, everything stays inside one regular dilute neighborhood, at the fixed finite grade and in the time window \eqref{eq:largeTPhysicsWindow} of Sec.~\ref{subsec:regime}, and to leading one-loop order in the normal directions.  As in Eq.~\eqref{eq:separations}, the index $r$ labels both a separation and the visit to a well that it spans, and a closed event sequence with $m$ events has exactly $m$ of each.  Two scales control the approximations,
\begin{equation}
 R:=\min_r\Re\tau_r\gg1,
 \qquad
 \omega_{\min}:=\min_v\omega_v,
 \label{eq:diluteScales}
\end{equation}
so that $R$ is the shortest separation in the configuration and $\omega_{\min}$ the slowest harmonic decay rate of Eq.~\eqref{eq:minima}, and every exponentially small error below is of the form $O(\ee^{-\omega_{\min}R})$. The reduction then rests on three regularity conditions.
\begin{itemize}[leftmargin=2em]
    \item[(1)] The separated translation modes stay independent.  Their Gram matrix \eqref{eq:GramMatrixPhysics}, formed with the complex-bilinear pairing of the complexified path space, remains invertible, so in particular the continued translation modes do not become isotropic.
    \item[(2)] The multi-instanton configuration is deformed only inside this regular region, so the split into collective and normal directions does not drop rank.
    \item[(3)] The reduced normal Hessian stays nondegenerate. On a simply connected patch free of zeros of its determinant, the normal rapid-decay generator and the branch of the square root of the determinant can be fixed once and continued without an extra sign, phase, or monodromy.\footnote{A rapid-decay cycle is an integration contour whose ends run into regions where the integrand is exponentially small, so that the exponential integral along it converges.  Since the normal directions enter only through a nondegenerate quadratic form, they carry exactly one such contour up to orientation, and we call it the {\it normal rapid-decay generator}.  Its period is proportional to the inverse square root of the determinant of that quadratic form.}
\end{itemize}
 We do not control the region $\tau_r\to0$ where two events merge. This is the main reason for working with a dilute gas of widely separated events, for which the only residual interaction is the exponentially small overlap of the instanton tails. Nor do we treat configurations at which the fluctuation operator becomes degenerate, which are the zeros of the determinant excluded by condition (3), or at which the split into collective and normal directions drops rank, which is excluded by condition (2). 

\subsection{Events and dilute gas approximation}
\label{subsec:cores}

The equation of motion is
\begin{equation}
 -\ddot x+V'(x)=0.
 \label{eq:EuclideanEOM}
\end{equation}
For equal-depth neighboring minima it has zero-energy heteroclinic solutions interpolating between adjacent wells.  These are the instanton profiles of Sec.~\ref{subsec:variables}, written here as $q_e(t-t_r)$ for an event centered at $t_r$.  Translating an isolated instanton is an exact zero mode.  In a multi-event configuration the relative instanton positions become weakly lifted quasi-zero modes because the exponentially small tails overlap, and it is these lifted modes that we need to integrate over.

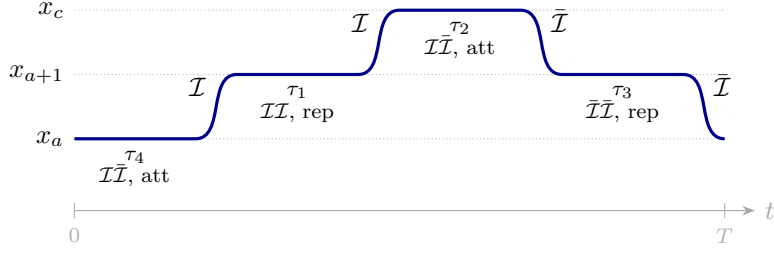
\begin{figure}[t]
\centering
\begin{tikzpicture}[>=Stealth,
  traj/.style={very thick,blue!55!black},
  wellline/.style={gray!55,densely dotted},
]
\begin{scope}
  \draw[->,gray!70] (0,-0.95) -- (9.0,-0.95) node[right] {\small$t$};
  \foreach \y/\l in {0.0/a,0.85/{a{+}1},1.7/c} {\draw[wellline] (0,\y) -- (8.60,\y); \node[left,font=\small] at (0,\y) {$x_{\l}$};}
  \draw[gray!60] (0,-1.05) -- (0,-0.85) node[pos=0,below,font=\scriptsize] {$0$};
  \draw[gray!60] (8.60,-1.05) -- (8.60,-0.85) node[pos=0,below,font=\scriptsize] {$T$};
  \draw[traj] (0,0.000) -- (1.600,0.000) to[out=0,in=180] (2.150,0.850) -- (3.750,0.850) to[out=0,in=180] (4.300,1.700) -- (5.900,1.700) to[out=0,in=180] (6.450,0.850) -- (8.050,0.850) to[out=0,in=180] (8.600,0.000);
  \node[font=\small,above left] at (1.875,0.425) {$\cI$};
  \node[font=\small,above left] at (4.025,1.275) {$\cI$};
  \node[font=\small,above right] at (6.175,1.275) {$\bar\cI$};
  \node[font=\small,above right] at (8.325,0.425) {$\bar\cI$};
  \node[below,font=\scriptsize,align=center] at (0.800,-0.020) {$\tau_4$\\[-1pt]$\cI\bar\cI$, att};
  \node[below,font=\scriptsize,align=center] at (2.950,0.830) {$\tau_1$\\[-1pt]$\cI\cI$, rep};
  \node[below,font=\scriptsize,align=center] at (5.100,1.680) {$\tau_2$\\[-1pt]$\cI\bar\cI$, att};
  \node[below,font=\scriptsize,align=center] at (7.250,0.830) {$\tau_3$\\[-1pt]$\bar\cI\bar\cI$, rep};
\end{scope}
\end{tikzpicture}
\caption{Trajectory $x(t)$ of the minimal event sequence \eqref{eq:primitiveWalk} of a two-barrier interval $I=[a,b]$, labeled as in Fig.~\ref{fig:wells}, on the circle of circumference $T$, with the endpoints $t=0$ and $t=T$ identified.  Each event is an instanton $\cI$ or an anti-instanton $\bar\cI$, and each separation $\tau_r$ spans one visit to a well.  At the outer wells the trajectory turns back and the separation is an attractive $\cI\bar\cI$ pair, while at the internal well it passes through twice, once as an $\cI\cI$ pair and once as a $\bar\cI\bar\cI$ pair, both repulsive.}
\label{fig:intervalSequence}
\end{figure}

For the interval $I=[a,b]$ of Eq.~\eqref{eq:interval}, the minimal closed event sequence crosses each barrier of $I$ exactly once in each direction.  It starts at the outer well $a$, runs out to the other outer well $c$, and returns,
\begin{equation}
 a\to a+1\to\cdots\to c\to c-1\to\cdots\to a.
 \label{eq:primitiveWalk}
\end{equation}
After removing the common translation of all events around the circle, the remaining collective coordinates are the long separations $\tau_r$, one for each visit to a well.  A visit occupies the separation between the two events that bound it, and the tail interaction across that separation is fixed by which of two types the visit is.  We fix both the names and the labels here once for the whole paper.
\begin{itemize}[leftmargin=2em]
\item \textbf{An $\cI\bar\cI$ pair, counted by $\mathbf R$ and labeled ${\rm att}$.} The incoming transition is immediately reversed, as in $v-1\to v\to v-1$, so the two events bounding the separation are an instanton and an anti-instanton, and the pair carries no net topological charge.  The two tails approach the minimum from the same side, and the leading interaction is attractive.  The outer wells $a$ and $c$ of Eq.~\eqref{eq:primitiveWalk} are visited this way.
\item \textbf{An $\cI\cI$ or $\bar\cI\bar\cI$ pair, counted by $\mathbf X$ and labeled ${\rm rep}$.} The trajectory enters through one barrier and leaves through the other, as in $v-1\to v\to v+1$, so the two bounding events carry the same charge.  The tails approach from opposite sides, and the leading interaction is repulsive.  Every internal well $a+1,\ldots,c-1$ of the interval is visited twice this way.
\end{itemize}
The minimal event sequence therefore has exactly two attractive separations, one at each outer well, and two repulsive separations per internal well. Only the attractive separations will turn out to carry a lateral ambiguity, so only the outer wells reproduce the lateral perturbative oscillations $u_v^\pm$ of Eq.~\eqref{eq:uDef}.  Figure~\ref{fig:intervalSequence} shows the minimal event sequence of a two-barrier interval with both kinds of visit.

Consider a general closed event sequence
\begin{equation}
 v_0\xrightarrow{e_1}v_1\xrightarrow{e_2}\cdots
 \xrightarrow{e_m}v_m=v_0,
 \label{eq:closedSequence}
\end{equation}
with event centers $t_r$ and separations $\tau_r=t_{r+1}-t_r$ as in Eq.~\eqref{eq:separations}.  We write $\omega_r:=\omega_{v_r}$ for the harmonic frequency of the well visited during the separation $r$.  A convenient periodized dilute gas ansatz is
\begin{equation}
 q^{(0)}(\boldsymbol\tau;t)
 =
 x_{v_0}
 +\sum_{n\in\mathbb Z}\sum_{r=1}^{m}
 \Bigl[q_{e_r}(t-nT-t_r)-x_{v_{r-1}}\Bigr],
 \label{eq:periodizedAnsatz}
\end{equation}
where the sum over $n$ periodizes the configuration on the circle of circumference $T$.  Because the separations are large, this ansatz is accurate in two complementary regions.  Near any one event the other events contribute only exponentially small tails, and along any one separation the configuration is exponentially close to the corresponding minimum.  Let $\sigma$ measure time from the start of the separation $r$, of length $\tau_r$ near the well $x_{v_r}$, and let $\xi$ denote the deviation from that minimum.  The tails reaching into the separation from the two events bounding it are then
\begin{equation}
 \xi_-\sim \alpha_-\ee^{-\omega_r\sigma},
 \qquad
 \xi_+\sim \alpha_+\ee^{\omega_r(\sigma-\tau_r)},
 \label{eq:tails}
\end{equation}
with $\alpha_\pm$ being the amplitudes of the two decaying tails.  Each tail on its own satisfies the linearized equation of motion, so the first term not canceled in the Euler--Lagrange equation is the overlap $\xi_-\xi_+=\alpha_-\alpha_+\ee^{-\omega_r\tau_r}$.  This product is independent of $\sigma$, so the mismatch is spread uniformly along the separation and is controlled by the separation length alone.  The estimate
\begin{equation}
 \left\|-\ddot q^{(0)}+V'(q^{(0)})\right\|_{L^2}
 =O(\ee^{-\omega_{\min} R})
 \label{eq:smallDefect}
\end{equation}
is therefore uniform in the configuration.  These configurations therefore fail to solve the equation of motion only by an exponentially small amount.  It is an approximate saddle manifold parametrized by the separations $\boldsymbol\tau$, and the residual tail interaction measured by Eq.~\eqref{eq:smallDefect} lifts the quasi-zero modes and generates the QZI exponent.

\subsection{Collective and normal fluctuations}
\label{subsec:collectiveNormal}

The fluctuation operator about the multi-event gas is
\begin{equation}
 \cO_{\boldsymbol\tau}
 =-\partial_t^2+V''(q^{(0)}(\boldsymbol\tau;t)).
 \label{eq:fluctuationOperator}
\end{equation}
Each isolated event has a translation zero mode $\dot q_e$.  In the standard treatment of a single instanton this zero mode is traded for a collective coordinate by the Faddeev--Popov procedure \cite{GervaisSakita,VZNS,Coleman1985,MarinoBook2015}, summarized in App.~\ref{app:collective}, and the construction below is its version for $m$ events on the complexified path space.  For a real monotone one-dimensional heteroclinic, $\dot q_e$ has no nodes.  By the Sturm--Liouville ordering, it is therefore the simple ground-state zero mode of the elementary Hessian, while the Hessian away from this translation mode has a nonzero gap.  This ordering holds in the real domain.  The QZI requires the analytic continuation of Eq.~\eqref{eq:fluctuationOperator} into the complex domain, so conditions (1) to (3) are imposed rather than derived.

For $m$ well-separated events, these translation modes become an $m$-dimensional near-zero block, lifted from zero due to the small overlap \eqref{eq:smallDefect}.  Choose localized right modes $\Phi_j$ and bilinear-dual localized modes $\Psi_j$ near the $j$th event, and define the Gram matrix,
\begin{equation}
 G_{ij}=\int_0^T\Psi_i(t)\Phi_j(t)\,\dd t
 =\delta_{ij}+O(\ee^{-\omega_{\min} R}).
 \label{eq:GramMatrixPhysics}
\end{equation}
The off-diagonal entries are exponentially small because the modes are localized at events separated by at least $R$.  The diagonal entries are the point at which condition (1) is needed, since on a complex branch the pairing is bilinear rather than Hermitian and $\langle\Psi_j,\Phi_j\rangle$ could in principle vanish.  Condition (1) excludes this.  Granting it, $G=I+O(\ee^{-\omega_{\min} R})$ is invertible for large enough $R$, and the explicit finite-rank projector onto the collective fluctuations is then given by
\begin{equation}
 P_R f
 =\sum_{i,j=1}^{m}\Phi_i(G^{-1})_{ij}
   \int_0^T\Psi_j(t)f(t)\,\dd t,
 \qquad
 Q_R:=1-P_R.
 \label{eq:ProjectorExplicit}
\end{equation}
The normal operator is defined as
\begin{equation}
 N_R:=Q_R\cO_{\boldsymbol\tau}Q_R.
 \label{eq:normalOperator}
\end{equation}
On each event we use the reduced inverse of the elementary heteroclinic Hessian, and on each separation the massive inverse $(-\partial_t^2+\omega_r^2)^{-1}$ is used. On a patch where the elementary reduced inverses and their analytic continuations have uniform bounds, patching with cutoffs of width $\ell_{\rm cut}$ gives an approximate inverse, or {\it parametrix}, $A_R$. Control of the constants on the complex patch is part of the regularity assumption, and the errors then obey the schematic estimate
\begin{equation}
 \|N_RA_R-1\|+\|A_RN_R-1\|
 \le C\left[\ell_{\rm cut}^{-1}+\ee^{-\omega_{\min}\ell_{\rm cut}}
 +\ee^{-\omega_{\min}(R-2\ell_{\rm cut})}\right].
 \label{eq:parametrixError}
\end{equation}
The first term wants a wide cutoff and the last a narrow one, and choosing $\ell_{\rm cut}=R/3$ balances the two exponentials, leaving a bound that is small when $R\gg1$. The Neumann series then yields a bounded inverse of $N_R$, as condition (3) requires. This explains how invertibility can be maintained under the stated regular-patch assumptions, and no positive spectral gap is assumed, since on a complex domain the operator is not self-adjoint.

The projected stationary equation
\begin{equation}
 Q_R[-\ddot q+V'(q)]=0,
 \qquad q=q^{(0)}+\eta,
 \label{eq:projectedStationary}
\end{equation}
can then be solved uniquely by the implicit-function/Newton argument,
\begin{equation}
 \eta_*(\boldsymbol\tau)=O(\ee^{-\omega_{\min} R}).
 \label{eq:normalCorrection}
\end{equation}
Having solved the normal directions in this way, the action splits, after a regulator is chosen, into the collective separations $\boldsymbol\tau$ and the normal coordinates $\zeta$ transverse to them,
\begin{equation}
 S_{\rm reg}(\boldsymbol\tau,\zeta)
 =S_{\rm coll}(\boldsymbol\tau)+Q_\perp(\boldsymbol\tau;\zeta),
 \label{eq:MorseSplit}
\end{equation}
with the normal quadratic form $Q_\perp$ nondegenerate by condition (3).  The Gaussian integration over $\zeta$ therefore produces a nonzero one-loop determinant factor and leaves the integral over $\boldsymbol\tau$.  In one dimension the regulated determinant can be represented by Gel'fand--Yaglom/Forman-type formulas \cite{GelfandYaglom,DunneDeterminants,BehtashInfinity,Forman1987,Forman1992Erratum}.  This factor can modify the scalar normalization and phase.  Once its square-root branch and one normal rapid-decay generator are fixed as in condition (3), however, it cannot alter the integer coefficients relating the collective contours, such as the Picard--Lefschetz relation between the two lateral contours of an $\cI\bar\cI$ separation.  The Stokes analysis can therefore be carried out entirely on the collective integral.

At any fixed finite grade the same construction applies sequence by sequence.  Let $w$ denote a finite ordered closed event sequence.  Write $\mathcal H_w^{\rm PI,tr}$ for the lattice of integration cycles on a neighborhood constructed by the normal exponential map around the multi-instanton configuration in the path space of $w$, and $\mathcal H_w^{\rm QZI,occ}$ for the corresponding lattice of collective-coordinate cycles, where the superscript ${\rm occ}$ means that separate occurrences of a well are kept distinct.  Both are lattices of rapid-decay cycles on the regular patch of this section, and neither is a global path-space construction.  On a simply connected regular dilute patch, after fixing one generator of the nondegenerate normal rapid-decay fiber, the cycle lattice factorizes as
\begin{equation}
 \mathcal H_{w}^{\rm PI,tr}
 \simeq
 \mathfrak L_{\perp,w}^{\rm rd}\otimes
 \mathcal H_{w}^{\rm QZI,occ}.
 \label{eq:physicsRegularDirectImage}
\end{equation}
Here $\mathfrak L_{\perp,w}^{\rm rd}$ is the rank-one normal factor and $\mathcal H_{w}^{\rm QZI,occ}$ keeps the occurrence data of the collective integral.  Equation~\eqref{eq:physicsRegularDirectImage} says that the Stokes jump of the regular patch sits in the collective QZI factor, while the normal directions contribute only a single nonvanishing period.  The reason is a rank count.  A rank-one lattice has a single generator, so a Stokes crossing can at most rescale it and has no second generator to admix, which leaves the collective factor as the only place a jump can act.  The factorization is limited to the dilute gas approximation, under the conditions listed at the start of this section and used to reach Eq.~\eqref{eq:MorseSplit}.

\subsection{Effective QZI action from zero-point energies and tail interactions}
\label{subsec:QZIaction}

Over the separation $r$ the trajectory sits near the minimum $x_v$ of the well $v$ it visits, with frequency $\omega_r=\omega_v$, and the fluctuations stem from the harmonic oscillator $\widehat H_r$ of that well.  For $\omega_r\tau_r\gg1$ the separation contributes the propagator of $\widehat H_r$ from the minimum back to it,
\begin{equation}
 K_r(0,0;\tau_r)
 =\left(\frac{\omega_r}{\pi\hbar}\right)^{1/2}
 \ee^{-\omega_r\tau_r/2}
 \left[1+O(\ee^{-2\omega_r\tau_r})\right].
 \label{eq:MehlerAsymptotic}
\end{equation}
We give the definition and brief derivation of this equation in App~\ref{app:harmonic}.  The factor $\ee^{-\omega_r\tau_r/2}$ is the decay of the oscillator ground state, so a long separation simply pays the zero-point energy $\tfrac12\hbar\omega_r$ of the visited well for its duration.  All wells are compared against one common reference factor $\ee^{-\omega_{\rm ref}\tau_r/2}$, where $\omega_{\rm ref}$ is an arbitrary fixed frequency whose only purpose is to remove the part shared by every separation.  Dividing Eq.~\eqref{eq:MehlerAsymptotic} by it leaves the relative factor
\begin{equation}
 \ee^{-\varrho_r\tau_r},
 \qquad
 \varrho_r=\frac{\omega_r-\omega_{\rm ref}}{2}
 \label{eq:rhoDef}
\end{equation}
for the lowest harmonic level of the well.  This is the term linear in $\tau_r$ in the QZI exponent below.  Even for minima of equal depth, wells of different curvature have different zero-point energies and hence different weights over long separations.

Using the two tails of Eq.~\eqref{eq:tails}, the energy in the quadratic overlap region is
\begin{align}
 \mathcal H_E
 &=\frac12\dot\xi^2-\frac12\omega_r^2\xi^2
 \nonumber\\
 &=-2\omega_r^2\alpha_-\alpha_+\ee^{-\omega_r\tau_r}
 +O(\ee^{-2\omega_r\tau_r}).
 \label{eq:EuclideanStress}
\end{align}
For an on-shell segment, varying the separation changes the action according to
\begin{equation}
 \frac{\partial S}{\partial\tau_r}=-\mathcal H_E.
 \label{eq:actionVariation}
\end{equation}
Integrating from infinite separation, where the interaction vanishes, gives
\begin{equation}
 S_{\rm int}(\tau_r)
 =-2\omega_r\alpha_-\alpha_+\ee^{-\omega_r\tau_r}
 +O(\ee^{-2\omega_r\tau_r}).
 \label{eq:tailInteraction}
\end{equation}
This is the classic instanton--anti-instanton interaction \cite{Bogomolny1980,ZinnJustin1981,ZinnJustinJentschuraI}, and App.~\ref{app:tail} derives it directly from the action of two overlapping profiles.  The sign of the product $\alpha_-\alpha_+$ distinguishes the two separation types of Sec.~\ref{subsec:cores}.  For a $\cI\bar\cI$ pair the trajectory enters the well and turns around, so both tails lie on the same side of the minimum and $\alpha_-\alpha_+>0$, which makes Eq.~\eqref{eq:tailInteraction} negative.  This is the attractive case labeled ${\rm att}$.  For a $\cI\cI$ or $\bar\cI\bar\cI$ pair the trajectory passes through, so the two tails lie on opposite sides of the minimum and $\alpha_-\alpha_+<0$, which makes Eq.~\eqref{eq:tailInteraction} positive.  This is the repulsive case labeled ${\rm rep}$.  Dividing by $\hbar$ gives the exponential interaction term in the QZI exponent, with $\lambda_r=-1$ in the first case and $\lambda_r=+1$ in the second.

At leading dilute order the collective-coordinate exponent has the form
\begin{equation}
 \cW_I(\boldsymbol\tau)
 =\sum_r\varrho_r\tau_r
 +\sum_r\lambda_r \Lambda_r\ee^{-\omega_r\tau_r},
 \qquad \lambda_r=\pm1,
 \label{eq:QZIaction}
\end{equation}
where $\lambda_r=-1$ for an attractive separation and $\lambda_r=+1$ for a repulsive one, and $\Lambda_r>0$ is the magnitude $2\omega_r|\alpha_-\alpha_+|/\hbar$ of the tail interaction \eqref{eq:tailInteraction}, so that $\Lambda_r\propto\hbar^{-1}$ at leading semiclassical order.  The first term of Eq.~\eqref{eq:QZIaction} is the relative zero-point energy of the intervening well and comes from Eq.~\eqref{eq:rhoDef}, while the second is the leading tail interaction and comes from Eq.~\eqref{eq:tailInteraction}.  Both were obtained from the regular dilute gas approximation of the path integral.

At fixed circumference $T$, the QZI has the form
\begin{equation}
 \mathscr I_I(T)
 =\mathcal N_I\int\prod_r\dd\tau_r\;
 \delta\left(T-\sum_r\tau_r\right)
 \ee^{-\cW_I(\boldsymbol\tau)},
 \label{eq:fixedTQZI}
\end{equation}
where $\mathcal N_I$ contains the event actions, the collective-coordinate Jacobian, and the one-loop factors that do not depend on the separations. The Stokes data of the sector are therefore those of the integral \eqref{eq:fixedTQZI}.

It remains to assemble these two terms with the normal reduction of Sec.~\ref{subsec:collectiveNormal}, which left the determinant and measure prefactor $\mathcal A_{\perp,I}(\boldsymbol\tau;\hbar)$ multiplying the classical collective weight $\ee^{-S_{\rm coll}(\boldsymbol\tau)/\hbar}$ of Eq.~\eqref{eq:MorseSplit}.  Each factor is now evaluated in the dilute regime.

The classical factor is a sequence of isolated events joined by nearly flat separations, so its action is the sum of the isolated event actions corrected by the pairwise tails of Eq.~\eqref{eq:tailInteraction},
\begin{equation}
 S_{\rm coll}(\boldsymbol\tau)
 =S_{\rm top}+\sum_r S_{\rm int}(\tau_r)
 +O(\ee^{-2\omega_{\min}R}),
 \qquad
 S_{\rm top}:=\sum_r S_{e_r},
 \label{eq:ScollDilute}
\end{equation}
with $S_{e_r}$ the action of the isolated event $e_r$.  Dividing by $\hbar$ turns the second piece into the interaction term of Eq.~\eqref{eq:QZIaction}.

The one-loop factor reduces because the dilute structure makes the fluctuation operator \eqref{eq:fluctuationOperator} block diagonal up to $O(\ee^{-\omega_{\min}R})$.  Near each event it is the elementary heteroclinic Hessian, which carries no separation dependence, and along each separation it is the harmonic operator $-\partial_t^2+\omega_r^2$ on an interval of length $\tau_r$, which is the same block structure already used to build the parametrix in Eq.~\eqref{eq:parametrixError}.  The regulated determinant factorizes accordingly, and by the Gel'fand--Yaglom/Forman formulas quoted in Sec.~\ref{subsec:collectiveNormal} the harmonic block on an interval of length $\tau_r$ is the propagator \eqref{eq:MehlerAsymptotic}.  Including the measure, this gives
\begin{equation}
 \mathcal A_{\perp,I}(\boldsymbol\tau;\hbar)
 =\mathcal J_I\Bigl[\prod_r a_{e_r}(\hbar)\Bigr]
 \prod_r K_r(0,0;\tau_r)
 \bigl[1+O(\ee^{-\omega_{\min}R})\bigr],
 \label{eq:AperpFactorized}
\end{equation}
where $a_{e_r}(\hbar)$ is the separation-independent one-loop factor of the isolated event $e_r$, and $\mathcal J_I$ is the collective-coordinate Jacobian of the change of variables from the path-integral measure along the near-zero block of Sec.~\ref{subsec:collectiveNormal} to $\prod_r\dd\tau_r$.  That Jacobian is built from the Gram matrix \eqref{eq:GramMatrixPhysics} and the norms of the elementary translation modes, so by $G=I+O(\ee^{-\omega_{\min}R})$ it too is separation independent to the accuracy used here.

Only the last product in Eq.~\eqref{eq:AperpFactorized} depends on the separations.  Inserting Eq.~\eqref{eq:MehlerAsymptotic} and splitting $\ee^{-\omega_r\tau_r/2}=\ee^{-\omega_{\rm ref}\tau_r/2}\ee^{-\varrho_r\tau_r}$ with $\varrho_r$ from Eq.~\eqref{eq:rhoDef} gives
\begin{equation}
 \prod_r K_r(0,0;\tau_r)
 =\Bigl[\prod_r\Bigl(\frac{\omega_r}{\pi\hbar}\Bigr)^{1/2}\Bigr]
 \ee^{-\omega_{\rm ref}T/2}
 \prod_r\ee^{-\varrho_r\tau_r},
 \label{eq:KernelProduct}
\end{equation}
where we used the sum $\sum_r\tau_r=T$.  The bracket is separation independent, the middle factor is the common factor already removed in passing to the shifted partition function of Eq.~\eqref{eq:shiftedTrace} and is therefore not carried below, and the remaining product is the zero-point term of Eq.~\eqref{eq:QZIaction}.  This is also why $\omega_{\rm ref}$ may be chosen arbitrarily and still cancels from everything physical.

Multiplying Eq.~\eqref{eq:AperpFactorized} by $\ee^{-S_{\rm coll}/\hbar}$ from Eq.~\eqref{eq:ScollDilute}, and collecting every separation-independent factor into
\begin{equation}
 \mathcal N_I
 :=\mathcal J_I\prod_r a_{e_r}(\hbar)
 \Bigl(\frac{\omega_r}{\pi\hbar}\Bigr)^{1/2}
 \ee^{-S_{\rm top}/\hbar},
 \label{eq:NIdef}
\end{equation}
the two exponents combine into $\cW_I$ and we obtain
\begin{equation}
 \mathcal A_{\perp,I}(\boldsymbol\tau;\hbar)
 \ee^{-S_{\rm coll}(\boldsymbol\tau)/\hbar}
 =\mathcal N_I\,
 \ee^{-\cW_I(\boldsymbol\tau)}
 \left[1+O(\ee^{-\omega_{\min}R})+O(\hbar)\right].
 \label{eq:PItoQZI}
\end{equation}
The $O(\ee^{-\omega_{\min}R})$ term collects the exponentially small corrections of the dilute gas approximation in Eqs.~\eqref{eq:ScollDilute} and \eqref{eq:AperpFactorized}, and the $O(\hbar)$ term the higher-loop corrections.  Comparison with Eq.~\eqref{eq:fixedTQZI} shows that $\mathcal N_I$ of Eq.~\eqref{eq:NIdef} is the normalization of the fixed-$T$ QZI.

One feature of the exponent \eqref{eq:QZIaction} deserves emphasis, because it makes the QZI critical point differ from the spectral saddle of Sec.~\ref{subsec:finiteT_WKB_period}. The linear term $\varrho_r\tau_r$ is generated by the one-loop fluctuations in the normal directions, whereas the classical collective action contains only the exponentially small tail interactions at this order. The two stationarity problems therefore differ generically. The exponent $\cW_I$ by itself need not have a finite stationary point, since all $\varrho_r$ vanish when the wells have equal frequencies, as in the double well. The separations are fixed by the constraint $\sum_r\tau_r=T$ of Eq.~\eqref{eq:fixedTQZI}, imposed through a Lagrange multiplier $s$, which is the Laplace variable of Sec.~\ref{subsec:laplaceFactorization}. It adds $s\tau_r$ to every separation, and with $\omega_r\mu_r:=s+\varrho_r$ stationarity of $\omega_r\mu_r\tau_r+\lambda_r\Lambda_r\ee^{-\omega_r\tau_r}$ gives
\begin{equation}
 \Lambda_r\ee^{-\omega_r\tau_r}= \lambda_r\mu_r,
 \qquad
 \Re\tau_r=\frac1{\omega_r}\log\frac1{|\hbar|}+O(1)
 \quad\text{for }0<c\le|\mu_r|\le C.
 \label{eq:QZICriticalInfinityPhysics}
\end{equation}
 A spectral pole has $\mu_r=\tfrac12-\epsilon/\omega_v$, but the condition $\epsilon=O(1)$ does not keep $\mu_r$ away from zero. Values at or near the poles of $\Gamma(\mu_r)$ must instead be handled by the meromorphic periods and their residues. At fixed $T$, on the other hand, the multiplier is set by $T$. For the double-well bion, for instance, $\tau^*=T/2$ and $\lambda\mu=\Lambda\ee^{-\omega T/2}$. Thus the logarithmic saddle estimate applies to a post-Laplace stationary point away from these poles, whereas constrained saddles can scale directly with $T$.
The infinite-separation boundary remains relevant to the dilute limit even when no finite stationary point of an individual separation exists. By this we mean the $\tau_r\to\infty$ end of the quasi-moduli space, at which the two events bounding a separation move apart without bound and the configuration ceases to be a single connected trajectory, degenerating into the two events joined by an infinite rest at the intervening minimum. For a repulsive pair this limit is a broken trajectory,\footnote{Each instanton solves the first-order equation $\dot x=\pm\sqrt{2(V-V_{\min})}$, and as $\tau_r\to\infty$ the configuration approaches a chain of two such solutions joined at the intervening minimum.  For a repulsive pair both legs solve the equation with the same sign, so the chain is monotone in $x$, whereas for an attractive pair the second leg retraces the first with the opposite sign.  No exact solution passes through the minimum, since $\sqrt{2(V-V_{\min})}$ vanishes linearly there and the approach takes infinite time.  This end of the quasi-moduli space is therefore a critical point at infinity of the QZI \cite{BehtashInfinity} rather than the boundary of a family of exact solutions.  On the spectral side the same limit is the approach to the double turning point at $x_v$, where the spectral curve $\dot x^2=2(V-E)$ acquires an ordinary double point, or node, as $E\to V_{\min}$ \cite{BehtashHomological}, and the logarithmic growth of the period in Eq.~\eqref{eq:generalLogPeriodPhysics} is the signature of this node.  The need for the BZJ continuation at attractive pairs comes from the sign of the tail interaction, as explained in Sec.~\ref{subsec:gammaPeriods}.} and for either kind of pair it is the opposite end of the quasi-moduli space from the region $\tau_r\to0$ where events merge, which is excluded from the dilute gas approximation. This is the standard critical-point-at-infinity mechanism of the QZI \cite{BehtashInfinity}, and the interaction threshold of Sec.~\ref{subsec:regime} motivates the lower time scale without a universal constrained-saddle assumption. The finite classical inverse-Laplace saddle of Sec.~\ref{subsec:finiteT_WKB_period} and this one-loop QZI saddle are thus distinct objects, even though both enter the same large-$T$ description.

\subsection{Laplace factorization and the spectral pole}
\label{subsec:laplaceFactorization}

At leading dilute order the fixed-$T$ constraint is the only coupling among the separation variables, so introducing a Laplace variable $s$ conjugate to $T$ decouples them,
\begin{align}
 \widetilde{\mathscr I}_I(s)
 &:=\int_0^\infty\dd T\;\ee^{-sT}\mathscr I_I(T)
 \nonumber\\
 &\mathrel{\text{``}{=}\text{''}}\mathcal N_I
 \int\prod_r\dd\tau_r\;
 \ee^{-\sum_r(s+\varrho_r)\tau_r}
 \exp\left[-\sum_r\lambda_r\Lambda_r\ee^{-\omega_r\tau_r}\right]
 \nonumber\\
 &=\mathcal N_I\prod_r J_r(s).
 \label{eq:LaplaceFactorization}
\end{align}
The delta function in Eq.~\eqref{eq:fixedTQZI} has been used here, and the exponent has factorized because $\cW_I$ in Eq.~\eqref{eq:QZIaction} is a sum of single-separation terms.  The quotation marks on the second equality mean that it holds only for the large-$T$ behavior.  The QZI represents the dilute-gas contribution to the path integral only in the window \eqref{eq:largeTPhysicsWindow}, where every separation is long, but the Laplace transform integrates over all $T>0$, including short periods at which the events overlap and the exponent \eqref{eq:QZIaction} no longer applies.  This mismatch is harmless.  Two functions of $T$ that agree for $T>T_0$ have Laplace transforms that differ by $\int_0^{T_0}\dd T\,\ee^{-sT}(\cdots)$, which is an entire function of $s$.  The poles of $\widetilde{\mathscr I}_I(s)$ and their residues are therefore unaffected, and they alone determine the large-$T$ behavior of $\mathscr I_I(T)$ through the inverse Laplace transform.  Therefore, for a single separation one has
\begin{equation}
 J_r(s)=\int_{\mathcal C_r}\dd\tau_r\;
 \exp\left[-\omega_r\mu_r(s)\tau_r
 -\lambda_r\Lambda_r\ee^{-\omega_r\tau_r}\right],
 \label{eq:oneGapIntegralTau}
\end{equation}
where again we note that $\mu_r(s)=(s+\varrho_r)/\omega_r$. Here $\mathcal C_r$ denotes the complexified integration contour selected by the BZJ prescription and its lateral analytic continuations, which are constructed below.

So far $s$ is a free transform variable and carries no spectral meaning.  This Laplace transform from $T$ to $s$ should not be confused with the spectral inverse Laplace transform \eqref{eq:Bromwich} from $\epsilon$ to $T$, which turns exact-WKB data into the partition function.  It is only a computational factorization of the QZI at fixed $T$, and $s$ acquires a spectral meaning once the transform is evaluated at a pole, since a level at $E-V_{\min}=\hbar\epsilon$ contributes $\ee^{-T\epsilon}$ to the shifted partition function of Eq.~\eqref{eq:shiftedTrace}, and removing the factor $\ee^{-\omega_{\rm ref}T/2}$ of Eq.~\eqref{eq:rhoDef} leaves $\exp[-T(\epsilon-\omega_{\rm ref}/2)]$.  Its Laplace transform has the pole
\begin{equation}
 s_\epsilon=\frac{\omega_{\rm ref}}2-\epsilon.
 \label{eq:sEnergyRelation}
\end{equation}
Using $\varrho_r$ from Eq.~\eqref{eq:rhoDef} in $\mu_r(s)$ makes $J_r$ depend on the separation only through the well $v=v_r$ that it spans,
\begin{equation}
 \mu_v(s_\epsilon)
 =\frac{s_\epsilon+\varrho_r}{\omega_r}
 =\frac12-\frac{\epsilon}{\omega_v}.
 \label{eq:muHarmonicDerivation}
\end{equation}
The quantity $\epsilon/\omega_v$ is the {\it harmonic level number} at well $v$, and the one-loop reduction of Secs.~\ref{subsec:QZIaction} and \ref{subsec:laplaceFactorization} gives Eq.~\eqref{eq:muHarmonicDerivation}.  The exact-WKB quantities depend on the energy through the local Weber level function $\mathsf n_v(\epsilon,\hbar)$ introduced in Sec.~\ref{subsec:questions}, which satisfies $\mathsf n_v=\epsilon/\omega_v+O(\hbar)$ and contains the anharmonic corrections of well $v$ to all orders in $\hbar$.  In the comparison with exact-WKB we use
\begin{equation}
 \mu_v=\frac12-\mathsf n_v(\epsilon,\hbar)
 \label{eq:muWeberIdentification}
\end{equation}
in place of Eq.~\eqref{eq:muHarmonicDerivation}. This replacement is not derived by the one-loop path-integral reduction, and we state it as a separate local Weber input. Its pole structure does follow from the path integral. Beyond one loop the propagator \eqref{eq:MehlerAsymptotic} over a long separation is replaced by the propagator of the full local Hamiltonian of well $v$, whose long-time expansion is a sum of exponentials $\ee^{-E_{v,k}\tau_r/\hbar}$ over the perturbative levels $E_{v,k}$ of the well. Its Laplace transform therefore has poles at those levels, which are the points $\mathsf n_v\in\mathbb Z+\tfrac12$, whereas $\Gamma(\mu_v)$ with the harmonic value \eqref{eq:muHarmonicDerivation} has them at $\epsilon/\omega_v\in\mathbb Z+\tfrac12$. Correct pole positions thus require Eq.~\eqref{eq:muWeberIdentification}. The whole single-separation period, and not only its poles, follows from the harmonic one by the substitution $\epsilon/\omega_v\to\mathsf n_v$ because near a double turning point the local Schr\"odinger equation is equivalent, order by order in $\hbar$, to the Weber equation with $\epsilon/\omega_v$ replaced by $\mathsf n_v$ \cite{AKTBenderWuII,TakeiWeber}. This is the same reduction that underlies the Weber-type description in exact-WKB \cite{SueishiExactWKBII}. In the present calculation this uniformization enters as local Weber input, and pole locations alone do not fix higher-loop path-integral residues or prefactors. Deriving the substitution and the corresponding measure corrections from the anharmonic path integral would remove that input. From here on, $s$ is always taken to be evaluated at the pole \eqref{eq:sEnergyRelation}.

\subsection{Gamma/Hankel periods from large separations}
\label{subsec:gammaPeriods}

Let us set
\begin{equation}
 z=\Lambda_r\ee^{-\omega_r\tau_r},
 \qquad
 \dd\tau_r=-\frac{\dd z}{\omega_r z}.
 \label{eq:ychange}
\end{equation}
Then Eq.~\eqref{eq:oneGapIntegralTau} becomes a Gamma-type integral,
\begin{equation}
 J_r
 =\frac{\Lambda_r^{-\mu_r}}{\omega_r}
 \int_{\mathcal C_z}z^{\mu_r-1}\ee^{-\lambda_r z}\dd z,
 \label{eq:GammaReduction}
\end{equation}
where $\mathcal C_z$ is the image of $\mathcal C_r$ under Eq.~\eqref{eq:ychange}.  We can now say which contour that is.

The change of coordinate in \eqref{eq:ychange} reverses orientation and sends large separations to small $z$, so a range of real separations $\tau_r\in(\tau_0,\infty)$ becomes the segment $z\in(0,z_0)$ with $z_0=\Lambda_r\ee^{-\omega_r\tau_0}$. The point $\tau_r=0$, where the two events merge, sits at the finite value $z=\Lambda_r$.  Since $\Lambda_r\propto\hbar^{-1}$, that locus goes to infinity in the semiclassical limit.  The dilute-gas integral can therefore be completed to a full Gamma period without ever extrapolating into the region where two events merge.  It remains to fix the far end of $\mathcal C_z$, which is decided by the sign $\lambda_r$, and to require integrability at the origin, which is the condition $\Re\mu_r>0$ with other values reached by analytic continuation.

In the notation of Sec.~\ref{subsec:cores}, we write $\Lambda_{{\rm att},v}$ and $\Lambda_{{\rm rep},v}$ for the interaction magnitudes of the attractive and repulsive separations at well $v$.
For a repulsive separation we have $\lambda_r=+1$, so the integrand of Eq.~\eqref{eq:GammaReduction} carries $\ee^{-z}$ and decays along the positive real axis.  The real contour is then admissible and $\mathcal C_z$ is the positive real axis run outward from the origin.  To see that only the weakly interacting region is used, first take a finite upper value $M$ for the scaled variable $z$.  The corresponding large-separation part is
\begin{equation}
 J_{v}^{\rm rep}(M)
 =\frac{\Lambda_{{\rm rep},v}^{-\mu_v}}{\omega_v}
 \int_0^M z^{\mu_v-1}\ee^{-z}\dd z
 =\frac{\Lambda_{{\rm rep},v}^{-\mu_v}}{\omega_v}\,
 \gamma(\mu_v,M),
 \label{eq:incompleteGamma}
\end{equation}
where $\gamma(\mu,M)$ is the lower incomplete Gamma function.  For fixed $M$ and $\Lambda_{{\rm rep},v}\sim\hbar^{-1}\to\infty$, the region $0\le z\le M$ corresponds to
\begin{equation}
 \tau_r\ge \frac1{\omega_v}\log\frac{\Lambda_{{\rm rep},v}}{M}\longrightarrow\infty,
\end{equation}
so it lies inside the dilute region, and $M$ stays far below the coalescence value $\Lambda_{{\rm rep},v}$.  After taking the semiclassical limit, we may send $M\to\infty$.  For $\Re\mu_v>0$ this gives
\begin{equation}
 J_v^{\rm rep}
 =\frac{1}{\omega_v}\Lambda_{{\rm rep},v}^{-\mu_v}\Gamma(\mu_v).
 \label{eq:JT}
\end{equation}
Other values of $\mu_v$ are reached by analytic continuation.  We always use this order of limits when the dilute QZI is represented by a Gamma period.

For an attractive separation, on the other hand, one has $\lambda_r=-1$, so the integrand of Eq.~\eqref{eq:GammaReduction} carries $\ee^{+z}$ and grows along the positive real axis.  The real contour is therefore not admissible, and this is the concrete reason for the BZJ prescription.  The BZJ prescription defines the same large-separation integral at the repulsive sign, with the contour just determined, and continues the result back to the attractive sign from above or below the Stokes direction.  In the $z$ plane, the continuation rotates the ray to the two sides of the negative real axis, that is $z=\ee^{\pm\ii\pi}w$ with $w>0$.  Under this substitution $\ee^{+z}$ becomes $\ee^{-w}$ while $z^{\mu_v-1}\dd z$ becomes $\ee^{\pm\ii\pi\mu_v}w^{\mu_v-1}\dd w$, so each rotated ray returns a Gamma period dressed by a phase.  Let $\mathsf c_+$ be the upper ray run outward from the origin and let $\mathsf c_-$ be the lower ray run inward toward the origin.  Then
\begin{align}
 J_{v,+}^{\rm att}
 &=\frac{1}{\omega_v}\Lambda_{{\rm att},v}^{-\mu_v}
 \ee^{+\ii\pi\mu_v}\Gamma(\mu_v),
 \label{eq:JNplus}
 \\
 J_{v,-}^{\rm att}
 &=-\frac{1}{\omega_v}\Lambda_{{\rm att},v}^{-\mu_v}
 \ee^{-\ii\pi\mu_v}\Gamma(\mu_v).
 \label{eq:JNminus}
\end{align}
The minus sign in the second line stands only for the contour orientation, since $\mathsf c_-$ runs inward.  The BZJ continuation also carries the attractive description past the short-distance region, which the same semiclassical approximation cannot cross directly \cite{SueishiExactWKBI,ZinnJustinJentschuraI}.  App.~\ref{app:dwExample} carries out these steps explicitly for the double-well bion.

\begin{figure}[t]
\centering
 \begin{tikzpicture}[
  >=Stealth,
  axis/.style={->,gray!70,thin},
  contour/.style={very thick,blue!55!black},
  mid arrow/.style={postaction={decorate,decoration={markings,mark=at position #1 with {\arrow{Stealth[length=2.4mm]}}}}},
  mid arrow/.default=0.55,
  cut/.style={decorate,decoration={zigzag,amplitude=0.6mm,segment length=1.6mm},gray!65!black},
]
\begin{scope}
  \fill[blue!7] (0,-0.32) rectangle (2.2,0.32);
  \draw[axis] (-1.0,0) -- (5.5,0) node[below left] {\small$\Re z$};
  \draw[axis] (0,-1.5) -- (0,1.5) node[below left] {\small$\Im z$};
  \draw[contour,mid arrow=0.28] (0,0) -- (5.2,0);
  \node[above,blue!55!black] at (1.1,0.32) {\small$\mathcal C_z$};
  \node[below,font=\scriptsize] at (1.1,-0.32) {$0\le z\le M$};
  \draw (2.2,-0.1) -- (2.2,0.1);
  \node[above] at (2.2,0.1) {\small$M$};
  \draw[fill=white] (3.9,0) circle (0.07);
  \node[above] at (3.9,0.1) {\small$\Lambda_{{\rm rep},v}$};
  \draw[->,gray] (3.7,-0.35) -- (4.8,-0.35);
  \node[below,font=\scriptsize] at (4.25,-0.38) {$\hbar\to0$};
  \fill (0,0) circle (0.05);
  \node[below left,font=\scriptsize] at (0,0) {$\tau_r\to\infty$};
  \node[font=\small] at (2.2,1.95) {(a) repulsive, $\lambda_r=+1$};
\end{scope}
\begin{scope}[xshift=10.2cm]
  \draw[axis] (-3.7,0) -- (2.7,0) node[below left] {\small$\Re z$};
  \draw[axis] (0,-1.5) -- (0,1.5) node[below left] {\small$\Im z$};
  \draw[gray,densely dotted,thick] (0,0) -- (2.4,0);
  \node[below,font=\scriptsize,gray!60!black] at (1.2,-0.05) {$\ee^{+z}$ grows};
  \draw[fill=white] (1.9,0) circle (0.07);
  \node[above] at (1.9,0.1) {\small$\Lambda_{{\rm att},v}$};
  \draw[cut] (0,0) -- (-3.5,0);
  \draw[contour,mid arrow=0.55] (-0.3031,0.175) -- (-3.5,0.175);
  \draw[contour,mid arrow=0.55] (-3.5,-0.175) -- (-0.3031,-0.175);
  \draw[contour,dashed] (-0.3031,-0.175) arc[start angle=-150,end angle=150,radius=0.35];
  \node[above,blue!55!black] at (-2.0,0.175) {\small$\mathsf c_+$};
  \node[below,blue!55!black] at (-2.0,-0.175) {\small$\mathsf c_-$};
  \node[font=\small] at (-2.0,1.0) {$H=\mathsf c_++\mathsf c_-$};
  \fill (0,0) circle (0.04);
  \node[font=\small] at (-0.5,1.95) {(b) attractive, $\lambda_r=-1$};
\end{scope}
\end{tikzpicture}
\caption{Integration contours of the single-separation integral \eqref{eq:GammaReduction} in the plane of the variable $z$ of Eq.~\eqref{eq:ychange}. The origin is the infinite-separation end $\tau_r\to\infty$, and the open circle marks the coalescence point $z=\Lambda_r$, which recedes to infinity as $\hbar\to0$. (a) Repulsive separation. The contour $\mathcal C_z$ is the positive real axis, first cut off at $M$ inside the dilute region (shaded) and completed only after the semiclassical limit, which gives Eq.~\eqref{eq:JT}. (b) Attractive separation. The positive real axis is not admissible, and the BZJ continuation rotates the contour onto the rays $\mathsf c_+$ and $\mathsf c_-$ on the two sides of the cut of $z^{\mu_v-1}$ along the negative real axis, which gives Eqs.~\eqref{eq:JNplus} and \eqref{eq:JNminus}. Together the two rays form the Hankel contour $H$ around the branch point, and neither ever reaches $z=\Lambda_{{\rm att},v}$.}
\label{fig:gammaHankelContours}
\end{figure}
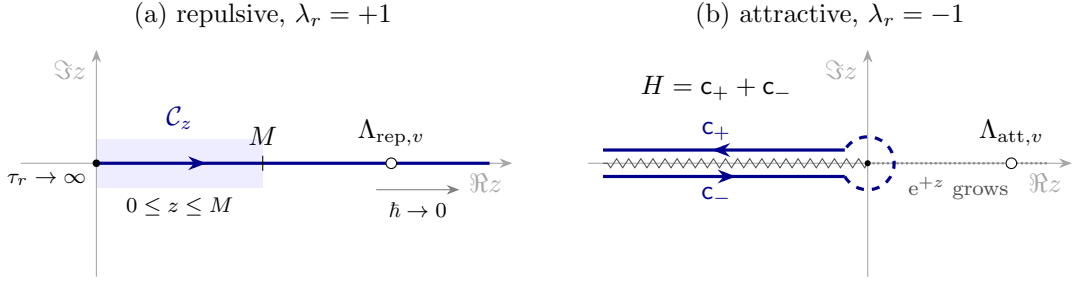

The two rays differ only by their side of the cut, so their combination closes into a cycle encircling the branch point of $z^{\mu_v-1}$ at the origin.  This is the Hankel contour $H$ and, with the orientation fixed, is given by $H=\mathsf c_+ + \mathsf c_-$.  Its period follows from Eqs.~\eqref{eq:JNplus} and \eqref{eq:JNminus} together with the Euler reflection formula.  Two features of $H$ matter below.  First, the cycle $H$ is constructed entirely from the two rays near $z=0$ and never reaches $z=\Lambda_{{\rm att},v}$, so it is an object of the dilute large-separation region.  Second, the lateral ambiguity of an attractive separation is exactly the freedom to add a multiple of $H$, which is the origin of the Stokes jump.  The contours of both cases are depicted in Fig.~\ref{fig:gammaHankelContours}.

For the minimal event sequence the completed post-Laplace QZI is therefore
\begin{equation}
 \widetilde{\mathscr I}_{I,\pm}(s)
 =J_{a,\pm}^{\rm att}(s)J_{c,\pm}^{\rm att}(s)
 \prod_{v=a+1}^{c-1}\bigl[J_v^{\rm rep}(s)\bigr]^2.
 \label{eq:QZIProduct}
\end{equation}
This is the result promised in Sec.~\ref{subsec:cores}.  Only the two attractive events carry a lateral choice, while the internal repulsive $\cI\cI$/$\bar\cI\bar\cI$ factors are identical on the two sides, which is the same asymmetry between outer and internal wells that the perturbative oscillations $u_v^\pm$ of Eq.~\eqref{eq:uDef} display in the exact-WKB coefficient.

The action and one-loop measure supply a raw event prefactor with classical factor $\prod_{e\in I}\ee^{-2S_{I,e}/\hbar}$. Converting it to the dressed spectral weight $\prod_{e\in I}B_e$ requires the elementary normalization of Sec.~\ref{subsec:weberMatching}. 

The infinite-separation boundary is $z=0$. On the regular dilute patch, Eq.~\eqref{eq:PItoQZI} identifies the collective integral with the local Gamma /Hankel model near that boundary. The resulting integer contour relation belongs to the reduced post-Laplace QZI and is inherited by that regular-patch description. 

\section{QZI realization of the DDP jump of connected coefficients}
\label{sec:QZIStokes}

From the path integral we now have the product \eqref{eq:QZIProduct} of single-separation Gamma and Hankel periods obtained in Sec.~\ref{sec:fullPI}, and from exact-WKB the connected coefficient $C_I^\pm$ of Eq.~\eqref{eq:spectralTarget}.  Comparing the two requires two local inputs that the one-loop QZI does not supply by itself, namely the local Weber level function in the exponents $\mu_v$, discussed after Eq.~\eqref{eq:muWeberIdentification}, and the absolute normalization of each single-separation period.  Both come from the same local Weber equation that describes every well in the large-$T$ regime of Sec.~\ref{subsec:largeTWeber}.  The normalization is fixed on the repulsive separations alone, so the attractive values, including the unit period of the Hankel cycle, are derived rather than imposed.  Once normalized, the periods of the two $\cI\bar\cI$ separations become the local perturbative oscillations $u_v^\pm$ and those of the internal $\cI\cI$/$\bar\cI\bar\cI$ separations assemble into the factor $\mathsf K_I$, so the two lateral QZI periods reproduce $C_I^+$ and $C_I^-$ separately.  Matching values in this way does not yet explain why the two laterals are related as they are.  For that we follow the integration cycles themselves rather than only their periods.  The two lateral cycles of an $\cI\bar\cI$ separation differ by exactly one Hankel cycle, and evaluating this integer relation with the normalized period map reproduces the complementary rule $u_v^-=1-u_v^+$ of Eq.~\eqref{eq:uComplement}.  Applied at both outer wells of the interval, it realizes the DDP Stokes jump of Eq.~\eqref{eq:DDPaction} on the connected coefficient as a Picard--Lefschetz relation among QZI cycles.

\subsection{Local Weber normalization and matching of the lateral coefficients}
\label{subsec:weberMatching}

Section~\ref{subsec:largeTWeber} motivates the Weber regime $\epsilon=O(1)$ from the scale of the classical inverse-Laplace saddle. In this regime each well has the local Weber description \eqref{eq:LocalWeberEquationPhysics}, and we use the exact-WKB input of Sec.~\ref{subsec:airyWeberDDP} to express the DDP action in that description. In that description the perturbative Voros symbol $A_v$ of Eq.~\eqref{eq:VorosSymbols} is controlled by a single local function $\mathsf n_v(\epsilon,\hbar)$, the {\it local Weber level function} of well $v$, through
\begin{equation}
 A_v=\ee^{2\pi\ii(-1)^v\mathsf n_v}.
 \label{eq:AWeber}
\end{equation}
The sign $(-1)^v$ is the alternating orientation of the $A$-cycles discussed after Eq.~\eqref{eq:DDPaction}, and $\mathsf n_v=\epsilon/\omega_v+O(\hbar)$ reduces to the harmonic level number of Eq.~\eqref{eq:muHarmonicDerivation}.\footnote{The function $\mathsf n_v$ has standard names in the literature.  Up to the orientation sign, $2\pi\hbar\,\mathsf n_v$ is the perturbative quantum A-period of well $v$, the all-orders WKB period around its classically allowed interval \cite{BasarDunneUnsalQG}.  It coincides with the function $B(E,g)$ of the exact quantization conditions of Refs.~\cite{ZinnJustinJentschuraI,JentschuraZinnJustinII}, with $g$ in the role of $\hbar$, and $\mathsf n_v-\tfrac12$ is the order of the local parabolic-cylinder function \cite{DunneUnsalUniform}.}  The perturbative levels of well $v$, at which $P_v=1+A_v$ vanishes, therefore sit at $\mathsf n_v\in\mathbb Z+\tfrac12$ for either parity of $v$.  Substituting Eq.~\eqref{eq:AWeber} into Eq.~\eqref{eq:uDef}, the parity dependence drops out and the local perturbative oscillations become
\begin{equation}
 u_v^+=\frac{\ee^{-\ii\pi\mathsf n_v}}{2\cos\pi\mathsf n_v},
 \qquad
 u_v^-=\frac{\ee^{+\ii\pi\mathsf n_v}}{2\cos\pi\mathsf n_v},
 \label{eq:uWeber}
\end{equation}
which obey $u_v^++u_v^-=1$ as required by Eq.~\eqref{eq:uComplement}.

The same local Weber equation supplies the connection matrix of well $v$, which relates the two local solutions entering the well to the two leaving it.  It is built from the elementary connection matrices at the two turning points of the well and the phase $\ee^{\mp\ii\pi\mathsf n_v}$ accumulated between them, and it uses no path-integral input,
\begin{equation}
 \Sigma_v^\pm=
 \begin{pmatrix}
 \ii u_v^\pm&r_v\\
 r_v&\ii u_v^\pm
 \end{pmatrix},
 \qquad
 r_v:=\frac{1}{2\cos\pi\mathsf n_v}.
 \label{eq:weberConnectionMatrix}
\end{equation}
We call the entries of $\Sigma_v^\pm$ the {\it local Weber amplitudes} of well $v$.  The diagonal entry, in which the solution leaves toward the side it came from, depends on the lateral choice.  The off-diagonal entry, in which it continues to the other side, does not.

Every raw QZI period of Sec.~\ref{subsec:gammaPeriods} at well $v$ carries the factor $\Lambda^{-\mu_v}/\omega_v$, where $\Lambda$ is the interaction magnitude of the separation.  This factor is set by the tail amplitudes $\alpha_\pm$ of Eq.~\eqref{eq:tails}, and it has no counterpart in the exact-WKB data.  The normalization of a single-separation period is a property of the local Weber equation of well $v$, not of the way the trajectory enters and leaves the well, so we look for one function
\begin{equation}
 \mathcal N_v(\Lambda)=\frac{\omega_v\Lambda^{\mu_v}}{2\pi}\,f_v(\mu_v)
 \label{eq:NvAnsatz}
\end{equation}
that applies to both kinds of separation at well $v$, with $\Lambda=\Lambda_{{\rm att},v}$ or $\Lambda_{{\rm rep},v}$.  We fix $f_v$ by a single condition on the repulsive separation, which carries no lateral choice, requiring its period to be the off-diagonal entry of Eq.~\eqref{eq:weberConnectionMatrix},
\begin{equation}
 \mathcal N_v(\Lambda_{{\rm rep},v})\,J_v^{\rm rep}=r_v.
 \label{eq:repNormalizationCondition}
\end{equation}
This is a Weber matching prescription rather than an evaluation of the path-integral prefactor $\mathcal N_I$, as stated in Sec.~\ref{sec:intro}.
With Eq.~\eqref{eq:JT}, the identification \eqref{eq:muWeberIdentification}, and Euler's reflection formula
\begin{equation}
 \Gamma(\mu)\Gamma(1-\mu)=\frac{\pi}{\sin\pi\mu},
 \qquad
 \sin(\pi\mu_v)=\cos(\pi\mathsf n_v),
 \label{eq:EulerReflection}
\end{equation}
this condition gives $f_v(\mu_v)=\Gamma(1-\mu_v)$, that is
\begin{equation}
 \mathcal N_v(\Lambda)
 =\frac{\omega_v\Lambda^{\mu_v}}{2\pi}\Gamma(1-\mu_v).
 \label{eq:RTWeberInput}
\end{equation}
No condition is imposed on the attractive separation.  Applying the same $\mathcal N_v$ to the rays of Eqs.~\eqref{eq:JNplus} and \eqref{eq:JNminus} gives
\begin{equation}
 \mathcal N_v(\Lambda_{{\rm att},v})\,J_{v,\pm}^{\rm att}=\ii u_v^\pm,
 \label{eq:attFromN}
\end{equation}
which are the diagonal entries of Eq.~\eqref{eq:weberConnectionMatrix}.  The sum of the two rays is the Hankel cycle $H=\mathsf c_++\mathsf c_-$ of Fig.~\ref{fig:gammaHankelContours}(b), which comes in from $-\infty$ below the negative real axis, encircles the origin counterclockwise, and returns to $-\infty$ above it.  On this cycle the attractive integral of Eq.~\eqref{eq:GammaReduction} is Hankel's representation of the reciprocal Gamma function,
\begin{equation}
 \int_H z^{\mu_v-1}\ee^{z}\,\dd z=\frac{2\pi\ii}{\Gamma(1-\mu_v)},
 \label{eq:HankelLoop}
\end{equation}
so that, with the prefactor $\Lambda_{{\rm att},v}^{-\mu_v}/\omega_v$ of Eq.~\eqref{eq:GammaReduction}, the raw Hankel period is
\begin{equation}
 J_{H,v}:=J_{v,+}^{\rm att}+J_{v,-}^{\rm att}
 =\frac{2\pi\ii}{\omega_v\Lambda_{{\rm att},v}^{\mu_v}\Gamma(1-\mu_v)}.
 \label{eq:rawHankelReflection}
\end{equation}
The same result follows by adding Eqs.~\eqref{eq:JNplus} and \eqref{eq:JNminus} and using the reflection formula \eqref{eq:EulerReflection}.  Therefore
\begin{equation}
 \mathcal N_v(\Lambda_{{\rm att},v})\,J_{H,v}=\ii .
 \label{eq:HankelFromN}
\end{equation}
The period of the Hankel cycle is thus a consequence of the normalization fixed on the repulsive separation, since the factor $\Gamma(1-\mu_v)$ in Eq.~\eqref{eq:RTWeberInput} cancels the factor $1/\Gamma(1-\mu_v)$ of Hankel's representation \eqref{eq:HankelLoop}.  In the same way, Eq.~\eqref{eq:attFromN} reproduces the lateral entries of the Weber connection matrix, although only its lateral-invariant entry was used as input.

The factor $\ii$ on the diagonal of Eq.~\eqref{eq:weberConnectionMatrix} is a phase convention of the connection matrix.  It is kept explicit at higher grade, where it enters a closed event sequence only through a power of $\ii$ fixed by the number of $\cI\bar\cI$ pairs.  For the minimal event sequence, which has exactly two $\cI\bar\cI$ pairs, it is simpler to remove it from each attractive period, and we set
\begin{equation}
 \cR_v^{\rm att}:=-\ii\,\mathcal N_v(\Lambda_{{\rm att},v}),
 \qquad
 \cR_v^{\rm rep}:=\mathcal N_v(\Lambda_{{\rm rep},v}).
 \label{eq:RNfromHankel}
\end{equation}
The product $(-\ii)^2=-1$ of the two removed phases is the sign with which a closed event sequence enters $\log\cZ_N^\pm$.  The interaction magnitudes $\Lambda_{{\rm att},v}$ and $\Lambda_{{\rm rep},v}$ cancel from every normalized period below, as they must.

To distinguish an integration cycle from the value of its integral, let $J_v^{\rm att}[\mathcal C]$ and $J_v^{\rm rep}[\mathcal C]$ denote the single-separation integral \eqref{eq:GammaReduction} at well $v$ evaluated on a contour $\mathcal C$, before the local Weber normalization is applied.  We define
\begin{equation}
 \mathcal P_v^{\rm QZI}(\mathcal C)
 :=
 \begin{cases}
  \cR_v^{\rm att}\,J_v^{\rm att}[\mathcal C],
  & \mathcal C\text{ a cycle of an attractive separation},\\[1mm]
  \cR_v^{\rm rep}\,J_v^{\rm rep}[\mathcal C],
  & \mathcal C\text{ a cycle of a repulsive separation}.
 \end{cases}
 \label{eq:normalizedQZIPeriodMap}
\end{equation}
Thus $\mathcal P_v^{\rm QZI}$ assigns to each single-separation QZI cycle its normalized period, which is a meromorphic function of $\mu_v$.  By Eqs.~\eqref{eq:HankelFromN} and \eqref{eq:RNfromHankel},
\begin{equation}
 \mathcal P_v^{\rm QZI}(H_v)=1.
 \label{eq:unitHankelPeriod}
\end{equation}

Written out with Eq.~\eqref{eq:EulerReflection}, the normalized periods are as follows.  For a repulsive separation, Eq.~\eqref{eq:repNormalizationCondition} reads
\begin{equation}
 \cR_v^{\rm rep}J_v^{\rm rep}=\frac{1}{2\pi}\Gamma(\mu_v)\Gamma(1-\mu_v) =\frac{1}{2\cos\pi\mathsf n_v}
 =r_v.
 \label{eq:rFactor}
\end{equation}
For the upper attractive ray,
\begin{align}
 \cR_v^{\rm att}J_{v,+}^{\rm att}
 &=-\ii\frac{\ee^{+\ii\pi\mu_v}}{2\pi}
 \Gamma(\mu_v)\Gamma(1-\mu_v)
 \nonumber\\
 &=-\ii\frac{\ee^{+\ii\pi(1/2-\mathsf n_v)}}{2\cos\pi\mathsf n_v}
 =\frac{\ee^{-\ii\pi\mathsf n_v}}{2\cos\pi\mathsf n_v}
 =u_v^+,
 \label{eq:uPlusFromGamma}
\end{align}
and for the lower attractive ray,
\begin{equation}
 \cR_v^{\rm att}J_{v,-}^{\rm att}
 =+\ii\frac{\ee^{-\ii\pi\mu_v}}{2\pi}
 \Gamma(\mu_v)\Gamma(1-\mu_v)
 =\frac{\ee^{+\ii\pi\mathsf n_v}}{2\cos\pi\mathsf n_v}
 =u_v^-.
 \label{eq:uMinusFromGamma}
\end{equation}
These are Eq.~\eqref{eq:attFromN} with the phase $\ii$ removed.  The two attractive values sum to one, in agreement with Eq.~\eqref{eq:unitHankelPeriod} and with the spectral relation \eqref{eq:uComplement}, and the repulsive value satisfies
\begin{equation}
 r_v^2=\frac{1}{4\cos^2\pi\mathsf n_v}=u_v^+u_v^-.
 \label{eq:rSquared}
\end{equation}
These identities are meromorphic in $\mu_v$, or equivalently in $\mathsf n_v$.  At $\mathsf n_v\in\mathbb Z+\tfrac12$, which are the perturbative levels of well $v$ and the poles of the Gamma factors, they are understood by analytic continuation from values of $\mu_v$ away from these poles, and spectral residues are taken only after this continuation.

Let $\widehat C_{I,\pm}^{\rm QZI}$ denote the normalized QZI coefficient of the minimal event sequence, obtained by applying $\mathcal P_v^{\rm QZI}$ to each single-separation factor of Eq.~\eqref{eq:QZIProduct} once the common barrier-weight factor $\prod_{e\in I}B_e$ and the overall normalization have been removed as in Sec.~\ref{subsec:gammaPeriods}.  The two outer wells contribute $u_a^\pm$ and $u_c^\pm$, and each internal well contributes the two repulsive separations of its $\cI\cI$ and $\bar\cI\bar\cI$ pairs, so that
\begin{align}
 \widehat C_{I,\pm}^{\rm QZI}
 &=u_a^\pm u_c^\pm
 \prod_{v=a+1}^{c-1}r_v^2 =u_a^\pm u_c^\pm
 \prod_{v=a+1}^{c-1}u_v^+u_v^-
 \nonumber\\
 &=\mathsf K_Iu_a^\pm u_c^\pm,
 \label{eq:QZIMatch}
\end{align}
with $\mathsf K_I$ the internal factor of Eq.~\eqref{eq:KI}.  Comparison with Eq.~\eqref{eq:spectralTarget} yields
\begin{equation}
\widehat C_{I,\pm}^{\rm QZI}=C_I^\pm.
 \label{eq:QZIequalsSpectral}
\end{equation}
The two lateral QZI periods therefore reproduce the two lateral exact-WKB connected coefficients separately.  On both sides the lateral dependence enters only through the $\cI\bar\cI$ factors, as anticipated after Eq.~\eqref{eq:spectralTarget}.

Equation~\eqref{eq:QZIequalsSpectral} uses the local Weber input \eqref{eq:muWeberIdentification}.  With the strictly one-loop exponent \eqref{eq:muHarmonicDerivation}, the same computation gives
\begin{equation}
 \widehat C_{I,\pm}^{\rm QZI,\,1\text{-loop}}
 =C_I^\pm\big|_{\mathsf n_v\to\epsilon/\omega_v},
 \label{eq:oneLoopMatch}
\end{equation}
with the substitution made at every well of the interval.  For $\epsilon$ in a compact set that avoids the perturbative levels, this is $C_I^\pm+O(\hbar)$.  Near a level the two sides have poles whose positions differ by $O(\hbar)$, so there the one-loop agreement holds in the variable $\mathsf n_v$ rather than uniformly in $\epsilon$.  The contour relations at the $\cI\bar\cI$ pairs derived below are not affected, because their coefficients are integers that do not depend on $\mu_v$, and the one-loop periods obey the complement rule and the DDP relation with $\mathsf n_v$ replaced by $\epsilon/\omega_v$.

With this input, Eq.~\eqref{eq:QZIequalsSpectral} is an off-shell equality, meaning that it holds as an identity of meromorphic functions of the energy before any residue at a level is taken, and it concerns the connected coefficient after the common barrier-weight factor has been removed.  By restoring that factor we define the {\it barrier-weighted connected spectral term}
\begin{equation}
 \mathcal F_{I,\pm}(E,\hbar)
 :=
 \left(\prod_{e\in I}B_e(E,\hbar)\right)C_I^\pm(E,\hbar).
 \label{eq:fullConnectedIntervalTermPhysics}
\end{equation}
The normalized comparison assigns that barrier-weight factor to the QZI contribution. The partition function is obtained only afterwards, by applying the logarithmic-resolvent derivative of Eq.~\eqref{eq:logResolvent} and then the Bromwich inversion \eqref{eq:Bromwich} with its residues,
\begin{equation}
 \mathcal F_{I,\pm}(E,\hbar)
 \ \longmapsto\
 -\partial_E\mathcal F_{I,\pm}(E,\hbar)
 \ \xrightarrow{\ {\rm Bromwich/residues}\ }\
 [\text{connected fixed-grade }T\text{-polynomial}]_\pm .
 \label{eq:connectedCoefficientToTracePhysics}
\end{equation}
The energy derivative therefore acts both on the normalized coefficient and on the energy dependence of the barrier weights.  Treating the barrier weights $B_e$ as formal grading variables, as in the coefficient extraction $[X]f$ of Sec.~\ref{subsec:intervalCoefficient}, is only an intermediate device.  For a multiplet of nearly degenerate levels split by tunneling from a single perturbative level, the coefficient linear in $T$ is the mean shift of the multiplet.  It is a useful scalar extracted only from the connected term.


\subsection{Stokes jump at a single \texorpdfstring{$\cI\bar{\cI}$}{instanton-anti-instanton} pair}
\label{subsec:singlePair}

Section~\ref{subsec:weberMatching} matched the values of the two lateral QZI periods with the two lateral exact-WKB coefficients.  We now show that the passage from one lateral value to the other is induced by an integer relation among the integration cycles themselves, and that this relation is the Picard--Lefschetz jump of an $\cI\bar\cI$ separation across the Stokes direction $\arg\hbar=0$.  Since the internal $\cI\cI$/$\bar\cI\bar\cI$ separations carry no lateral choice, it suffices to consider one attractive separation, at a well $v$ that may be either outer well of the interval or, at higher grade, any well at which the trajectory turns back.  For the rest of this subsection the well is fixed, so we drop the subscript $v$.

The rays $\mathsf c_+$ and $\mathsf c_-$ of Sec.~\ref{subsec:gammaPeriods} are ordinary Gamma contours only when $z^{\mu-1}$ has a genuine branch point at $z=0$, that is, for $\mu\notin\mathbb Z$.  The physical perturbative levels are not of this type.  At $\mathsf n=n+\tfrac12$ with $n=0,1,2,\ldots$ one has $\mu=-n$ by Eq.~\eqref{eq:muWeberIdentification}, and $\Gamma(\mu)$ in Eqs.~\eqref{eq:JNplus} and \eqref{eq:JNminus} has a pole there.  This pole is not an artifact of the contour.  By Eq.~\eqref{eq:uWeber} it is the pole of $u^\pm$ at a zero of $P_v=1+A_v$, that is, at a perturbative level of the well, which was divided out of the quantization function in Eq.~\eqref{eq:normalizedZ}.  The QZI thus sees the perturbative levels of a well as the poles of its Gamma periods.  At these values we define the cycles by a small noninteger shift $\mu\to\mu+\delta$, derive every contour relation at $\delta\neq0$, and continue $\delta\to0$ only in the periods.  The relations derived below have constant integer coefficients, so the continuation does not affect them.

With the orientations fixed in Sec.~\ref{subsec:gammaPeriods}, $\mathsf c_+$ runs from the origin to $-\infty$ just above the negative real axis and $\mathsf c_-$ runs from $-\infty$ back to the origin just below it.  Their sum is the Hankel contour,
\begin{equation}
 H=\mathsf c_+ + \mathsf c_-,
 \label{eq:HankelAsRays}
\end{equation}
and the normalized period map \eqref{eq:normalizedQZIPeriodMap} assigns to the three cycles the values found in Eqs.~\eqref{eq:unitHankelPeriod}, \eqref{eq:uPlusFromGamma}, and \eqref{eq:uMinusFromGamma},
\begin{equation}
 \mathcal P^{\rm QZI}(H)=1,
 \qquad
 \mathcal P^{\rm QZI}(\mathsf c_+)=u^+,
 \qquad
 \mathcal P^{\rm QZI}(\mathsf c_-)=u^-.
 \label{eq:normalizedRayPeriods}
\end{equation}
Here $H$ and $\mathsf c_\pm$ are integration cycles, whereas $1$ and $u^\pm$ are their normalized period values.  Since $\mathcal P^{\rm QZI}$ is linear, applying it to the cycle relation \eqref{eq:HankelAsRays} gives
\begin{equation}
 1=u^+ + u^-,
 \qquad\text{or equivalently}\qquad
 u^-=1-u^+.
 \label{eq:periodImageComplement}
\end{equation}
Since the normalization was fixed on the repulsive separation alone, the unit period of $H$ is itself derived, from Eq.~\eqref{eq:HankelFromN}, and the spectral complement rule \eqref{eq:uComplement} is the period image of a relation among QZI cycles rather than an independent input.

It is useful to separate the two lateral values into a lateral-invariant part and a part that changes sign.  From Eq.~\eqref{eq:uWeber},
\begin{equation}
 u^\pm=\frac12\mp\frac{\ii}{2}\tan\pi\mathsf n,
 \qquad
 u^+-u^-=-\ii\tan\pi\mathsf n.
 \label{eq:uSplit}
\end{equation}
The lateral-invariant part is half the Hankel period.  The part that changes sign is purely imaginary for real $\mathsf n$, and it is the imaginary ambiguity of the $\cI\bar\cI$ amplitude familiar from the BZJ prescription \cite{ZinnJustinJentschuraI}.  In the complete spectral problem it is not observable, because the upper and lower quantization conditions of Sec.~\ref{subsec:lateralQC} are related by the DDP action \eqref{eq:DDPaction} and describe the same levels.  This parallels the cancellation, in the BZJ analysis, between the imaginary part of the $\cI\bar\cI$ amplitude and the Borel ambiguity of perturbation theory.  The QZI adds the contour origin of the ambiguity, since the two lateral rays are the two halves of the single cycle $H$.

\begin{figure}[t]
\centering
 \begin{tikzpicture}[
  >=Stealth,
  axis/.style={->,gray!70,thin},
  contour/.style={very thick,blue!55!black},
  minusH/.style={very thick,orange!85!black},
  mid arrow/.style={postaction={decorate,decoration={markings,mark=at position #1 with {\arrow{Stealth[length=2.2mm]}}}}},
  mid arrow/.default=0.55,
  cut/.style={decorate,decoration={zigzag,amplitude=0.5mm,segment length=1.4mm},gray!65!black},
]
\newcommand{\zplane}{%
  \draw[axis] (-2.4,0) -- (0.95,0);
  \draw[axis] (0,-0.9) -- (0,0.95);
  \draw[cut] (0,0) -- (-2.3,0);
  \fill (0,0) circle (0.035);
}
\begin{scope}
  \zplane
  \node[draw,inner sep=1.5pt,font=\scriptsize] at (0.7,0.72) {$z$};
  \draw[contour,mid arrow] (-0.277,0.16) -- (-2.2,0.16);
  \node[above,blue!55!black,font=\small] at (-1.25,0.16) {$\mathsf c_+$};
  \node[font=\small] at (-0.75,-1.25) {(a) $\mathsf c_+$};
\end{scope}
\node[font=\Large] at (1.35,0) {$-$};
\begin{scope}[xshift=4.2cm]
  \zplane
  \draw[gray,densely dotted,thick] (0,0) -- (0.85,0);
  \draw[contour,mid arrow] (-0.277,0.16) -- (-2.2,0.16);
  \draw[contour,mid arrow] (-2.2,-0.16) -- (-0.277,-0.16);
  \draw[contour,mid arrow=0.3] (-0.277,-0.16) arc[start angle=-150,end angle=150,radius=0.32];
  \fill[gray!65!black] (0.32,0) circle (0.055);
  \node[below right,font=\scriptsize,gray!65!black] at (0.28,-0.02) {$z>0$};
  \node[font=\small] at (-0.75,-1.25) {(b) $H$};
\end{scope}
\node[font=\Large] at (5.55,0) {$=$};
\begin{scope}[xshift=8.4cm]
  \zplane
  \draw[contour,mid arrow] (-0.277,0.32) -- (-2.2,0.32);
  \draw[minusH,mid arrow] (-2.2,0.16) -- (-0.277,0.16);
  \draw[minusH,dashed,mid arrow=0.3] (-0.277,0.16) arc[start angle=150,end angle=-150,radius=0.32];
  \draw[minusH,mid arrow] (-0.277,-0.16) -- (-2.2,-0.16);
  \node[above,gray!40!black,font=\scriptsize] at (-1.25,0.32) {cancel};
  \node[font=\small] at (-0.75,-1.25) {(c) $\mathsf c_+ + (-H)$};
\end{scope}
\node[font=\Large] at (9.75,0) {$=$};
\begin{scope}[xshift=12.6cm]
  \zplane
  \draw[contour,mid arrow] (-0.277,-0.16) -- (-2.2,-0.16);
  \node[below,blue!55!black,font=\small] at (-1.25,-0.16) {$\widetilde{\mathsf c}_-$};
  \node[font=\small] at (-0.75,-1.25) {(d) $\widetilde{\mathsf c}_-=-\mathsf c_-$};
\end{scope}
\end{tikzpicture}
\caption{Illustration of Eq.~\eqref{eq:PLone} in the plane of the variable $z$ of Eq.~\eqref{eq:ychange}, with the cut of $z^{\mu-1}$ along the negative real axis drawn as a zigzag line. (a) The upper ray $\mathsf c_+$, assigned to the $+$ coefficient in the adopted exact-WKB connection convention. (b) The Hankel cycle $H=\mathsf c_++\mathsf c_-$ of Eq.~\eqref{eq:HankelAsRays}. Its small arc is the only piece of any of these cycles that crosses the growing direction $z>0$ of $\ee^{+z}$, at the point marked in gray. (c) Subtracting $H$ reverses each of its pieces (orange). The two upper rays cancel, and the small arc (dashed) shrinks onto the origin, where both rays start, with vanishing integral for $\Re\mu>0$. Only the lower ray run outward survives. (d) The orientation-reversed lower ray of Eq.~\eqref{eq:orientedLowerRay}, assigned to the $-$ coefficient in the same convention and written in the common orientation.}
\label{fig:PLone}
\end{figure}
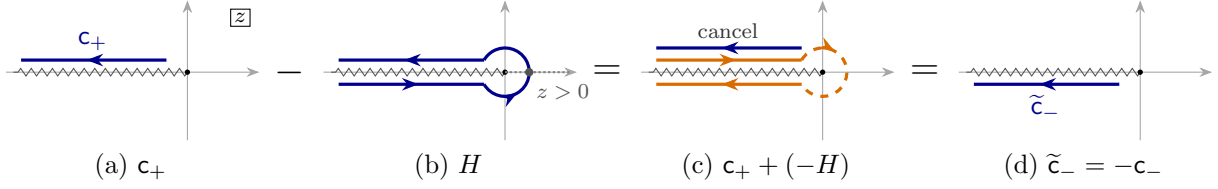

For the attractive sign the integrand of Eq.~\eqref{eq:GammaReduction} grows along the positive real axis and decays only toward the left half-plane. Its direction of fastest decay, the negative real axis, coincides with the cut of $z^{\mu-1}$, so the steepest-descent contour of an $\cI\bar\cI$ separation lies on a Stokes line and each lateral resummation selects one side of it. The upper ray reproduces $u^+$ and the inward lower ray $u^-$ in the chosen period convention. Their coefficient labels are matched to $Q^+$ and $Q^-$ in the exact-WKB convention of Ref.~\cite{SueishiExactWKBI}, because a ray label alone does not determine a physical Borel lateral without the continuation path and orientation. Crossing the Stokes direction therefore replaces the contour $\mathsf c_+$ by $\mathsf c_-$. To write this replacement as a map on a fixed lattice of cycles, we compare the two lateral contours in a common orientation and define the orientation-reversed lower ray
\begin{equation}
 \widetilde{\mathsf c}_-:=-\mathsf c_-,
 \label{eq:orientedLowerRay}
\end{equation}
which runs outward from the origin like $\mathsf c_+$.  Solving Eq.~\eqref{eq:HankelAsRays} for it gives
\begin{equation}
 \widetilde{\mathsf c}_-=\mathsf c_+-H.
 \label{eq:PLone}
\end{equation}
The upper contour is thus carried into the lower one by subtracting exactly one Hankel cycle, as shown step by step in Fig.~\ref{fig:PLone}.

The coefficient of $H$ counts how often the difference $\mathsf c_+-\widetilde{\mathsf c}_-$ crosses the growing direction $z>0$.  The small arc of $H$ around the origin crosses it once and neither ray meets it, so the magnitude of the coefficient is one as a matter of topology, and its sign is fixed by the orientations chosen above.  Being an integer, it cannot be changed by the local Weber normalization of Sec.~\ref{subsec:weberMatching} or by the one-loop determinant of Sec.~\ref{subsec:collectiveNormal}, which multiply periods by nonvanishing factors but do not act on the cycles.

In the ordered basis $(H,\mathsf c)$, where $\mathsf c$ denotes the attractive ray in the common orientation, $\mathsf c_+$ before the crossing and $\widetilde{\mathsf c}_-$ after it, Eq.~\eqref{eq:PLone} reads
\begin{equation}
 H\mapsto H,
 \qquad
 \mathsf c\mapsto\mathsf c-H,
 \qquad
 S_{\cI\bar\cI}=
 \begin{pmatrix}1&0\\ -1&1\end{pmatrix}
 =\mathbf 1+\mathsf N_{\cI\bar\cI},
 \qquad \mathsf N_{\cI\bar\cI}^{\,2}=0.
 \label{eq:PLmatrix}
\end{equation}
The Hankel cycle is fixed because it closes around the branch point and is the same on both sides of the Stokes direction, in agreement with its lateral-invariant period. The matrix is unipotent with a rank-one nilpotent part, which is the form a Picard--Lefschetz transformation takes when a single cycle is added \cite{Witten2010,TanizakiKoike}. We call $S_{\cI\bar\cI}$ the {\it local Picard--Lefschetz transformation} of the reduced post-Laplace QZI. The physical crossing from $0_+$ to $0_-$ has coefficient one, so only $S_{\cI\bar\cI}$ itself enters, and the path-integral argument needs no continuous interpolation of it.

In the common orientation the reversed lower ray $\widetilde{\mathsf c}_-$ has period $-u^-$ rather than the physical $u^-$.  This sign is purely orientational and never reaches a physical amplitude.  The minimal event sequence of an interval contains exactly two $\cI\bar\cI$ pairs, so its lower cycle involves the product of two reversed lower rays, in which the two signs cancel, as Eq.~\eqref{eq:physicalProduct} below makes explicit.

Finally, the jump is located entirely at the infinite-separation end of the quasi-moduli space.  Both rays start at $z=0$, which by Eq.~\eqref{eq:ychange} is the limit $\Re\tau_r\to+\infty$, and $H$ consists of the two rays joined by a small arc around that point.  None of the three cycles reaches the coalescence point $z=\Lambda_{\rm att}$, so the Stokes jump does not involve the region where the two events merge, and it depends on the $\cI\bar\cI$ interaction only through the tail term of the QZI action \eqref{eq:QZIaction}.  This is the concrete content of the remark at the end of Sec.~\ref{subsec:gammaPeriods} that the jump is inherited from the path integral.

\subsection{Stokes jump at the two outer-well \texorpdfstring{$\cI\bar{\cI}$}{instanton-anti-instanton} pairs}
\label{subsec:bothPairs}

We now return to the full minimal event sequence of Eq.~\eqref{eq:primitiveWalk}.  By the Laplace factorization \eqref{eq:LaplaceFactorization}, the post-Laplace QZI \eqref{eq:QZIProduct} is a product of single-separation integrals in independent variables, so its integration cycle is a product of single-separation cycles.  Let $\mathcal C_v^{\rm rep}$ denote the positive real contour of a repulsive separation at well $v$, drawn in Fig.~\ref{fig:gammaHankelContours}(a).  The internal wells contribute the common cycle
\begin{equation}
 \Xi_I:=\prod_{v=a+1}^{c-1}\bigl(\mathcal C_v^{\rm rep}\bigr)^2,
 \label{eq:internalCycle}
\end{equation}
where the square encodes the $\cI\cI$ and $\bar\cI\bar\cI$ separations at each internal well, and the complete lateral cycles of the interval are
\begin{equation}
 \Gamma_{I,+}=\Xi_I\otimes\mathsf c_{a,+}\mathsf c_{c,+},
 \qquad
 \Gamma_{I,-}=\Xi_I\otimes\mathsf c_{a,-}\mathsf c_{c,-}.
 \label{eq:intervalCycles}
\end{equation}
Here juxtaposition and $\otimes$ both denote the product contour in the space of the independent separations.  The normalized period map extends multiplicatively to such products, and this property was used to build $\widehat C_{I,\pm}^{\rm QZI}$ in Eq.~\eqref{eq:QZIMatch}.  In particular $\mathcal P^{\rm QZI}(\Xi_I)=\prod_vr_v^2=\mathsf K_I$ by Eq.~\eqref{eq:rSquared}, and Eqs.~\eqref{eq:QZIMatch} and \eqref{eq:QZIequalsSpectral} state that $\mathcal P^{\rm QZI}(\Gamma_{I,\pm})=C_I^\pm$.

Two features of Eq.~\eqref{eq:intervalCycles} are used in the rest of the argument.  First, the internal cycle $\Xi_I$ is the same on both sides.  A repulsive separation has an admissible real contour, as shown in Sec.~\ref{subsec:gammaPeriods}, so no Stokes line passes through it and it carries no lateral choice.  Second, the two outer factors change independently, because after the Laplace transform the separations of the two $\cI\bar\cI$ pairs are independent variables.  The lateral change of the interval is therefore the product of the two local Picard--Lefschetz transformations of Sec.~\ref{subsec:singlePair}, one at $a$ and one at $c$, with $\Xi_I$ as a spectator.

To apply Eq.~\eqref{eq:PLone} we pass to the common orientation.  The two orientation signs cancel in the lower cycle,
\begin{equation}
 \mathsf c_{a,-}\mathsf c_{c,-}
 =\bigl(-\widetilde{\mathsf c}_{a,-}\bigr)\bigl(-\widetilde{\mathsf c}_{c,-}\bigr)
 =\widetilde{\mathsf c}_{a,-}\widetilde{\mathsf c}_{c,-},
 \label{eq:physicalProduct}
\end{equation}
which is the cancellation anticipated at the end of Sec.~\ref{subsec:singlePair}.  Substituting Eq.~\eqref{eq:PLone} at $a$ and at $c$ then expresses the lower cycle of the interval through upper and Hankel cycles,
\begin{align}
 \Gamma_{I,-}
 &=\Xi_I\otimes(\mathsf c_{a,+}-H_a)(\mathsf c_{c,+}-H_c)
 \nonumber\\
 &=\Xi_I\otimes\bigl(
 \mathsf c_{a,+}\mathsf c_{c,+}
 -H_a\mathsf c_{c,+}
 -\mathsf c_{a,+}H_c
 +H_aH_c\bigr).
 \label{eq:fourContours}
\end{align}
Equivalently, $\Gamma_{I,-}=(S_{\cI\bar\cI,a}\otimes S_{\cI\bar\cI,c})\,\Gamma_{I,+}$, with the matrix of Eq.~\eqref{eq:PLmatrix} acting at each outer well.  In the basis $(\mathsf c_{a,+}\mathsf c_{c,+},\,H_a\mathsf c_{c,+},\,\mathsf c_{a,+}H_c,\,H_aH_c)$ the integer coefficient vector is $(+1,-1,-1,+1)$.  The four terms have a simple interpretation.  The first is the upper cycle itself.  In each of the two middle terms one $\cI\bar\cI$ separation has been replaced by its Hankel cycle while the other stays on its upper ray, and in the last term both have been replaced.  Since each Hankel cycle is lateral invariant, the terms are graded by how many of the two $\cI\bar\cI$ separations have been closed around the infinite-separation point $z=0$.

Taking periods with $\mathcal P^{\rm QZI}(H_v)=1$, and writing $u_v:=u_v^+$ for brevity, gives
\begin{equation}
 \mathcal P^{\rm QZI}(\Gamma_{I,-})
 =\mathsf K_Iu_au_c-\mathsf K_Iu_c-\mathsf K_Iu_a+\mathsf K_I.
 \label{eq:fourSourcePeriods}
\end{equation}
In terms of the exact-WKB variables, Eq.~\eqref{eq:uComplement} gives
\begin{align}
 C_I^-
 &=\mathsf K_Iu_a^-u_c^-
 =\mathsf K_I(1-u_a)(1-u_c)
 \nonumber\\
 &=\mathsf K_I-\mathsf K_Iu_a-\mathsf K_Iu_c+\mathsf K_Iu_au_c.
 \label{eq:DDPFiniteDifference}
\end{align}
The same four terms appear with the same integer coefficients.  Hence
\begin{equation}
 \mathcal P^{\rm QZI}\bigl((S_{\cI\bar\cI,a}\otimes S_{\cI\bar\cI,c})\,\Gamma_{I,+}\bigr)=C_I^-.
 \label{eq:primitiveIntertwining}
\end{equation}
Equation~\eqref{eq:primitiveIntertwining} says more than Eq.~\eqref{eq:QZIequalsSpectral}. There the lower coefficient was obtained by evaluating the lower rays directly. Here it follows from the upper cycle by an integer contour relation, so the lateral change $C_I^+\to C_I^-$ required by the DDP Stokes automorphism is obtained from the reduced QZI contour relation under the stated Weber matching and regular-patch assumptions.

The same structure fixes the size of the lateral ambiguity of the connected coefficient.  Subtracting Eq.~\eqref{eq:fourContours} from $\Gamma_{I,+}$ leaves $\Gamma_{I,+}-\Gamma_{I,-}=\Xi_I\otimes(H_a\mathsf c_{c,+}+\mathsf c_{a,+}H_c-H_aH_c)$, whose period is, by Eq.~\eqref{eq:uSplit},
\begin{align}
 C_I^+-C_I^-
 &=\mathsf K_I\bigl(u_a^++u_c^+-1\bigr)
 \nonumber\\
 &=-\frac{\ii}{2}\mathsf K_I\bigl(\tan\pi\mathsf n_a+\tan\pi\mathsf n_c\bigr).
 \label{eq:intervalLateralDifference}
\end{align}
The product term $\mathsf c_{a,+}\mathsf c_{c,+}$ has dropped out, so at this grade the lateral differences of the two $\cI\bar\cI$ pairs simply add, each weighted by the common internal factor.  For real energies $\mathsf K_I=\prod_v(4\cos^2\pi\mathsf n_v)^{-1}$ is real and the difference is purely imaginary, while the lateral average $\tfrac12(C_I^++C_I^-)=\tfrac14\mathsf K_I(1-\tan\pi\mathsf n_a\tan\pi\mathsf n_c)$ is real.  This extends to the connected coefficient the single-pair decomposition below Eq.~\eqref{eq:uSplit} into a real lateral-invariant part and a purely imaginary part that changes sign.

The contour relation is linear and acts separately at the two outer wells. The lateral change of $C_I$ looks quadratic in the $u_v$ only because the period of a product of cycles is the product of their periods. In a common formal chart, $u_v^+\to1-u_v^+$ exchanges the two explicit coefficient factors. Compatibility with the full quantization function, including the perturbative factor $P_N$, is described in Sec.~\ref{subsec:nonlinearDDP}, where the exact-WKB relation \eqref{eq:QStokesPhysics} is taken as input. 


\section{Ordered event sequences and spectral information loss at higher fixed grade}
\label{sec:higher_order_projection}

Sections~\ref{sec:fullPI} and \ref{sec:QZIStokes} treated the minimal event sequence of a single interval, which crosses each barrier once in each direction and therefore has a fixed ordering.  At higher bion grade a closed event sequence may cross the same barrier several times and revisit the same well, and event sequences with the same crossings can differ in the order in which they occur.  This is the situation of question~(3) in Sec.~\ref{subsec:questions}.  The local components are unchanged.  Every visit to a well is still either an $\cI\bar\cI$ pair or an $\cI\cI$/$\bar\cI\bar\cI$ pair, and after the local Weber normalization of Sec.~\ref{subsec:weberMatching} the first contributes $u_v^\pm$ and the second $r_v$.  In this section we combine these normalized periods into a transfer matrix acting on oriented tunneling events, show that its determinant reproduces the exact-WKB function $\cZ_N^\pm$, and use its trace expansion to determine which features of an ordered event sequence survive in the scalar spectral coefficient.  We then check that the Stokes jump of Sec.~\ref{sec:QZIStokes} respects this loss of information, and close by explaining how the nonlinear form of the DDP action arises from the linear contour relation.

\subsection{Transfer matrix of consecutive tunneling events}
\label{subsec:transferMatrix}

At a well $v$ the local Weber equation has two independent WKB solutions, which at an internal well may be taken to describe motion toward the left and toward the right neighbor.  The connection across the well is therefore described by a $2\times2$ matrix.  With
\begin{equation}
 \mathsf S_+=\begin{pmatrix}1&\ii\\0&1\end{pmatrix},\qquad
 \mathsf S_-=\begin{pmatrix}1&0\\\ii&1\end{pmatrix},\qquad
 a_v^\pm:=\frac{u_v^\pm}{r_v},
\end{equation}
the normalized matrix that carries the solutions on the left of well $v$ to those on its right is
\begin{align}
 \mathsf W_v^\pm
 &=(\mathsf S_-)^{-1}
 \operatorname{diag}\bigl(a_v^\pm,(a_v^\pm)^{-1}\bigr)\mathsf S_+
 \notag\\
 &=
 \begin{pmatrix}
 u_v^\pm/r_v&\ii u_v^\pm/r_v\\
 -\ii u_v^\pm/r_v&(r_v^2+(u_v^\pm)^2)/(u_v^\pm r_v)
 \end{pmatrix}
 =
 \frac1{r_v}
 \begin{pmatrix}
 u_v^\pm&\ii u_v^\pm\\
 -\ii u_v^\pm&1
 \end{pmatrix}.
 \label{eq:physicalWeberTransferPhysics}
\end{align}
In the last step we used $r_v^2=u_v^\pm(1-u_v^\pm)$, which follows from Eqs.~\eqref{eq:uComplement} and \eqref{eq:rSquared}, and the same relation shows that $\det\mathsf W_v^\pm=1$.  By Eq.~\eqref{eq:uWeber} and the definition of $r_v$ in Eq.~\eqref{eq:weberConnectionMatrix}, $a_v^\pm=\ee^{\mp\ii\pi\mathsf n_v}$, so the diagonal factor is the WKB phase accumulated across the well, half of the perturbative period of Eq.~\eqref{eq:AWeber}, while $\mathsf S_\pm$ are the elementary connection matrices at the two turning points of the well.

The same information is more transparent as a connection matrix $\Sigma$ that maps the two local solutions entering the well to the two leaving it.  For a unit-determinant matrix
\begin{equation}
 W=\begin{pmatrix}\alpha&\beta\\ \gamma&\delta\end{pmatrix},
 \qquad \det W=1,
\end{equation}
one has
\begin{equation}
 \Sigma=\frac1{\delta}
 \begin{pmatrix}
 -\gamma&1\\
 1&\beta
 \end{pmatrix},
 \label{eq:transferToScatteringPhysics}
\end{equation}
and substituting Eq.~\eqref{eq:physicalWeberTransferPhysics} gives
\begin{equation}
 \Sigma_v^\pm=
 \begin{pmatrix}
 \ii u_v^\pm&r_v\\
 r_v&\ii u_v^\pm
 \end{pmatrix}.
 \label{eq:physicalWeberScatteringPhysics}
\end{equation}
The diagonal entry, in which the trajectory leaves toward the neighbor it came from, is the $\cI\bar\cI$ factor $\ii u_v^\pm$.  The off-diagonal entry, in which it continues to the other neighbor, is the $\cI\cI$/$\bar\cI\bar\cI$ factor $r_v$.  This is the matrix \eqref{eq:weberConnectionMatrix} used in Sec.~\ref{subsec:weberMatching}, and its entries are the periods \eqref{eq:attFromN} and \eqref{eq:repNormalizationCondition} of the attractive and repulsive separations under the single normalization $\mathcal N_v$.  The factor $\ii$ is a phase convention of the connection matrix and enters a closed event sequence only through the power $\ii^{R(w)}$ found below.  Since the evolution is in Euclidean time, $\Sigma_v^\pm$ is complex symmetric but not unitary, and ``entering'' and ``leaving'' refer only to the order of events in time.

\begin{figure}[t]
\centering
\begin{tikzpicture}[>=Stealth]
\begin{scope}[shift={(0,0)}]
  \draw[gray!55,densely dotted] (0,0.0) -- (2.50,0.0);
  \node[left,font=\scriptsize] at (0,0.0) {$x_{v-1}$};
  \draw[gray!55,densely dotted] (0,0.62) -- (2.50,0.62);
  \node[left,font=\scriptsize] at (0,0.62) {$x_v$};
  \draw[gray!55,densely dotted] (0,1.24) -- (2.50,1.24);
  \node[left,font=\scriptsize] at (0,1.24) {$x_{v+1}$};
  \draw[very thick,blue!55!black] (0,0.000) -- (0.425,0.000) to[out=0,in=180] (0.825,0.620) -- (1.675,0.620) to[out=0,in=180] (2.075,0.000) -- (2.500,0.000);
  \node[above,font=\scriptsize] at (1.250,0.620) {$\ii u_v^\pm$};
  \node[font=\small,anchor=west] at (-0.6,1.95) {(a) $q=p$, an $\cI\bar\cI$ pair};
\end{scope}
  \node[font=\scriptsize,above left] at (0.625,0.310) {$d$};
  \node[font=\scriptsize,above right] at (1.875,0.310) {$d'$};
\begin{scope}[shift={(4.2,0)}]
  \draw[gray!55,densely dotted] (0,0.0) -- (2.50,0.0);
  \node[left,font=\scriptsize] at (0,0.0) {$x_{v-1}$};
  \draw[gray!55,densely dotted] (0,0.62) -- (2.50,0.62);
  \node[left,font=\scriptsize] at (0,0.62) {$x_v$};
  \draw[gray!55,densely dotted] (0,1.24) -- (2.50,1.24);
  \node[left,font=\scriptsize] at (0,1.24) {$x_{v+1}$};
  \draw[very thick,blue!55!black] (0,0.000) -- (0.425,0.000) to[out=0,in=180] (0.825,0.620) -- (1.675,0.620) to[out=0,in=180] (2.075,1.240) -- (2.500,1.240);
  \node[below,font=\scriptsize] at (1.250,0.620) {$r_v$};
  \node[font=\small,anchor=west] at (-0.6,1.95) {(b) $q\neq p$, an $\cI\cI$ pair};
\end{scope}
  \node[font=\scriptsize,above left] at (4.825,0.310) {$d$};
  \node[font=\scriptsize,above left] at (6.075,0.930) {$d'$};
\begin{scope}[shift={(0,-3.1)}]
  \draw[gray!55,densely dotted] (0,0.0) -- (5.00,0.0);
  \node[left,font=\scriptsize] at (0,0.0) {$x_{v-1}$};
  \draw[gray!55,densely dotted] (0,0.62) -- (5.00,0.62);
  \node[left,font=\scriptsize] at (0,0.62) {$x_v$};
  \draw[gray!55,densely dotted] (0,1.24) -- (5.00,1.24);
  \node[left,font=\scriptsize] at (0,1.24) {$x_{v+1}$};
  \draw[very thick,blue!55!black] (0,0.000) -- (0.425,0.000) to[out=0,in=180] (0.825,0.620) -- (1.675,0.620) to[out=0,in=180] (2.075,1.240) -- (2.925,1.240) to[out=0,in=180] (3.325,0.620) -- (4.175,0.620) to[out=0,in=180] (4.575,0.000) -- (5.000,0.000);
  \node[below,font=\scriptsize] at (1.250,0.620) {$r_v$};
  \node[below,font=\scriptsize] at (3.750,0.620) {$r_v$};
  \node[font=\small,anchor=west] at (-0.6,1.95) {(c) minimal event sequence of $I$};
\end{scope}
\begin{scope}[shift={(6.9,-3.1)}]
  \draw[gray!55,densely dotted] (0,0.0) -- (2.50,0.0);
  \node[left,font=\scriptsize] at (0,0.0) {$x_{v-1}$};
  \draw[gray!55,densely dotted] (0,0.62) -- (2.50,0.62);
  \node[left,font=\scriptsize] at (0,0.62) {$x_v$};
  \draw[gray!55,densely dotted] (0,1.24) -- (2.50,1.24);
  \node[left,font=\scriptsize] at (0,1.24) {$x_{v+1}$};
  \draw[very thick,blue!55!black] (0,0.000) -- (0.425,0.000) to[out=0,in=180] (0.825,0.620) -- (1.675,0.620) to[out=0,in=180] (2.075,0.000) -- (2.500,0.000);
  \node[above,font=\scriptsize] at (1.250,0.620) {$\ii u_v^\pm$};
  \node[font=\small,anchor=west] at (-0.6,1.95) {(d) minimal event sequences of the two subintervals};
\end{scope}
\begin{scope}[shift={(10.3,-3.1)}]
  \draw[gray!55,densely dotted] (0,0.0) -- (2.50,0.0);
  \draw[gray!55,densely dotted] (0,0.62) -- (2.50,0.62);
  \draw[gray!55,densely dotted] (0,1.24) -- (2.50,1.24);
  \draw[very thick,blue!55!black] (0,1.240) -- (0.425,1.240) to[out=0,in=180] (0.825,0.620) -- (1.675,0.620) to[out=0,in=180] (2.075,1.240) -- (2.500,1.240);
  \node[below,font=\scriptsize] at (1.250,0.620) {$\ii u_v^\pm$};
\end{scope}
  \node[font=\small] at (9.85,-2.48) {$\times$};
  \node[font=\small,align=left,anchor=west] at (8.0,0.62) {$\mathbb T^\pm_{d,d'}=\sqrt{B_{\{v,q\}}}\times
  \begin{cases}\ii u_v^\pm,& q=p,\\ r_v,& q\neq p,\end{cases}$\\[3pt] for $d=(p\to v)$ and $d'=(v\to q)$};
  \node[font=\small] at (6.3,-4.0) {at an internal well $v$ of $I$, the sum of (c) and (d) gives $r_v^2+(u_v^\pm)^2=u_v^\pm$};
\end{tikzpicture}
\caption{The transfer matrix \eqref{eq:transferMatrix} and the determinant identity \eqref{eq:determinantIdentity}, drawn as trajectories through three neighboring wells.  (a), (b) The two entries of $\mathbb T^\pm_{d,d'}$ at a well $v$ for consecutive events $d=(p\to v)$ and $d'=(v\to q)$.  If the trajectory returns to the well it came from, the visit is an $\cI\bar\cI$ pair and contributes $\ii u_v^\pm$.  If it continues to the other neighbor, the visit is an $\cI\cI$ or $\bar\cI\bar\cI$ pair and contributes $r_v$.  Each event also carries the square root of the barrier weight of the barrier it crosses.  (c), (d) The two ways in which $\det(1-\mathbb T^\pm)$ covers the barriers of an interval $I$ at an internal well $v$, with the factors at the outer wells of $I$ omitted.  The minimal event sequence of $I$ passes through $v$ twice and gives $r_v^2$, while the minimal event sequences of the two subintervals on either side of $v$ both turn back at $v$ and give $(u_v^\pm)^2$ once the signs are included.  By Eq.~\eqref{eq:cutIdentity} the two alternatives add up to the factor $u_v^\pm$ of Eq.~\eqref{eq:intervalActivity}.}
\label{fig:transferEntries}
\end{figure}
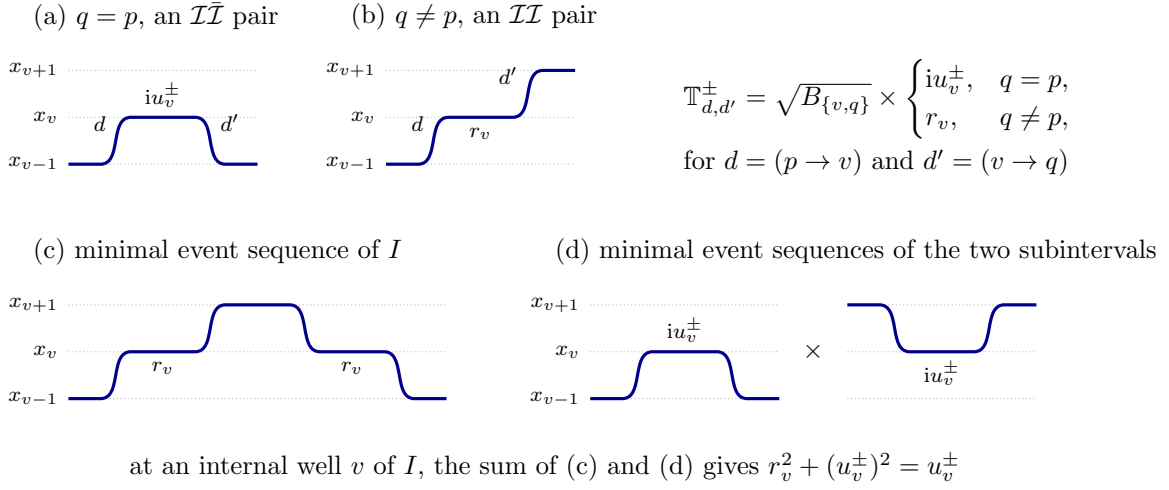

To follow an event sequence through several wells, we label each tunneling event by the two wells it connects and the direction in which it crosses the barrier between them, $d=(p\to v)$ with $p=v\pm1$.  An instanton crossing barrier $e$ from left to right is the event $(e\to e+1)$, and the corresponding anti-instanton is $(e+1\to e)$.  Two events
\begin{equation}
 d=(p\to v),
 \qquad d'=(v\to q)
\end{equation}
are consecutive when the second leaves the well that the first enters.  Let $B_{\{v,q\}}$ denote the barrier weight of the barrier between the neighboring wells $v$ and $q$.  Attaching to each event the square root of the barrier weight of the barrier it crosses, the {\it transfer matrix} between consecutive events is
\begin{equation}
 \mathbb T_{d,d'}^\pm
 =\sqrt{B_{\{v,q\}}}
 \begin{cases}
 \ii u_v^\pm,&q=p\quad\text{($\cI\bar\cI$)},\\[1mm]
 r_v,&q\neq p\quad\text{($\cI\cI$/$\bar\cI\bar\cI$)}.
 \end{cases}
 \label{eq:transferMatrix}
\end{equation}
The square root assigns half of a barrier weight to each crossing, in agreement with Eq.~\eqref{eq:BfromInstantonPair}, and the two entries are drawn in Fig.~\ref{fig:transferEntries}(a) and (b).  A closed event sequence crosses each barrier equally often in the two directions, so its weight contains integer powers of the barrier weights $B_e$.

The transfer matrix keeps no memory of the order of events beyond the current one.  To see this, let $E_{d',d}$ be the matrix with a single unit entry that maps the basis vector of event $d$ to that of event $d'$, and define
\begin{equation}
 \mathcal A_{d\to d'}^\pm
 :=\mathbb T_{d,d'}^\pm E_{d',d}.
\end{equation}
Along a sequence of consecutive events $(d_0,d_1,\ldots,d_L)$ all intermediate labels cancel,
\begin{equation}
 \mathcal A_{d_{L-1}\to d_L}^\pm\cdots
 \mathcal A_{d_0\to d_1}^\pm
 =\left(\prod_{j=0}^{L-1}\mathbb T_{d_j,d_{j+1}}^\pm\right)
 E_{d_L,d_0}.
 \label{eq:physicalChannelTelescopingPhysics}
\end{equation}
For a closed event sequence, $d_L=d_0$, the product is the scalar weight of the sequence times the projector onto its first event.  The two-dimensional local space at each well therefore stores nothing about the sequence beyond the product of its local factors.

For the minimal closed event sequence \eqref{eq:primitiveWalk} of an interval $I=[a,b]$, the two outer wells are $\cI\bar\cI$ pairs, which give $\ii^2=-1$, and each internal well is visited once by an $\cI\cI$ pair and once by a $\bar\cI\bar\cI$ pair.  Its weight is
\begin{equation}
 -\left(\prod_{e=a}^{b}B_e\right)
 u_a^\pm u_{b+1}^\pm
 \prod_{v=a+1}^{b}r_v^2.
 \label{eq:simpleCycleWeight}
\end{equation}
This differs from the interval weight $w_I^\pm$ of Eq.~\eqref{eq:intervalActivity} by the factor $r_v^2$ in place of $u_v^\pm$ at each internal well, and the determinant of $1-\mathbb T^\pm$ accounts for the difference.  A closed event sequence that uses each oriented event at most once is the minimal event sequence of an interval, so $\det(1-\mathbb T^\pm)$ expands into products of minimal event sequences that share no event, each contributing its weight with an extra minus sign.  At an internal well $v$ of $I$ the barriers of $I$ can be covered either by the minimal event sequence of $I$, which passes through $v$ and contributes $r_v^2$, or by the minimal event sequences of the two subintervals on either side of $v$, which both turn back at $v$ and together contribute $(u_v^\pm)^2$ once the two extra signs are included, as shown in Fig.~\ref{fig:transferEntries}(c) and (d).  Since
\begin{equation}
 r_v^2+(u_v^\pm)^2
 =u_v^\pm(1-u_v^\pm)+(u_v^\pm)^2
 =u_v^\pm,
 \label{eq:cutIdentity}
\end{equation}
the sum over these alternatives restores the factor $u_v^\pm$ of Eq.~\eqref{eq:intervalActivity} at every internal well.  This is the inclusion and exclusion of Eq.~\eqref{eq:internalFactorDerivation} read in reverse, and it gives
\begin{equation}
 \cZ_N^\pm=\det(1-\mathbb T^\pm).
 \label{eq:determinantIdentity}
\end{equation}
The transfer matrix built from the normalized QZI periods thus reproduces the exact-WKB function of Eq.~\eqref{eq:normalizedZ} exactly.  The identity uses only $u_v^++u_v^-=1$ and $r_v^2=u_v^+u_v^-$, so it holds for the exact resummed Voros symbols wherever the input quantization condition does, whereas identifying an individual event sequence with a product of QZI periods requires the regular dilute, one-loop approximation.  Taking the logarithm turns the determinant into a sum over all closed sequences of consecutive events, now with repetitions allowed,
\begin{equation}
 \log\cZ_N^\pm
 =-\sum_{r\ge1}\frac1r\Tr(\mathbb T^\pm)^r.
 \label{eq:traceLog}
\end{equation}
A term of $\Tr(\mathbb T^\pm)^r$ is a closed event sequence of $r$ events together with a choice of its first event.  Choosing the first event fixes the time origin, and the sum over the $r$ choices divided by $r$ removes that choice again.  Equation~\eqref{eq:traceLog} is therefore the connected expansion of $\log\cZ_N^\pm$ over ordered closed event sequences, as used below. 

\subsection{Dependence of the scalar weight on a closed event sequence}
\label{subsec:wordWeights}

Write a closed event sequence of bion grade $k$ as the list of wells it visits,
\begin{equation}
 w=(v_0,v_1,\ldots,v_{2k-1}),
 \qquad v_{2k}=v_0,
 \label{eq:word}
\end{equation}
with $|v_{j+1}-v_j|=1$ and indices read cyclically.  Each consecutive pair $(v_j,v_{j+1})$ is one tunneling event.  Identifying barrier $e$ with the pair of wells $\{e,e+1\}$ it separates, we define
\begin{align}
 k_e(w)&:=\frac12\#\{j:\{v_j,v_{j+1}\}=e\},
 \label{eq:ke}\\
 R_v(w)&:=\#\{j:v_j=v,\;v_{j-1}=v_{j+1}\},
 \label{eq:Rv}\\
 X_v(w)&:=\#\{j:v_j=v,\;v_{j-1}\neq v_{j+1}\}.
 \label{eq:Xv}
\end{align}
Thus $k_e$ is the number of times barrier $e$ is crossed in each direction, $R_v$ counts the $\cI\bar\cI$ pairs at well $v$, and $X_v$ counts the $\cI\cI$/$\bar\cI\bar\cI$ pairs there.  These form the data $(\mathbf k,\mathbf R,\mathbf X)$ of question~(3) in Sec.~\ref{subsec:questions}.

Multiplying the local factors of Eq.~\eqref{eq:transferMatrix} along the sequence gives its leading scalar weight.  Since the factors $u_v^\pm$ and $r_v$ are the normalized periods \eqref{eq:normalizedQZIPeriodMap} of its single separations, we call this weight the {\it period} of $w$ and write
\begin{equation}
 \operatorname{Per}_{w,\pm}
 =
 \bB^{\mathbf k(w)}
 \ii^{R(w)}
 \prod_v(u_v^\pm)^{R_v(w)}r_v^{X_v(w)},
 \qquad
 \bB^{\mathbf k}:=\prod_eB_e^{k_e},
 \label{eq:rawWordWeight}
\end{equation}
where $R(w)=\sum_vR_v(w)$.  Each $\cI\bar\cI$ pair reverses the direction of tunneling, and a closed event sequence reverses direction an even number of times, so $R(w)$ is even.  Each $X_v$ is even as well, because a closed event sequence that passes through well $v$ from left to right must pass through it from right to left equally often.  Hence, using $r_v^2=u_v^+u_v^-$,
\begin{equation}
 \operatorname{Per}_{w,\pm}
 =
 (-1)^{R(w)/2}\bB^{\mathbf k(w)}
 \prod_v(u_v^\pm)^{R_v(w)}
 (u_v^+u_v^-)^{X_v(w)/2}.
 \label{eq:reducedWordWeight}
\end{equation}
The order of events no longer appears.  The weight depends on the sequence only through $(\mathbf k,\mathbf R,\mathbf X)$.

The weight resolves exactly these counts.  Every arrival at $v$ is the first event of either an $\cI\bar\cI$ pair or an $\cI\cI$/$\bar\cI\bar\cI$ pair at $v$, so
\begin{equation}
 R_v+X_v=k_{v-1}+k_v,
 \qquad k_0=k_N=0.
 \label{eq:arrivalIdentity}
\end{equation}
For either lateral choice write $z_v:=u_v^\pm$, so that $u_v^+u_v^-=z_v(1-z_v)$.  Summing Eq.~\eqref{eq:arrivalIdentity} over $v$ gives
\begin{equation}
 R(w)+X(w)=2|\mathbf k|,
 \qquad X(w):=\sum_vX_v,
 \qquad |\mathbf k|:=\sum_ek_e,
 \label{eq:RXtotal}
\end{equation}
and therefore
\begin{equation}
 (-1)^{R/2}\prod_v(1-z_v)^{X_v/2}
 =(-1)^{|\mathbf k|}\prod_v(z_v-1)^{X_v/2}.
 \label{eq:wordSignDerivation}
\end{equation}
Equation~\eqref{eq:reducedWordWeight} thus becomes
\begin{equation}
 \operatorname{Per}_{w,\pm}
 =(-1)^{|\mathbf k|}\bB^{\mathbf k}
 \prod_v
 z_v^{\,k_{v-1}+k_v-X_v/2}(z_v-1)^{X_v/2}.
 \label{eq:countPolynomial}
\end{equation}
Different barrier multiplicities are separated by $\bB^{\mathbf k}$. At fixed $\mathbf k$, each well contributes a polynomial $z^{m-b}(z-1)^b$ with $m=k_{v-1}+k_v$ fixed and $b=X_v/2$, which vanishes to order $m-b$ at $z=0$, so different values of $b$ give linearly independent polynomials. Since $B_e$ and $z_v$ are independent variables, two closed event sequences have the same leading scalar weight exactly when their $(\mathbf k,\mathbf R,\mathbf X)$ agree. The scalar weight resolves the count data completely when $B_e$ and $z_v$ are independent formal variables. For a particular potential they are functions of $(E,\hbar)$, and their relations can identify further event sequences. In a reflection-symmetric triple well, for example, $B_1=B_2$ and $u_1^\pm=u_3^\pm$ already identify the two elementary left and right excursions. This is a further physical specialization, distinct from equal formal count data.

Section~\ref{subsec:tripleHigher} gives a grade-six triple-well pair and an earlier primitive grade-five pair on four consecutive wells. Equal count data is not the only source of information loss. Repeated occurrences are grouped before the scalar is evaluated, and the physical integration cycle occupies only one combination of QZI cycles. We aim to describe both next.

\subsection{Projection from ordered event sequences to spectral data}
\label{subsec:projection}

Let $\Pi_{K,\pm}$ denote the map that takes each event sequence through total bion grade $K$ to its normalized connected scalar contribution.  In the trace expansion \eqref{eq:traceLog} an event sequence of grade $k$ has $2k$ events and carries the factor $-1/(2k)$.  If $\mathsf s(w)$ cyclic shifts map $w$ to itself, the trace contains $2k/\mathsf s(w)$ distinct choices of first event, so the cyclic class of $w$ receives the familiar symmetry factor
\begin{equation}
 -\frac{1}{2k}\frac{2k}{\mathsf s(w)}
 =-\frac{1}{\mathsf s(w)}.
 \label{eq:rootingFactorDerivation}
\end{equation}
The contribution of a single event sequence with a chosen first event is therefore
\begin{equation}
 \Pi_{K,\pm}(w)
 =-\frac{1}{2k(w)}\operatorname{Per}_{w,\pm}.
 \label{eq:PiPhysics}
\end{equation}

The contour relation $\widetilde{\mathsf c}_-=\mathsf c_+-H$ at an $\cI\bar\cI$ pair has integer coefficients, and so does its product over several pairs, as in Eq.~\eqref{eq:fourContours}.  We note that rational factors such as $1/(2k)$ and $1/\mathsf s(w)$ enter only when one passes from event sequences with a chosen first event to the connected trace.  The fractional coefficients of $\log\cZ_N^\pm$ are therefore trace weights, while the underlying Picard--Lefschetz coefficients remain integers.

Within a fixed count class $\sigma=(\mathbf k,\mathbf R,\mathbf X)$, let $W_\sigma$ denote the set of event sequences realizing it.  All of them have the same scalar image, so any combination
\begin{equation}
 \sum_{w\in W_\sigma}c_w\,w,
 \qquad
 \sum_{w\in W_\sigma}c_w=0,
 \label{eq:invisibleCombination}
\end{equation}
is invisible to the leading spectral coefficient.

\begin{figure}[t]
\centering
 \begin{tikzpicture}[>=Stealth,
  sp/.style={font=\normalsize,inner sep=3pt},
  map/.style={->,thick,black!75},
  st/.style={->,thick,gray!65!black},
  lost/.style={font=\scriptsize\itshape,gray!35!black,align=center},
  ml/.style={font=\small,fill=white,inner sep=1.5pt},
]
\node[sp] (o1) at (0,0) {$\mathcal H_{\rm occ}^{(K)}$};
\node[sp] (g1) at (4,0) {$\mathcal H_{\rm grp}^{(K)}$};
\node[sp] (p1) at (8,0) {$\mathcal H_{\rm phys}^{(K)}$};
\node[sp] (s1) at (12,0) {$\mathcal O_{\rm spec}^{(K,+)}$};
\node[sp] (o2) at (0,-2.4) {$\mathcal H_{\rm occ}^{(K)}$};
\node[sp] (g2) at (4,-2.4) {$\mathcal H_{\rm grp}^{(K)}$};
\node[sp] (p2) at (8,-2.4) {$\mathcal H_{\rm phys}^{(K)}$};
\node[sp] (s2) at (12,-2.4) {$\mathcal O_{\rm spec}^{(K,-)}$};
\draw[map] (o1) -- node[ml,above]{$\mathsf G^{(K)}$} (g1);
\draw[map] (g1) -- node[ml,above]{$n^{(K)}$} (p1);
\draw[map] (p1) -- node[ml,above]{$\Pi_{K,+}$} (s1);
\draw[map] (o2) -- node[ml,below]{$\mathsf G^{(K)}$} (g2);
\draw[map] (g2) -- node[ml,below]{$n^{(K)}$} (p2);
\draw[map] (p2) -- node[ml,below]{$\Pi_{K,-}$} (s2);
\draw[st] (o1) -- node[ml,left]{$\cS_{\rm PI}^{(K)}$} (o2);
\draw[st] (g1) -- (g2);
\draw[st] (p1) -- node[ml,left]{$\cS_{\rm phys}^{(K)}$} (p2);
\draw[st] (s1) -- node[ml,left]{$\cS_{\rm DDP}^{\rm obs}$} (s2);
\node[lost] at (2,0.95) {loses occurrence and\\ logarithmic-sheet labels};
\node[lost] at (6,0.95) {loses grouped combinations\\ not in the physical cycle};
\node[lost] at (10,0.95) {loses the ordering\\ within a count class};
\node[font=\small,gray!25!black] at (6,-3.45) {invisible combinations:\quad $\mathcal V_{\rm grp}^{(K)}\subseteq\mathcal V_{\rm coeff}^{(K)}\subseteq\mathcal V_{\rm spec}^{(K,\pm)}$};
\node[font=\scriptsize,gray!65!black,anchor=west] at (12.25,-1.2) {Stokes jump};
\end{tikzpicture}
\caption{The three stages of information loss of Sec.~\ref{subsec:projection} and their compatibility with the Stokes jump of Sec.~\ref{subsec:stokesCovariance}. The upper and lower rows are the two lateral sides. Along each row, the grouping map $\mathsf G^{(K)}$, the coefficient selection $n^{(K)}$, and the scalar evaluation $\Pi_{K,\pm}$ compose to the spectral projection $O_{K,\pm}$ of Eq.~\eqref{eq:standardPhysicalObservationPhysics}, and the invisible combinations after each stage form the nested sets of Eq.~\eqref{eq:physicalKernelFiltrationPhysics}. The vertical arrows are the Picard--Lefschetz continuation, the product of the local transformations $S_{\cI\bar\cI}$ at all $\cI\bar\cI$ pairs, and its DDP image on the scalar coefficients. All squares commute under the assumptions on physical-cycle selection and its compatibility with the Stokes jump stated in Secs.~\ref{subsec:projection} and \ref{subsec:stokesCovariance}. }
\label{fig:threeStages}
\end{figure}
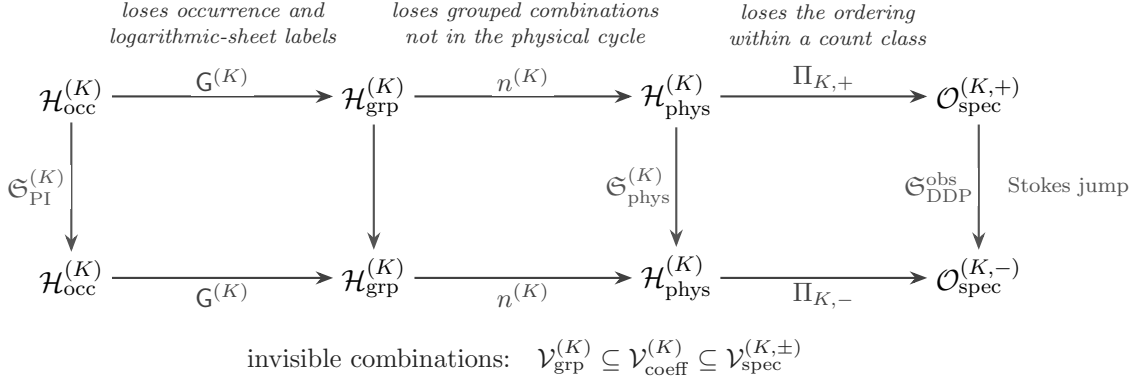

Equal scalar weights within a count class are the last of three stages at which information about an event sequence is lost.  To describe all three, we recall from Sec.~\ref{subsec:collectiveNormal} that $\mathcal H_w^{\rm QZI,occ}$ is the lattice of collective-coordinate cycles of the sequence $w$ in which repeated occurrences of a well are kept distinct.  Let $\mathcal W_{\le K}$ be the finite set of event sequences through total bion grade $K$, and collect these lattices into
\begin{equation}
 \mathcal H_{\rm occ}^{(K)}
 :=\bigoplus_{w\in\mathcal W_{\le K}}\mathcal H_w^{\rm QZI,occ}.
 \label{eq:gradeWiseQZIModulesPhysics}
\end{equation}
The first stage is the grouping of occurrences.  Within one sequence, repeated visits to the same well by the same kind of pair give separation integrals of the same form.  They are distinguished only by their occurrence label and by the logarithmic sheet of the complex separation variable, which by Eq.~\eqref{eq:vacuumWindingShiftPhysics} is an integer winding around the perturbative $A$-cycle of the well.  The scalar coefficient does not depend on how these labels are distributed among the occurrences.  The grouping map $\mathsf G_w:\mathcal H_w^{\rm QZI,occ}\to\mathcal H_w^{\rm QZI,grp}$ forgets this distribution, and we write
\begin{equation}
 \mathsf G^{(K)}:=\bigoplus_{w\in\mathcal W_{\le K}}\mathsf G_w,
 \qquad
 \mathcal H_{\rm grp}^{(K)}
 :=\bigoplus_{w\in\mathcal W_{\le K}}\mathcal H_w^{\rm QZI,grp}.
 \label{eq:gradeWiseGroupingPhysics}
\end{equation}
The second stage is the selection of the physical cycle. The physical integration cycle of an event sequence is one particular integer combination of its grouped cycles. For the minimal event sequence this combination is explicit, namely the cycle $\Gamma_{I,\pm}$ of Eq.~\eqref{eq:intervalCycles}. For arbitrary event sequences we assume that, on the regular dilute region, a locally constant primitive integer covector $n_w^{\mathsf T}$ defines the coefficient map onto the chosen rank-one channel generated by $\Gamma_w^{\rm phys}$. For the Stokes comparison we additionally require this coefficient map to be compatible with continuation, as specified in Sec.~\ref{subsec:stokesCovariance}. Local constancy and primitivity alone do not establish that compatibility. With it we set
\begin{equation}
 \mathcal H_{\rm phys}^{(K)}
 :=\bigoplus_{w\in\mathcal W_{\le K}}\mathbb Z\,\Gamma_w^{\rm phys},
 \qquad
 n^{(K)}:=\bigoplus_{w\in\mathcal W_{\le K}}n_w^{\mathsf T}
 :\mathcal H_{\rm grp}^{(K)}\longrightarrow\mathcal H_{\rm phys}^{(K)}.
 \label{eq:physicalCoefficientMapPhysics}
\end{equation}
The third stage is the scalar evaluation.  Extending Eq.~\eqref{eq:PiPhysics} linearly gives a map $\Pi_{K,\pm}:\mathcal H_{\rm phys}^{(K)}\to\mathcal O_{\rm spec}^{(K,\pm)}$ into the normalized connected spectral coefficients through grade $K$.  The composite of the three stages is the spectral projection
\begin{equation}
 O_{K,\pm}
 :=\Pi_{K,\pm}\circ n^{(K)}\circ\mathsf G^{(K)}:
 \mathcal H_{\rm occ}^{(K)}
 \longrightarrow
 \mathcal O_{\rm spec}^{(K,\pm)}.
 \label{eq:standardPhysicalObservationPhysics}
\end{equation}
The combinations of occurrence-resolved cycles that are invisible after each stage are
\begin{equation}
 \mathcal V_{\rm grp}^{(K)}:=\ker\mathsf G^{(K)},
 \qquad
 \mathcal V_{\rm coeff}^{(K)}:=\ker\bigl(n^{(K)}\circ\mathsf G^{(K)}\bigr),
 \qquad
 \mathcal V_{\rm spec}^{(K,\pm)}:=\ker O_{K,\pm},
 \label{eq:threePhysicalKernelsPhysics}
\end{equation}
and since each of these composites is obtained from the previous one by applying one more map,
\begin{equation}
 \mathcal V_{\rm grp}^{(K)}
 \subseteq\mathcal V_{\rm coeff}^{(K)}
 \subseteq\mathcal V_{\rm spec}^{(K,\pm)}.
 \label{eq:physicalKernelFiltrationPhysics}
\end{equation}
The first stage loses the occurrence and branch labels, the second loses the grouped combinations that the physical cycle does not occupy, and the third loses the ordering within a count class.  A lift of a zero-sum combination \eqref{eq:invisibleCombination} to $\mathcal H_{\rm occ}^{(K)}$ lies in $\mathcal V_{\rm spec}^{(K,\pm)}$ but not in $\mathcal V_{\rm coeff}^{(K)}$, so the last inclusion is strict whenever a count class contains more than one sequence.  Once a combination is invisible to the scalar coefficient, no further processing of that coefficient can recover it.  The explicit triple-well degeneracy concerns the third stage and does not use the coefficient-selection assumption, which is needed only to describe the second stage for arbitrary event sequences.  The three stages, together with the Stokes continuation discussed next, are summarized in Fig.~\ref{fig:threeStages}.

\subsection{Compatibility of the spectral projection with the Stokes jump}
\label{subsec:stokesCovariance}

We note that the Stokes jump of Sec.~\ref{sec:QZIStokes} should be compatible with these three maps, in the sense that it must not turn an invisible combination into a visible one.  Across the Stokes direction each $\cI\bar\cI$ separation of an event sequence changes as in Eq.~\eqref{eq:PLone},
\begin{equation}
 \mathsf c_{v,+}\longmapsto\mathsf c_{v,+}-H_v
 =\widetilde{\mathsf c}_{v,-},
 \label{eq:historyLocalPL}
\end{equation}
while the contours of the $\cI\cI$/$\bar\cI\bar\cI$ separations are unchanged, exactly as for the minimal event sequence in Sec.~\ref{subsec:bothPairs}.  An event sequence $w$ contains $R(w)$ $\cI\bar\cI$ pairs, and $R(w)$ is even, as shown after Eq.~\eqref{eq:rawWordWeight}.  Replacing every reversed lower ray by the inward-running lower ray $\mathsf c_{v,-}=-\widetilde{\mathsf c}_{v,-}$ therefore produces the overall sign
\begin{equation}
 (-1)^{R(w)}=1,
 \label{eq:lowerOrientationCancellation}
\end{equation}
which generalizes Eq.~\eqref{eq:physicalProduct}.  The product of the local transformations $S_{\cI\bar\cI}$ of Eq.~\eqref{eq:PLmatrix} over all $\cI\bar\cI$ pairs of $w$ thus carries the upper physical cycle of $w$ to the lower physical cycle of the \emph{same} event sequence $w$.  The barrier multiplicities and the numbers of $\cI\bar\cI$ and $\cI\cI$/$\bar\cI\bar\cI$ pairs are unchanged, so each count class is carried to the corresponding count class on the other lateral side, and a zero-sum combination remains zero-sum.

Let $\cS_{\rm phys}^{(K)}$ denote this continuation on $\mathcal H_{\rm phys}^{(K)}$, which keeps the sequence label.  Then the invisible combinations of the third stage are carried into those on the opposite lateral side,
\begin{equation}
 \cS_{\rm phys}^{(K)}\!\left(\ker\Pi_{K,+}\right)
 =\ker\Pi_{K,-},
 \label{eq:kernelPreservation}
\end{equation}
and the scalar coefficients obey
\begin{equation}
 \Pi_{K,-}\,\cS_{\rm phys}^{(K)}
 =\cS_{\rm DDP}^{\rm obs}\,\Pi_{K,+}.
 \label{eq:intertwining}
\end{equation}
Here $\cS_{\rm DDP}^{\rm obs}$ is the DDP-induced map on normalized connected coefficients, which sends each $\cI\bar\cI$ factor $u_v^+$ to $u_v^-=1-u_v^+$ while keeping the barrier weights $B_e$ and each $\cI\cI$/$\bar\cI\bar\cI$ factor $u_v^+u_v^-$ fixed.  It is the lateral change $C_I^+\to C_I^-$ of Eq.~\eqref{eq:primitiveIntertwining}, applied sequence by sequence.

For the full three-stage diagram, grouping must be an equivariant quotient of the occurrence-resolved local model, and the coefficient map must intertwine continuation. If $\cS_{\rm grp}^{(K)}$ denotes the induced continuation on the grouped cycles, the required conditions are
\[
 \mathsf G^{(K)}\cS_{\rm PI}^{(K)}
 =\cS_{\rm grp}^{(K)}\mathsf G^{(K)},\qquad
 n^{(K)}\cS_{\rm grp}^{(K)}
 =\cS_{\rm phys}^{(K)}n^{(K)}.
\]
These conditions are part of the assumption that physical-cycle selection is compatible with the Stokes jump. Under these conditions and the scalar intertwining above, composition yields
\begin{equation}
 O_{K,-}\,\cS_{\rm PI}^{(K)}
 =
 \cS_{\rm DDP}^{\rm obs}\,
 O_{K,+},
 \qquad
 \cS_{\rm PI}^{(K)}\!\left(\mathcal V_{\rm spec}^{(K,+)}\right)
 =\mathcal V_{\rm spec}^{(K,-)}.
 \label{eq:fullStandardIntertwiningPhysics}
\end{equation}
 Under these assumptions on physical-cycle selection, all squares of the three-stage diagram in Fig.~\ref{fig:threeStages} commute at fixed grade, that is, the spectral projection is compatible with the Stokes jump. The formal local-amplitude model preserves the zero-sum relations within each count class independently of those hypotheses. Interpreting the kernel identities after physical specialization also requires that the specialization relations survive the DDP action on the Voros symbols. An $\cI\bar\cI$ factor changes from $u_v^+$ to $u_v^-$, while the $\cI\cI$/$\bar\cI\bar\cI$ factors and the barrier weights stay unchanged, so two event sequences that are indistinguishable on one lateral side remain indistinguishable on the other. The Stokes jump does not restore the ordering information. These conclusions concern scalar observables built from the local Weber amplitudes of Eq.~\eqref{eq:transferMatrix}.

\subsection{Origin of the nonlinear DDP factor}
\label{subsec:nonlinearDDP}

The local transformation at each $\cI\bar\cI$ pair is linear, whereas the DDP action \eqref{eq:DDPaction} multiplies the perturbative symbols by factors $(1+B_e)^{\pm1}$.  To separate the two, introduce the barrier-adapted variables
\begin{equation}
 Y_0:=\prod_{i=1}^N A_i,
 \qquad
 Y_j:=\left(\prod_{i=1}^jA_i\right)^{(-1)^j},
 \quad j=1,\ldots,N-1.
 \label{eq:edgeCharactersPhysics}
\end{equation}
In the product $\prod_{i\le j}\cS_0^{\rm DDP}[A_i]$ each factor $1+B_m$ with $m<j$ appears once from well $m$ and once from well $m+1$ with opposite exponents, so only $1+B_j$ survives, and with $B_0=B_N=0$ the DDP action becomes
\begin{equation}
 \cS_0^{\rm DDP}[Y_0]=Y_0,
 \qquad
 \cS_0^{\rm DDP}[Y_j]=Y_j(1+B_j).
 \label{eq:edgeCharacterDDPPhysics}
\end{equation}
The alternating exponent $(-1)^v$ of Eq.~\eqref{eq:DDPaction}, which is the intersection pairing of the $A$-cycle of well $v$ with the adjacent $B$-cycles, has been absorbed into the alternating powers in $Y_j$, and each barrier now contributes a single factor $1+B_j$.

The exact-WKB input is the Stokes relation between the two lateral quantization functions,
\begin{equation}
 \cS_0^{\rm DDP}[Q_N^+]=Q_N^-,
 \label{eq:QStokesPhysics}
\end{equation}
which gives, in a common formal chart with $B$ fixed,
\[
 \log\cZ_N^-(A,B)=\log\cZ_N^+(\mathfrak S A,B)
 +\log\frac{P_N(\mathfrak S A)}{P_N(A)}.
 \]
The logarithm is its formal expansion at $B=0$. For a single barrier,
 \[
 [B_e]\log\frac{P_N(\mathfrak S A)}{P_N(A)}
 =1-u_e^+-u_{e+1}^+=C_{[e]}^- -C_{[e]}^+.
\]
The linear variation of $\log P_v$ is $(-1)^v A_v(B_{v-1}+B_v)/(1+A_v)$, and adding the two incident wells gives this identity. At higher grade, lower-grade terms also contribute after $A\mapsto\mathfrak S A$. Thus the action is trivial on the associated graded quotient without making the explicit lateral coefficients equal. This relation between the two lateral coefficients is a consequence of the exact-WKB input \eqref{eq:QStokesPhysics}, and the full Voros-symbol action is not derived from the QZI.

\section{Double- and triple-well cases}
\label{sec:physical_cases}

We provide two simple examples to check the general $N$-well construction of Secs.~\ref{sec:QZIStokes} and \ref{sec:higher_order_projection}. The double well shows how the expansion in the classical bion weight reproduces the ordinary one-instanton splitting, and it allows the QZI/DDP matching of Sec.~\ref{sec:QZIStokes} to be checked at every fixed bion grade. The triple well is the smallest case with an internal well. It tests the four-term relation \eqref{eq:fourContours} with a nontrivial internal factor $\Xi_I$ and exhibits the grade-six triple-well ordering degeneracy of Sec.~\ref{sec:higher_order_projection}.

\subsection{Instanton splitting and fixed-grade QZI/DDP matching in the double well}
\label{subsec:doubleWell}

We start with the symmetric quartic double well
\begin{equation}
 V(x)=\frac12x^2(1-x)^2,
 \qquad
 S_I=\frac16,
 \qquad
 q:=\exp\!\left(-\frac{2S_I}{\hbar}\right)
 =\exp\!\left(-\frac{1}{3\hbar}\right).
 \label{eq:dwExamplePotential}
\end{equation}
The natural small parameter of the partition function and of the reduced determinant is therefore the \emph{closed} classical bion weight $q$.  Nevertheless the individual energy levels know about the one-instanton scale $q^{1/2}=\ee^{-S_I/\hbar}$.

At leading Weber-type order, multiplying the two quantization conditions of the double well for states even and odd under the reflection $x\to1-x$ gives, up to a nonvanishing local normalization factor \cite{ZinnJustinJentschuraI},
\begin{equation}
 D_{{\rm pair}, \eta}(\epsilon,\hbar)
 =\frac{1}{\Gamma(\frac12-\epsilon)^2}
 -\frac{q}{2\pi}
 \left(\frac{2}{\hbar}\right)^{2\epsilon}
 \ee^{ \eta\ii\pi(2\epsilon-1)}
 +O(\hbar q),
 \qquad \eta=\pm1.
 \label{eq:dwPairDetExample}
\end{equation}
Here $\eta=\pm1$ is the local double-well logarithm-branch label. In the adopted common Weber chart, $D_{{\rm pair},\eta}$ matches $Q_2^{-\eta}$, that is, $\eta=-1$ corresponds to the $+$ coefficient and $\eta=+1$ to the $-$ coefficient, as follows from the phases in Eq.~(88) of Ref.~\cite{SueishiExactWKBI}. This branch label is distinct from the triple-well coefficient label $\varsigma$ below and from the parity of an eigenstate, and the branch labels of this section are collected in Table~\ref{tab:branchLabels}. Thus the two functions $D_{{\rm pair},\eta}$ are the Weber-type forms of the two lateral quantization functions of Eq.~\eqref{eq:Q2example}. At $q=0$ the perturbative level
\begin{equation}
 \theta_n=n+\frac12
\end{equation}
is a double zero.  Writing $\epsilon=\theta_n+x$ and using
\begin{equation}
 \frac{1}{\Gamma(-n-x)}
 =(-1)^{n+1}n!\,x+O(x^2),
\end{equation}
Eq.~\eqref{eq:dwPairDetExample} becomes, to the first nontrivial order,
\begin{equation}
 (n!)^2x^2
 -\frac{q}{2\pi}
 \left(\frac{2}{\hbar}\right)^{2n+1}
 +\cdots=0.
 \label{eq:dwDoubleZeroLocal}
\end{equation}
The two roots therefore separate by
\begin{equation}
 \Delta_n
 =\frac{2^{n+1}}{n!\sqrt\pi}\,
 \hbar^{-n-\frac12}\ee^{-S_I/\hbar}
 \left[1+O(\hbar)\right].
 \label{eq:dwInstantonSplittingExample}
\end{equation}
Thus the one-instanton scale is not inserted by hand into the expansion in the classical bion weight.  It appears automatically because the $O(q)$ correction resolves a perturbative double zero into two roots that are power series in $q^{1/2}$,
\begin{equation}
 q=\ee^{-2S_I/\hbar}
 \quad\Longrightarrow\quad
 \epsilon_{n,1}-\epsilon_{n,2}\sim q^{1/2}=\ee^{-S_I/\hbar}.
 \label{eq:bionToInstantonMechanism}
\end{equation}
The quantization function and the partition function are graded by the number of bions, whereas the individual roots can carry half-integer powers of $q$ when perturbative levels are degenerate.

The same example also tests the lateral QZI/DDP matching beyond the leading splitting. In the double well both wells are outer wells and there is no internal well, so $\mathsf K_I=1$ in Eq.~\eqref{eq:spectralTarget} and the whole lateral dependence comes from the two $\cI\bar\cI$ pairs. The common $O(q)$ shift of the two levels is analytic in $q$ and contains the familiar lateral ambiguity. With the average $\bar\epsilon_{n,\eta}:=(\epsilon_{n,1}^{(\eta)}+\epsilon_{n,2}^{(\eta)})/2$, the digamma function $\psi$, and Euler's constant $\gamma_E$, one finds
\begin{equation}
 \bar\epsilon_{n, \eta}-\theta_n
 =\frac{q}{2\pi(n!)^2}
 \left(\frac{2}{\hbar}\right)^{2n+1}
 \left[
 \log\frac{2}{\hbar}-\psi(n+1)+ \eta\ii\pi
 \right]
 +O(q^2).
 \label{eq:dwCommonShiftExample}
\end{equation}
Its imaginary part is canceled by the opposite lateral ambiguity of the perturbative Borel sum, which is the standard BZJ/resurgent cancellation \cite{SueishiExactWKBI,ZinnJustinJentschuraI}.  In terms of the lateral Borel sums $\mathcal S_{0\pm}$ and the DDP automorphism of Sec.~\ref{subsec:lateralQC}, the ambiguity-free real resummation is the median prescription
\begin{equation}
 \mathcal S_{\rm med}
 =\mathcal S_{0+}\circ\bigl(\cS_0^{\rm DDP}\bigr)^{-1/2}
 =\mathcal S_{0-}\circ\bigl(\cS_0^{\rm DDP}\bigr)^{+1/2}.
 \label{eq:dwMedianExample}
\end{equation}
The lateral DDP data and the real spectrum are two representations of the same resurgent level splitting, not separate nonperturbative effects.

\begin{table}[ht]
\centering
\begin{tabular}{@{}lll@{}}
\toprule
Quantity & $+$ coefficient & $-$ coefficient\\
\midrule
Exact-WKB function & $Q^+$ & $Q^-$\\
Normalized Gamma ray & $\mathsf c_+$ (upper, outward) & $\mathsf c_-$ (lower, inward)\\
Double-well branch & $\eta=-1$ & $\eta=+1$\\
Triple-well branch & $\varsigma=+1$ & $\varsigma=-1$\\
Double-well phase & $\ee^{-\ii\pi(2\epsilon-1)}$ & $\ee^{+\ii\pi(2\epsilon-1)}$\\
\bottomrule
\end{tabular}
\caption{Distinct local branch labels and the adopted common Weber coefficient convention. The table fixes the internal matching, while a physical continuation requires its stated path and cycle orientation.}
\label{tab:branchLabels}
\end{table}

\paragraph{All fixed bion grades.}
The one-bion comparison extends without changing the local normalization.  Let $Z_{\rm SHO}$ denote the common perturbative harmonic-oscillator reference partition function of a single well, by which all expressions below are normalized.  In the normalization of Eq.~\eqref{eq:dwExamplePotential}, the two-event QZI has the large-$T$ form \cite{DersySchwartz}
\begin{equation}
 \mathscr I^{\rm QZI}_{2, \eta}(T)
 =T-2\gamma_E-2\log\frac{2}{\hbar}
 -2 \eta\pi\ii+O(Te^{-T}),
 \label{eq:dwTwoEventQZIExample}
\end{equation}
and the corresponding normalized path-integral contribution is
\begin{equation}
  \frac{\widetilde Z^{\rm PI}_{2,\eta}}{Z_{\rm SHO}}
 =\frac{q}{\pi\hbar}
 \left[
 T^2-2T\left(\log\frac2\hbar+\gamma_E+ \eta\pi\ii\right)
 \right]
 +O(qTe^{-T})+O(\hbar q).
 \label{eq:dwOneBionPolynomialExample}
\end{equation}
The prefactor $\sigma_1^2=1/(\pi\hbar)$ is independently obtained from the equal-frequency instanton determinant and translation measure of App.~\ref{app:collective}. At the ground pair, $\Delta_0^2=4q/(\pi\hbar)$ and $\bar\epsilon_{0,\eta}-\tfrac12=(q/\pi\hbar)(\log(2/\hbar)+\gamma_E+\eta\pi\ii)$ at the retained order. Consequently,
\[
 2\ee^{-T(\bar\epsilon_{0,\eta}-1/2)}\cosh(T\Delta_0/2)
 =2+\frac{q}{\pi\hbar}\left[T^2-2T\bigl(\log(2/\hbar)+\gamma_E+\eta\pi\ii\bigr)\right]
 +O(q^2).
\]
This verifies the one-bion polynomial and its absolute one-loop prefactor for the symmetric double well. 

More generally, at $2k$ instanton events, or bion grade $k$, the BZJ-continued QZI can be written as the Mellin--Barnes residue
\begin{equation}
 \mathscr I^{\rm QZI}_{2k, \eta}(T)
 =\Res_{u=0}\left[
 \Gamma(u)^{2k}
 \kappa_{2k}^{-2ku}
 \ee^{- \eta 2k\pi\ii u}
 \right]
 +O(\ee^{-T}),
 \qquad
 \kappa_{2k}:=\frac2\hbar \ee^{-T/(2k)}.
 \label{eq:dwGeneralQZIExample}
\end{equation}
Including the collective-coordinate measure and the elementary one-instanton normalization gives
\begin{equation}
  \frac{\widetilde Z^{\rm PI}_{2k,\eta}}{Z_{\rm SHO}}
 =q^k\frac{T}{k}\sigma_1^{2k}
 \mathscr I^{\rm QZI}_{2k, \eta}(T)
 +O(q^ke^{-T})+O(\hbar q^k),
 \qquad
 \sigma_1^2=\frac1{\pi\hbar}.
 \label{eq:dwGeneralPIExample}
\end{equation}
After writing $\Gamma(u)^{2k}=\Gamma(1+u)^{2k}/u^{2k}$ and setting $u=-x$, the QZI residue is generated by
\begin{equation}
 \Gamma(1-x)^{2k}
 \left(\frac2\hbar\right)^{2kx}
 \ee^{-Tx}\ee^{ \eta2k\pi\ii x},
 \label{eq:dwCommonGeneratorExample}
\end{equation}
which is the same function that generates the residue expansion of the spectral partition function built from Eq.~\eqref{eq:dwPairDetExample}.  Hence, at every fixed $k\ge1$,
\begin{equation}
  \frac{\widetilde Z_{{\rm pair},\eta}^{(k),{\rm spec}}}{Z_{\rm SHO}}
 = \frac{\widetilde Z^{\rm PI}_{2k,\eta}}{Z_{\rm SHO}}
 +O(q^ke^{-T})+O(\hbar q^k).
 \label{eq:dwAllFixedGradeMatchExample}
\end{equation}
The equality is coefficientwise through the complete leading large-$T$ polynomial $T^{2k},T^{2k-1},\ldots,T$, including the lateral imaginary part. For compactness we write $R^{\rm spec}_{k,\eta}(T)$ for the coefficient of $q^k$ in the normalized spectral partition function $\widetilde Z_{{\rm pair},\eta}^{\rm spec}/Z_{\rm SHO}$ after stripping the explicit factor $q^k$.
In particular,
\begin{equation}
 R^{\rm spec}_{k,\eta=+1}(T)-R^{\rm spec}_{k, \eta=-1}(T)
 =\frac{T}{k}\sigma_1^{2k}
 \left[
 \mathscr I^{\rm QZI}_{2k,\eta=+1}(T)
 -\mathscr I^{\rm QZI}_{2k, \eta=-1}(T)
 \right]
 +O(\ee^{-T}).
 \label{eq:dwAllFixedGradeJumpExample}
\end{equation}
 Since $\eta=+1$ corresponds to the $-$ coefficient, the left-hand side is the lower minus the upper coefficient, that is, the discontinuity $\operatorname{Disc}_0$ of Sec.~\ref{subsec:questions}. With this identification, in the double well the reduced Picard--Lefschetz jump of the QZI and the corresponding spectral-coefficient jump coincide not only at one bion but at every fixed bion grade. A closed double-well event sequence of grade $k$ simply alternates between the two wells, so it is fixed up to the choice of its first event and there is no ordering degeneracy of the type found in the triple well below. This case also shows that the lateral ambiguity belongs to the complexified QZI contour attached to the real multi-instanton sector. It need not be interpreted as a genuinely complex full-action saddle suddenly acquiring a nonzero coefficient \cite{DersySchwartz}.

\paragraph{Strict zero-temperature limit.}

The fixed-grade large-$T$ operation used in the derivation must still be distinguished from the strict physical limit at fixed $q$.  If the two exact levels are written as
\begin{equation}
 \epsilon_{\rm high}(q)=\bar\epsilon(q)+\frac{\Delta(q)}2,
 \qquad
 \epsilon_{\rm low}(q)=\bar\epsilon(q)-\frac{\Delta(q)}2,
\end{equation}
then their exact contribution to the partition function is
\begin{equation}
 \ee^{-T\epsilon_{\rm high}}+\ee^{-T\epsilon_{\rm low}}
 =2\ee^{-T\bar\epsilon(q)}
 \cosh\!\left(\frac{T\Delta(q)}2\right).
 \label{eq:dwCoshExample}
\end{equation}
A finite expansion in integer powers of the classical bion weight $q$ ceases to be uniform when $T\Delta=O(1)$, with $\Delta\sim q^{1/2}$.  At fixed $q$ one must therefore resum the nonperturbative grades before taking $T\to\infty$, and the lower exponential in Eq.~\eqref{eq:dwCoshExample} then dominates and selects the energy $\bar\epsilon-|\Delta|/2$.  Thus the same one-instanton scale $q^{1/2}$ that resolves the spectral double zero also reappears directly in the strict ground-state limit.

\subsection{Two-barrier matching and Stokes jump at the \texorpdfstring{$\cI\bar{\cI}$}{instanton-anti-instanton} pairs in the triple well}
\label{subsec:tripleWell}

For a two-barrier case, we take the bosonic symmetric triple well \cite{DunneTripleWell2020}
\begin{equation}
 V(x)=\frac18x^2(x^2-1)^2,
 \qquad
 \omega_{\rm in}=\frac12,
 \qquad
 \omega_{\rm out}=1,
 \qquad
 S_I=\frac18,
 \label{eq:twExamplePotential}
\end{equation}
and write
\begin{equation}
 q_{\rm tw}:=\ee^{-1/(4\hbar)}.
\end{equation}
This is the classical bion weight $\mathfrak q_e=\ee^{-2S_I/\hbar}$ of either barrier, the triple-well analog of $q$ in Eq.~\eqref{eq:dwExamplePotential}, and both barrier weights reduce to it at leading exponential order.
Let us name the left, central, and right wells $1$, $2$, and $3$.  Wells $1$ and $3$ have frequency $\omega_{\rm out}$ and are the outer wells of the two-barrier interval $I=[1,2]$ of Eq.~\eqref{eq:interval}, while well $2$ has frequency $\omega_{\rm in}$ and is its only internal well.  The minimal event sequence of $I$, started at the central well, is
\begin{equation}
 2\longrightarrow3\longrightarrow2\longrightarrow1\longrightarrow2.
 \label{eq:twPrimitiveHistoryExample}
\end{equation}
It has an $\cI\bar\cI$ pair at each outer well and an $\cI\cI$ pair and a $\bar\cI\bar\cI$ pair at the central well.  It is therefore the smallest case in which the internal cycle $\Xi_I$ of Eq.~\eqref{eq:internalCycle} is nontrivial.

At leading Weber-type order the connected spectral coefficient of the barrier weight $B_1B_2$ can be written as
\begin{equation}
 \mathcal C_{I,\varsigma}^{\rm spec}(\epsilon)
 =\frac{q_{\rm tw}^2}{(2\pi)^2}
 2^{-2\epsilon}\hbar^{-6\epsilon}\ee^{-\varsigma2\pi\ii\epsilon}
 \Gamma\!\left(\frac12-\epsilon\right)^2
 \Gamma\!\left(\frac12-2\epsilon\right)^2
 \left[1+O(\hbar)\right].
 \label{eq:twSpectralGammaExample}
\end{equation}
We test in App~\ref{app:twNumerics} the coefficient \eqref{eq:twSpectralGammaExample} against the numerically computed spectrum through the splitting of the outer-well ground-state doublet that it predicts. The QZI over the four separations of the sequence \eqref{eq:twPrimitiveHistoryExample} gives the Mellin--Barnes kernel
\begin{equation}
 \mathcal K_{\varsigma}^{\rm MB}(s)
 =2^{3-2\epsilon}\hbar^{2-6\epsilon}
 \ee^{\varsigma2\pi\ii\mu_{\rm out}}
 \Gamma(\mu_{\rm out})^2\Gamma(\mu_{\rm in})^2,
 \qquad
 s=\frac14-\epsilon,
 \quad
 \mu_{\rm out}=\frac12-\epsilon,
 \quad
 \mu_{\rm in}=\frac12-2\epsilon.
 \label{eq:twMBKernelExample}
\end{equation}
Here $\mu_{\rm out}$ and $\mu_{\rm in}$ are the parameters $\mu_v=\tfrac12-\mathsf n_v$ of Eq.~\eqref{eq:muWeberIdentification} at the outer and central wells, evaluated at the harmonic values $\mathsf n_v=\epsilon/\omega_v$.  The two descriptions obey the off-shell relation
\begin{equation}
 \mathcal C_{I,\varsigma}^{\rm spec}(\epsilon)
 =-\mathcal N_4\,
 \mathcal K_{\varsigma}^{\rm MB}\!\left(\frac14-\epsilon\right)
 \left[1+O(\hbar)\right],
 \qquad
 \mathcal N_4
 =\frac{q_{\rm tw}^2}{32\pi^2\hbar^2}
 =\mathcal N_2^2.
 \label{eq:twOffshellMatchExample}
\end{equation}
Here
\begin{equation}
 \mathcal N_2=\frac{q_{\rm tw}}{2^{5/2}\pi\hbar}
\end{equation}
is the elementary one-barrier normalization.  We do not introduce any additional two-barrier normalization, since once the one-barrier local block is normalized, the connected two-barrier normalization factorizes as $\mathcal N_4=\mathcal N_2^2$.  The residue map $s=1/4-\epsilon$ then gives equality of the complete leading residue polynomial, not only of the largest power of $T$.

With the stated elementary normalization, the off-shell equality fixes the relative two-barrier normalization. We next examine the reduced Stokes geometry at the two attractive endpoint pairs. This is the situation of Sec.~\ref{subsec:bothPairs} with $a=1$ and $c=3$. The two repulsive separations at the central well make up $\Xi_I=(\mathcal C_2^{\rm rep})^2$, which carries no lateral choice, and the single-pair relation \eqref{eq:PLone} holds at each outer well. Equation~\eqref{eq:fourContours} then gives the lower cycle of the interval as
\begin{equation}
 \Gamma_{I,-}
 =\Xi_I\otimes
 \left(
 \mathsf c_{1,+}\mathsf c_{3,+}
 -H_1\mathsf c_{3,+}
 -\mathsf c_{1,+}H_3
 +H_1H_3
 \right),
 \label{eq:twTensorPLEExample}
\end{equation}
with integer coefficient vector $(+1,-1,-1,+1)$ in the basis of Sec.~\ref{subsec:bothPairs}.  The exact-WKB internal factor of Eq.~\eqref{eq:KI} is $\mathsf K_I=A_2/(1+A_2)^2=\mathcal P^{\rm QZI}(\Xi_I)$, and since wells $1$ and $3$ are odd, Eq.~\eqref{eq:uDef} gives $u_1=A_1/(1+A_1)$ and $u_3=A_3/(1+A_3)$ for the upper outer-well factors.  Equation~\eqref{eq:DDPFiniteDifference} becomes
\begin{equation}
 C_I^-=\mathsf K_I-\mathsf K_Iu_1-\mathsf K_Iu_3+\mathsf K_Iu_1u_3,
 \label{eq:twFourTermsExample}
\end{equation}
which is term by term the period of Eq.~\eqref{eq:twTensorPLEExample}.  Thus the four-term DDP structure is the period image of two independent local Picard--Lefschetz transformations, one at each $\cI\bar\cI$ pair, and the factor $(1-u_1)(1-u_3)$, although quadratic in the Voros variables, does not require a nonlinear contour transformation.  Since the potential \eqref{eq:twExamplePotential} is even, $\mathsf n_1=\mathsf n_3$, and Eq.~\eqref{eq:intervalLateralDifference} reduces to $C_I^+-C_I^-=-\ii\,\mathsf K_I\tan\pi\mathsf n_1$, twice the contribution of a single $\cI\bar\cI$ pair.

\subsection{Ordering degeneracy in the triple well at higher grade}
\label{subsec:tripleHigher}

The minimal event sequence of $I$ has a fixed ordering.  Ordering degeneracy first appears when higher-grade excursions are included, as anticipated in Sec.~\ref{sec:higher_order_projection}.  We start an event sequence at the central well and define the two {\it elementary closed excursions}
\begin{equation}
 \mathsf x:2\to1\to2,
 \qquad
 \mathsf y:2\to3\to2.
 \label{eq:twABExcursionsExample}
\end{equation}
At total grade six, we consider
\begin{equation}
 \varpi_1=\mathsf{yyxyyx},
 \qquad
 \varpi_2=\mathsf{yxyyyx}.
 \label{eq:twTwoWordsExample}
\end{equation}
They are not related by a cyclic shift or a reversal, since between consecutive $\mathsf x$ excursions $\varpi_1$ makes runs of two and two $\mathsf y$ excursions while $\varpi_2$ makes runs of one and three, so they are genuinely different event sequences, as Fig.~\ref{fig:twoWords} shows.  Nevertheless both have
\begin{equation}
 \mathbf k=(2,4),
 \qquad
 \mathbf R=(2,2,4),
 \qquad
 \mathbf X=(0,4,0).
 \label{eq:twSameCountsExample}
\end{equation}
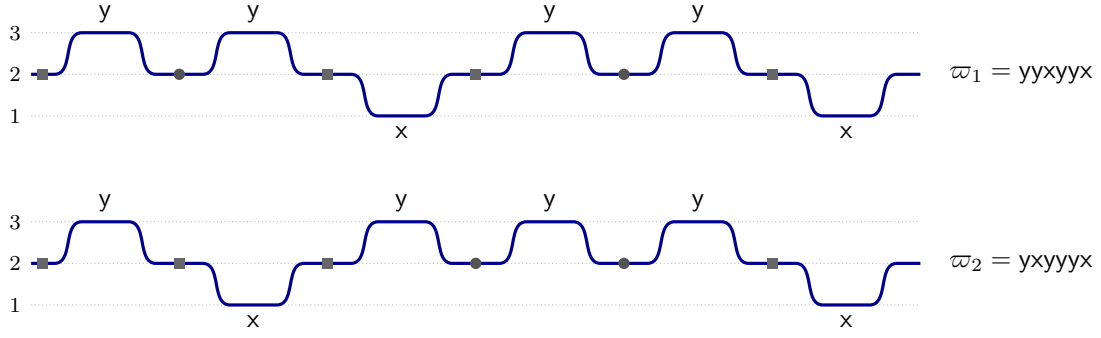
\begin{figure}[t]
\centering
 \begin{tikzpicture}[>=Stealth]
\begin{scope}[yshift=0cm]
  \foreach \y/\l in {0.0/1,0.55/2,1.1/3} {\draw[gray!55,densely dotted] (0,\y) -- (11.76,\y); \node[left,font=\scriptsize] at (0,\y) {$\l$};}
  \draw[very thick,blue!55!black] (0,0.550) -- (0.310,0.550) to[out=0,in=180] (0.670,1.100) -- (1.290,1.100) to[out=0,in=180] (1.650,0.550) -- (2.270,0.550) to[out=0,in=180] (2.630,1.100) -- (3.250,1.100) to[out=0,in=180] (3.610,0.550) -- (4.230,0.550) to[out=0,in=180] (4.590,0.000) -- (5.210,0.000) to[out=0,in=180] (5.570,0.550) -- (6.190,0.550) to[out=0,in=180] (6.550,1.100) -- (7.170,1.100) to[out=0,in=180] (7.530,0.550) -- (8.150,0.550) to[out=0,in=180] (8.510,1.100) -- (9.130,1.100) to[out=0,in=180] (9.490,0.550) -- (10.110,0.550) to[out=0,in=180] (10.470,0.000) -- (11.090,0.000) to[out=0,in=180] (11.450,0.550) -- (11.760,0.550);
  \fill[black!60] (0.085,0.480) rectangle (0.225,0.620);
  \node[above,font=\small] at (0.980,1.100) {$\mathsf y$};
  \fill[gray!65!black] (1.960,0.550) circle (0.075);
  \node[above,font=\small] at (2.940,1.100) {$\mathsf y$};
  \fill[black!60] (3.850,0.480) rectangle (3.990,0.620);
  \node[below,font=\small] at (4.900,0.000) {$\mathsf x$};
  \fill[black!60] (5.810,0.480) rectangle (5.950,0.620);
  \node[above,font=\small] at (6.860,1.100) {$\mathsf y$};
  \fill[gray!65!black] (7.840,0.550) circle (0.075);
  \node[above,font=\small] at (8.820,1.100) {$\mathsf y$};
  \fill[black!60] (9.730,0.480) rectangle (9.870,0.620);
  \node[below,font=\small] at (10.780,0.000) {$\mathsf x$};
  \node[right,font=\small] at (12.01,0.55) {$\varpi_1=\mathsf{yyxyyx}$};
\end{scope}
\begin{scope}[yshift=-2.5cm]
  \foreach \y/\l in {0.0/1,0.55/2,1.1/3} {\draw[gray!55,densely dotted] (0,\y) -- (11.76,\y); \node[left,font=\scriptsize] at (0,\y) {$\l$};}
  \draw[very thick,blue!55!black] (0,0.550) -- (0.310,0.550) to[out=0,in=180] (0.670,1.100) -- (1.290,1.100) to[out=0,in=180] (1.650,0.550) -- (2.270,0.550) to[out=0,in=180] (2.630,0.000) -- (3.250,0.000) to[out=0,in=180] (3.610,0.550) -- (4.230,0.550) to[out=0,in=180] (4.590,1.100) -- (5.210,1.100) to[out=0,in=180] (5.570,0.550) -- (6.190,0.550) to[out=0,in=180] (6.550,1.100) -- (7.170,1.100) to[out=0,in=180] (7.530,0.550) -- (8.150,0.550) to[out=0,in=180] (8.510,1.100) -- (9.130,1.100) to[out=0,in=180] (9.490,0.550) -- (10.110,0.550) to[out=0,in=180] (10.470,0.000) -- (11.090,0.000) to[out=0,in=180] (11.450,0.550) -- (11.760,0.550);
  \fill[black!60] (0.085,0.480) rectangle (0.225,0.620);
  \node[above,font=\small] at (0.980,1.100) {$\mathsf y$};
  \fill[black!60] (1.890,0.480) rectangle (2.030,0.620);
  \node[below,font=\small] at (2.940,0.000) {$\mathsf x$};
  \fill[black!60] (3.850,0.480) rectangle (3.990,0.620);
  \node[above,font=\small] at (4.900,1.100) {$\mathsf y$};
  \fill[gray!65!black] (5.880,0.550) circle (0.075);
  \node[above,font=\small] at (6.860,1.100) {$\mathsf y$};
  \fill[gray!65!black] (7.840,0.550) circle (0.075);
  \node[above,font=\small] at (8.820,1.100) {$\mathsf y$};
  \fill[black!60] (9.730,0.480) rectangle (9.870,0.620);
  \node[below,font=\small] at (10.780,0.000) {$\mathsf x$};
  \node[right,font=\small] at (12.01,0.55) {$\varpi_2=\mathsf{yxyyyx}$};
\end{scope}
\end{tikzpicture}
\caption{The two grade-six event sequences of Eq.~\eqref{eq:twTwoWordsExample}, drawn as trajectories over one period with the wells $1$, $2$, $3$ of the triple well on the vertical axis. Each excursion $\mathsf x$ or $\mathsf y$ of Eq.~\eqref{eq:twABExcursionsExample} is labeled at the outer well it reaches, where the trajectory always turns back. At the central well a dot marks an $\cI\bar\cI$ pair, where the trajectory turns back, and a square marks an $\cI\cI$ or $\bar\cI\bar\cI$ pair, where it passes through. The two half plateaus at the ends join into a single separation on the circle, marked once at the left. Both sequences have two dots and four squares at the central well and the same numbers of excursions to each outer well, which is the equality of counts \eqref{eq:twSameCountsExample}, although the two orderings are different.}
\label{fig:twoWords}
\end{figure}
Therefore Eq.~\eqref{eq:reducedWordWeight} gives
\begin{equation}
 \operatorname{Per}_{\varpi_1,\varsigma}
 =\operatorname{Per}_{\varpi_2,\varsigma}
\end{equation}
although the ordered event sequences are different.  If the two event sequences start with the same tunneling event, let $\mathcal A_{\varsigma}(w)$ denote the ordered product of the operators $\mathcal A_{d\to d'}^\pm$ of Sec.~\ref{subsec:transferMatrix} along the sequence, with the lateral sign matched to $\varsigma=\pm1$.  Equation~\eqref{eq:physicalChannelTelescopingPhysics} then gives
\begin{equation}
 \mathcal A_{\varsigma}(\varpi_1)=\mathcal A_{\varsigma}(\varpi_2),
 \label{eq:tripleSameRootPhysicalAlias}
\end{equation}
because the products of local Weber amplitudes are identical and the same projector onto the first event remains.  The two event sequences are therefore identified even before the trace is taken.

Equal rooted products must be distinguished from their cyclic contributions to $\log\cZ_N$. Here $\varpi_1=(\mathsf{yyx})^2$ is a twice-traversed grade-three word, so its cyclic contribution is $-\operatorname{Per}/2$, whereas the primitive $\varpi_2$ contributes $-\operatorname{Per}$. A primitive triple-well pair is $\mathsf{xyxyyyy}$ and $\mathsf{xyyxyyy}$ at grade seven.

For four consecutive wells an earlier primitive pair at grade five is
\[
 w_A=(1,2,3,2,3,2,3,4,3,2),\qquad
 w_B=(1,2,3,2,3,4,3,2,3,2).
\]
Both have $\mathbf k=(1,3,1)$, $\mathbf R=(1,2,2,1)$ and $\mathbf X=(0,2,2,0)$. Each visits wells 1 and 4 once, but their cyclic distance is three steps in $w_A$ and five in $w_B$, so no rotation or reversal identifies them. Both are primitive. Exhaustive enumeration below grade five is finite for all chains, since a grade-$K$ closed event sequence uses at most $K$ barriers, and it yields no earlier equal-count pair. The first triple-well pair is at grade six, while its first primitive pair is at grade seven. These findings refer to the formal products of local amplitudes within each count class.

The consequence is not restricted to the triple well. Every potential with $N\ge3$ wells contains three consecutive wells, and the same pair of grade-six event sequences can be placed there with all other barrier multiplicities zero. Hence, within the regular dilute, leading one-loop, local Weber approximation, the transfer-matrix description of Sec.~\ref{subsec:transferMatrix} does not distinguish all ordered event sequences for any $N\ge3$ once grade six is included. This degeneracy does not use the coefficient-selection assumption of Sec.~\ref{subsec:projection}. The two ordered event sequences are distinct configurations in path space, but at this order their dilute path-integral contributions coincide as well, since the fixed-$T$ QZI \eqref{eq:fixedTQZI} is symmetric under permutations of the separations and its prefactor depends only on which events occur. The degeneracy is therefore a property of the leading dilute, one-loop amplitudes, which the scalar spectral coefficient inherits, and the Stokes jump changes the local $\cI\bar\cI$ factors without restoring the ordering information. Contributions that depend on the ordering require effects beyond this approximation, such as finite-separation corrections or interactions among more than two events, or observables other than the partition function, such as time-ordered correlators of $x(t)$.

The two examples thus test complementary aspects of the same $N$-well construction.  The double well tests the grading and the all-fixed-grade spectral/QZI match, while the triple well tests the multi-barrier factorization and the loss of ordering information.

\section{Conclusions and outlook}
\label{sec:conclusions}

We have compared exact-WKB/DDP Stokes data with the multi-instanton expansion of a one-dimensional potential with $N$ equal-depth wells in the fixed-grade large-$T$ regime. The classical inverse-Laplace saddle obeys the Legendre relation with the finite-period Euclidean orbit and motivates the Weber energy scale. The QZI reduces to Gamma/Hankel periods in the stated dilute limit, although uniform dominance of a single saddle of the full Bromwich integral is not established. After the harmonic level number of each well is replaced by its local Weber level function and one elementary Weber normalization is fixed on the repulsive separations, the periods of the two $\cI\bar\cI$ separations in the minimal event sequence of a consecutive barrier interval reproduce the outer-well factors $u_a^\pm$ and $u_c^\pm$, and the $\cI\cI$/$\bar\cI\bar\cI$ separations at the internal wells give the lateral-invariant factor $\mathsf K_I$. The lower contour of each $\cI\bar\cI$ separation differs from the upper one by exactly one Hankel cycle, $\widetilde{\mathsf c}_-=\mathsf c_+-H$, and taking periods turns this linear relation into the exact-WKB complement rule $u^-=1-u^+$. Combining the relations at the two $\cI\bar\cI$ pairs reproduces the four-term DDP change of the connected coefficient, including its purely imaginary lateral difference at real energy.

The same QZI is obtained from the path integral on a regular dilute neighborhood of the widely separated multi-instanton configuration. The fluctuations transverse to the quasi-zero modes remain nondegenerate and contribute a single Gaussian factor with one rapid-decay cycle, so the nontrivial Stokes jump is carried by the collective quasi-zero modes. This separates the classical inverse-Laplace saddle from the constrained or post-Laplace QZI stationary problem. The general absolute normalization is fixed by the local Weber connection matrix rather than computed from the path integral, while the symmetric double well gives an independent one-loop check.

At higher fixed grade, the map from ordered event sequences to normalized scalar coefficient products is not one-to-one. We built a transfer matrix acting on oriented tunneling events from the local Weber amplitudes. Its determinant reproduces the normalized exact-WKB function $\cZ_N^\pm$, and its trace expansion generates every closed event sequence, once for each choice of its first event, together with a scalar weight. That weight depends only on the counts $(\mathbf k,\mathbf R,\mathbf X)$ of barrier crossings, $\cI\bar\cI$ pairs, and $\cI\cI$/$\bar\cI\bar\cI$ pairs. In the QZI the ordering information is lost in three stages, namely the grouping of repeated occurrences of a well and of their logarithmic sheets, the selection of the physical cycle among the grouped cycles, and the scalar evaluation, which identifies all event sequences with the same counts. For arbitrary event sequences, physical-cycle selection and compatibility with Stokes continuation are assumed. Under these hypotheses the spectral projection, composed of all three stages, intertwines the Picard--Lefschetz continuation of the QZI cycles with the DDP action on the scalar coefficients. The DDP relation between the normalized lateral coefficients follows from the exact-WKB relation \eqref{eq:QStokesPhysics}, which is taken as input and includes the perturbative factor, whereas the reduced endpoint contour identity alone does not derive the full factors $(1+B_e)^{\pm1}$. The loss of ordering information itself does not rely on the coefficient-selection assumption. Already at grade six, two inequivalent event sequences in the triple-well case give the same product of local Weber amplitudes, and the same example embeds in every potential with $N\ge3$ wells through any three consecutive wells. Since the leading dilute QZI depends on the separations only through their types, the two event sequences also have equal dilute path-integral contributions, so the degeneracy is already present at this order of the path integral and is inherited by the spectral coefficient.

The double- and triple-well calculations are meant to check the general construction here. In the double well, the expansion in the classical bion weight reproduces the ordinary one-instanton splitting by resolving a perturbative double zero, and the QZI/DDP match holds at every fixed bion grade. In the triple well the minimal event sequence of the two-barrier interval reproduces the four-term DDP change with a nontrivial internal factor, while higher grade gives the triple-well ordering example, with the repetition-factor qualification stated above. The connected two-barrier coefficient also predicts the splitting of the outer-well ground-state doublet, and this prediction agrees with the numerically computed spectrum as $\hbar\to0$. Several extensions are left for future work. Distinguishing event sequences with equal counts requires going beyond the leading dilute approximation, through finite-separation or multi-event interactions, or using time-ordered correlators rather than the partition function. The coefficient-selection assumption of Sec.~\ref{subsec:projection} should be derived from the path integral for arbitrary event sequences, as it was for the minimal event sequence, and the replacement of the harmonic level $\epsilon/\omega_v$ by the local Weber level function $\mathsf n_v$ should be derived from the anharmonic loop expansion of the path integral. Relaxing the equal-depth assumption, or identifying the outer barriers of the $N$-well potential to obtain a periodic one, would test whether the transfer matrix and the three stages of information loss persist when neighboring wells have different perturbative levels or when event sequences can wind around the potential. It would also be natural to extend the analysis to the supersymmetric, deformed, and $\mathcal{PT}$-symmetric potentials of Refs.~\cite{KamataExactWKB,KamataPT,BehtashHTA}, where hidden topological angles change the phases of the $\cI\bar\cI$ periods, and to determine how these phases enter the local Picard--Lefschetz transformation.

A further direction comes from the geometric picture recalled in Sec.~\ref{sec:intro}.  In holomorphic Floer theory the path integral can be regarded as an infinite-dimensional exponential integral \cite{KS-HFT}, and the local Picard--Lefschetz transformation $S_{\cI\bar\cI}$ has the form of the period image of a mutation by a single vanishing cycle in a Fukaya category \cite{SeidelVC,SeidelDef}.  Since the three stages of information loss boil down to the data of periods, it would be interesting to know whether a lift of the QZI cycles to objects of the Fukaya category attached to the WKB differential \cite{BridgelandSmith,HKK} keeps the ordering of event sequences that the scalar spectrum forgets, and whether the Kontsevich--Soibelman wall-crossing formula \cite{KSWallCrossing}, through its relation to cluster-algebra mutations \cite{IwakiNakanishi}, then expresses the combined DDP transformation of all barriers as an ordered product of elementary mutations at the individual barriers.

\section*{Acknowledgements}
A. Behtash thanks Mithat \"Unsal for useful discussions. S.~Kamata is supported by JSPS KAKENHI Grant Nos.~22H05118 and 25K07298. The authors acknowledge using an internally developed agentic workflow in Stacks built on Claude Opus 5.5, Opus 5, Opus 4.8, and GPT-5.6 Sol for proofreading the final manuscript and verifying some of the mathematical calculations.

\appendix

\section{Notation}
\label{app:notation}

Table~\ref{tab:notation} lists the main symbols, grouped by the part of the paper in which they are introduced, together with the equation or section where each is defined.

{\small
\begin{longtable}{@{}p{3.3cm}p{9.9cm}l@{}}
\caption{Main notation.}\label{tab:notation}\\
\toprule
Symbol & Meaning & Defined in\\
\midrule
\endfirsthead
\toprule
Symbol & Meaning & Defined in\\
\midrule
\endhead
\bottomrule
\endfoot
\multicolumn{3}{@{}l}{\emph{Setup}}\\
$x_v$, $\omega_v$, $V_{\min}$ & Minimum of well $v$, its harmonic frequency, and the common depth & Eq.~\eqref{eq:minima}\\
$v$, $e$ & Well label $v=1,\ldots,N$ and barrier label $e=1,\ldots,N-1$, barrier $e$ between wells $e$ and $e+1$ & Fig.~\ref{fig:wells}\\
$I=[a,b]$, $c$ & Consecutive barrier interval, with outer wells $a$ and $c=b+1$ & Eq.~\eqref{eq:interval}\\
$T$, $Z$, $\widetilde Z$ & Euclidean period, partition function, and partition function with $V_{\min}$ removed & Eqs.~\eqref{eq:thermaltrace}, \eqref{eq:shiftedTrace}\\
$E$, $\epsilon$ & Energy and scaled energy, $E=V_{\min}+\hbar\epsilon$ & Eq.~\eqref{eq:scaledEnergy}\\
$\gamma_v$, $\delta_e$ & Perturbative ($A$-) cycle of well $v$ and nonperturbative ($B$-) cycle of barrier $e$ & Eq.~\eqref{eq:VorosSymbols}\\
$A_v$, $B_e$ & Borel-resummed Voros symbols, with $B_e$ the barrier weight & Eq.~\eqref{eq:VorosSymbols}\\
$S_{I,e}$ & Action of one instanton across barrier $e$ & Eq.~\eqref{eq:openToClosedAction}\\
$\pm$ & Lateral Borel resummation at $\arg\hbar=0_\pm$ & Sec.~\ref{subsec:variables}\\
$\cI$, $\bar\cI$ & Instanton and anti-instanton events & Sec.~\ref{subsec:variables}\\
$t_r$, $\tau_r$ & Event centers and separations, where $r$ labels a separation and the visit it spans & Eq.~\eqref{eq:separations}\\
$K$ & Total nonperturbative (bion) grade, the number of barrier weights in a term & Sec.~\ref{subsec:regime}\\
$\Delta_{\rm np}$, $T_{\rm QZI}$, $T_*$ & Scaled nonperturbative level shift or splitting and the time-window bounds & Eq.~\eqref{eq:largeTPhysicsWindow}\\
\midrule
\multicolumn{3}{@{}l}{\emph{Exact-WKB}}\\
$\cS_0^{\rm DDP}$ & DDP Stokes automorphism across $\arg\hbar=0$ & Eq.~\eqref{eq:DDPaction}\\
$Q_N^\pm$ & Lateral quantization functions & Eqs.~\eqref{eq:Qplus}, \eqref{eq:QZInus}\\
$P_v$, $P_N$, $\cZ_N^\pm$ & $P_v=1+A_v$, $P_N=\prod_vP_v$, and $\cZ_N^\pm=Q_N^\pm/P_N$ & Eqs.~\eqref{eq:PN}, \eqref{eq:normalizedZ}\\
$u_v^\pm$ & Local perturbative oscillation of well $v$, with $u_v^-=1-u_v^+$ & Eq.~\eqref{eq:uDef}\\
$C_I^\pm$, $\mathsf K_I$ & Connected coefficient of interval $I$ and its lateral-invariant internal factor & Eqs.~\eqref{eq:CIconnected}, \eqref{eq:KI}\\
$D$ & Any quantization function with the exact levels as zeros & Sec.~\ref{subsec:lateralQC}\\
$\mathsf n_v$ & Local Weber level function, $\mathsf n_v=\epsilon/\omega_v+O(\hbar)$ & Eq.~\eqref{eq:AWeber}\\
$r_v$ & Lateral-invariant pass-through factor, $r_v^2=u_v^+u_v^-$ & Eq.~\eqref{eq:weberConnectionMatrix}\\
$\mathfrak q_e$, $U_{W,e}$ & Classical bion weight of barrier $e$ and its local Weber dressing & Eq.~\eqref{eq:generalAWBarrierPhysics}\\
\midrule
\multicolumn{3}{@{}l}{\emph{Path integral and QZI}}\\
$R$, $\omega_{\min}$ & Shortest separation and slowest decay rate, which control the dilute errors & Eq.~\eqref{eq:diluteScales}\\
$\alpha_\pm$ & Amplitudes of the two tails entering a separation & Eq.~\eqref{eq:tails}\\
$\varrho_r$, $\omega_{\rm ref}$ & Relative zero-point energy of a separation and the reference frequency & Eq.~\eqref{eq:rhoDef}\\
$\lambda_r$, $\Lambda_r$ & Sign ($-1$ attractive, $+1$ repulsive) and magnitude of the tail interaction & Eq.~\eqref{eq:QZIaction}\\
$\cW_I$ & QZI exponent of the minimal event sequence of $I$ & Eq.~\eqref{eq:QZIaction}\\
$\mathscr I_I(T)$, $\mathcal N_I$ & Fixed-$T$ QZI and its normalization & Eqs.~\eqref{eq:fixedTQZI}, \eqref{eq:NIdef}\\
$s$, $\mu_v$ & Laplace variable conjugate to $T$ and Gamma exponent, $\mu_v=\tfrac12-\mathsf n_v$ & Eqs.~\eqref{eq:LaplaceFactorization}, \eqref{eq:muWeberIdentification}\\
$J_v^{\rm rep}$, $J_{v,\pm}^{\rm att}$ & Single-separation periods of repulsive and attractive separations & Eqs.~\eqref{eq:JT}, \eqref{eq:JNplus}\\
$\mathsf c_\pm$, $\widetilde{\mathsf c}_-$, $H$ & Upper and lower rays, reversed lower ray, and Hankel cycle $H=\mathsf c_++\mathsf c_-$ & Eqs.~\eqref{eq:HankelAsRays}, \eqref{eq:orientedLowerRay}\\
$\mathcal N_v$ & Local Weber normalization of the periods at well $v$ & Eq.~\eqref{eq:NvAnsatz}\\
$\cR_v^{\rm att}$, $\cR_v^{\rm rep}$ & Normalizations of attractive and repulsive separations, with the connection-matrix phase removed & Eq.~\eqref{eq:RNfromHankel}\\
$\mathcal P^{\rm QZI}$ & Normalized period map from QZI cycles to their values & Eq.~\eqref{eq:normalizedQZIPeriodMap}\\
$\Xi_I$, $\Gamma_{I,\pm}$ & Internal cycle and the two lateral cycles of interval $I$ & Eqs.~\eqref{eq:internalCycle}, \eqref{eq:intervalCycles}\\
$S_{\cI\bar\cI}$ & Local Picard--Lefschetz transformation at an $\cI\bar\cI$ pair & Eq.~\eqref{eq:PLmatrix}\\
\midrule
\multicolumn{3}{@{}l}{\emph{Higher fixed grade}}\\
$\mathbb T^\pm$ & Transfer matrix between consecutive tunneling events & Eq.~\eqref{eq:transferMatrix}\\
$w$ & Closed event sequence, written as the list of wells it visits & Eq.~\eqref{eq:word}\\
$(\mathbf k,\mathbf R,\mathbf X)$ & Barrier crossings, $\cI\bar\cI$ pairs, and $\cI\cI$/$\bar\cI\bar\cI$ pairs of $w$ & Eqs.~\eqref{eq:ke}--\eqref{eq:Xv}\\
$\operatorname{Per}_{w,\pm}$ & Leading scalar weight (period) of $w$ & Eq.~\eqref{eq:rawWordWeight}\\
$\mathsf G^{(K)}$, $n^{(K)}$, $\Pi_{K,\pm}$ & Grouping, coefficient selection, and scalar evaluation & Sec.~\ref{subsec:projection}\\
$O_{K,\pm}$ & Spectral projection, the composite of the three stages & Eq.~\eqref{eq:standardPhysicalObservationPhysics}\\
$\mathcal V_{\rm grp}$, $\mathcal V_{\rm coeff}$, $\mathcal V_{\rm spec}$ & Combinations invisible after each stage & Eq.~\eqref{eq:threePhysicalKernelsPhysics}\\
\end{longtable}
}

\section{The QZI in standard form}
\label{app:QZI}

This appendix rederives the formulas of Sec.~\ref{sec:fullPI} in the language of the standard instanton calculus, for readers who prefer that route to the regular-patch formulation used there.  We note that nothing here is new.  The aim is to make each factor of the QZI \eqref{eq:fixedTQZI} explicit and to point to the literature for each step.  We work with real configurations and continue to the attractive sign only at the end, as the BZJ prescription requires.

The steps are standard, and several references treat them in more detail.  The Faddeev--Popov treatment of the translation mode and the evaluation of the one-instanton determinant are given in Refs.~\cite{Coleman1985,VZNS} and in the textbooks \cite{MarinoBook2015,ZinnJustinBook}, and the collective-coordinate method itself goes back to Ref.~\cite{GervaisSakita}.  The Gel'fand--Yaglom formula \cite{GelfandYaglom} and its generalizations are reviewed in Ref.~\cite{DunneDeterminants}, and the regulated version used for the normal directions in Sec.~\ref{subsec:collectiveNormal} is due to Forman \cite{Forman1987,Forman1992Erratum}.  The instanton--anti-instanton interaction, the separation integral of an $\cI\bar\cI$ pair, and its continuation from the repulsive sign are derived in Refs.~\cite{Bogomolny1980,ZinnJustin1981,ZinnJustin1983}.  Refs.~\cite{ZinnJustinJentschuraI,JentschuraZinnJustinII} extend them to all multi-instanton sectors of the double well and the periodic potential and obtain the resulting generalized Bohr--Sommerfeld conditions.  Refs.~\cite{BehtashInfinity,BehtashPL} interpret the separation integral as a critical point at infinity and the BZJ continuation as a choice of Lefschetz thimble, and Ref.~\cite{FujimoriMultiBion} studies multi-bion QZIs to all orders.  Bions in triple-well and multi-well potentials, including the separations at internal wells, are discussed in Ref.~\cite{DunneTripleWell2020}, and the relation between QZIs and uniform WKB is developed in Ref.~\cite{DunneUnsalUniform}.

\subsection{Collective coordinates and the one-loop measure}
\label{app:collective}

Consider first a single event $e$, an instanton profile $q_e(t-t_0)$ in a long time interval.  Its translation mode $\dot q_e$ is an exact zero mode of the fluctuation operator \eqref{eq:fluctuationOperator}, so the Gaussian integral over fluctuations diverges in that direction.  The Faddeev--Popov procedure trades this direction for the collective coordinate $t_0$ \cite{GervaisSakita,VZNS,Coleman1985}.  With
\begin{equation}
 F(t_0):=\int\dd t\;\dot q_e(t-t_0)\bigl[x(t)-q_e(t-t_0)\bigr],
\end{equation}
one inserts $1=\int\dd t_0\,|\partial_{t_0}F|\,\delta(F)$ into the path integral.  At leading order $\partial_{t_0}F=\|\dot q_e\|^2$, and the zero-energy condition $\tfrac12\dot q_e^2=V(q_e)-V_{\min}$ gives
\begin{equation}
 \|\dot q_e\|^2=\int\dd t\;\dot q_e^2=S_e,
\end{equation}
the action of the event.  Expanding the fluctuation in normalized eigenmodes of the fluctuation operator, with measure $\prod_n\dd c_n/(2\pi\hbar)^{1/2}$, the zero-mode coefficient is $c_0=F/\|\dot q_e\|$, and the $\delta$ function together with the factor $\|\dot q_e\|^2$ leaves
\begin{equation}
 \dd t_0\,\Bigl(\frac{S_e}{2\pi\hbar}\Bigr)^{1/2}
 \Bigl[\frac{{\det}'\bigl(-\partial_t^2+V''(q_e)\bigr)}{\det\bigl(-\partial_t^2+\omega^2\bigr)}\Bigr]^{-1/2}
 \ee^{-S_e/\hbar},
 \label{eq:appSingleEvent}
\end{equation}
measured relative to the harmonic oscillator of the well, where ${\det}'$ omits the zero eigenvalue.  For a barrier with the same frequency $\omega$ on both sides and $\dot q_e\sim A_e\ee^{-\omega|t-t_0|}$ as $|t-t_0|\to\infty$, the ratio of determinants equals $S_e/(2\omega A_e^2)$ \cite{Coleman1985,MarinoBook2015,ZinnJustinBook}, so that the amplitude per unit time of an isolated event is
\begin{equation}
 K_e=A_e\Bigl(\frac{\omega}{\pi\hbar}\Bigr)^{1/2}.
 \label{eq:appKe}
\end{equation}
For the double well of Eq.~\eqref{eq:dwExamplePotential} one has $\omega=A_e=1$, so $K_e=(\pi\hbar)^{-1/2}$, which is the factor $\sigma_1$ of Eq.~\eqref{eq:dwGeneralPIExample}.

For $m$ events in the dilute gas the same procedure is applied to each event, with the zero modes replaced by the localized modes $\Phi_j$ of Sec.~\ref{subsec:collectiveNormal}.  Since their Gram matrix \eqref{eq:GramMatrixPhysics} is the identity up to $O(\ee^{-\omega_{\min}R})$, the Faddeev--Popov factor is the product of single-event factors, and one collective coordinate per event remains.  One combination of them is the common translation around the circle, removed in Sec.~\ref{subsec:cores}, and the others are traded for the separations $\tau_r$ subject to $\sum_r\tau_r=T$.  This is the content of the Jacobian $\mathcal J_I$ and the event factors $a_{e_r}$ in Eq.~\eqref{eq:AperpFactorized}.

\subsection{Harmonic oscillator propagator over a separation}
\label{app:harmonic}

Near the minimum $x_v$ visited by the separation $r$, the shifted Hamiltonian $\widehat H-V_{\min}$ is approximated by the harmonic oscillator
\begin{equation}
 \widehat H_r=-\frac{\hbar^2}{2}\partial_x^2+\frac12\omega_r^2(x-x_v)^2,
 \qquad \omega_r=\omega_v .
\end{equation}
We write $K_r(y_f,y_i;\tau)=\langle x_v+y_f|\ee^{-\tau\widehat H_r/\hbar}|x_v+y_i\rangle$ for its propagator, where $y_i$ and $y_f$ are the initial and final displacements from the minimum.  The two events that bound the separation leave the trajectory at the minimum up to their exponentially small tails, so the separation contributes the propagator with both displacements set to zero.  Its exact form follows from the Gel'fand--Yaglom method \cite{GelfandYaglom,DunneDeterminants}.  The spectral (functional) determinant of $-\partial_t^2+\omega_r^2$ on $[0,\tau_r]$ is proportional to $y(\tau_r)$, where $y$ solves $\ddot y=\omega_r^2y$ with $y(0)=0$ and $\dot y(0)=1$.  Hence $y(\tau_r)=\sinh(\omega_r\tau_r)/\omega_r$ and
\begin{equation}
 K_r(0,0;\tau_r)=\bigl(2\pi\hbar\,y(\tau_r)\bigr)^{-1/2}
 =\left(\frac{\omega_r}{2\pi\hbar\sinh(\omega_r\tau_r)}\right)^{1/2}.
 \label{eq:MehlerExact}
\end{equation}
The large-separation form \eqref{eq:MehlerAsymptotic} is clearest in the eigenstates $\psi_n$ of $\widehat H_r$, with energies $E_n=\hbar\omega_r(n+\tfrac12)$.  In this basis
\begin{equation}
 K_r(0,0;\tau_r)=\sum_{n\ge0}|\psi_n(0)|^2\,\ee^{-E_n\tau_r/\hbar},
\end{equation}
and for $\omega_r\tau_r\gg1$ only the ground state survives, with corrections of order $\ee^{-2\omega_r\tau_r}$ because $\psi_1(0)=0$.  The whole $\tau_r$ dependence is therefore $\ee^{-E_0\tau_r/\hbar}=\ee^{-\omega_r\tau_r/2}$, and the prefactor in Eq.~\eqref{eq:MehlerAsymptotic} is $|\psi_0(0)|^2=(\omega_r/\pi\hbar)^{1/2}$, from $\psi_0(y)=(\omega_r/\pi\hbar)^{1/4}\ee^{-\omega_r y^2/2\hbar}$.

\subsection{Tail interaction from the action of two profiles}
\label{app:tail}

Let two consecutive events, centered at $t_1<t_2$ with $\tau=t_2-t_1\gg1$, bound a visit to the well $x_v$ of frequency $\omega_v$, and consider their sum $x=q_1+q_2-x_v$, where $q_1$ and $q_2$ are the two isolated profiles.  With $U:=V-V_{\min}$, the action of an isolated event is $S[q_j]=\int\dd t\,\bigl(\tfrac12\dot q_j^2+U(q_j)\bigr)$.  Split the time axis at a point $t_m$ inside the separation.  For $t<t_m$ one has $x=q_1+\xi_+$ with $\xi_+:=q_2-x_v$ small.  Expanding to first order in $\xi_+$ and integrating by parts with the equation of motion $\ddot q_1=U'(q_1)$,
\begin{equation}
 \int_{-\infty}^{t_m}\!\dd t\,\Bigl(\tfrac12\dot x^2+U(x)\Bigr)
 =\int_{-\infty}^{t_m}\!\dd t\,\Bigl(\tfrac12\dot q_1^2+U(q_1)\Bigr)
 +\dot q_1(t_m)\,\xi_+(t_m)
 +\frac12\int_{-\infty}^{t_m}\!\dd t\,\bigl(\dot\xi_+^2+\omega_v^2\xi_+^2\bigr)
 +O(\ee^{-2\omega\tau}),
\end{equation}
where $\omega$ is the smallest frequency involved.  The last integral is the part of $S[q_2]$ to the left of $t_m$, because $q_2$ is harmonic there, up to a correction near the core of $q_1$ that is quadratic in the tail and therefore of order $\ee^{-2\omega\tau}$.  The same computation for $t>t_m$, with $\xi_-:=q_1-x_v$, gives the boundary term $-\dot q_2(t_m)\,\xi_-(t_m)$.  In the overlap region $\dot q_1=\dot\xi_-$ and $\dot q_2=\dot\xi_+$, so adding the two halves gives
\begin{equation}
 S[x]-S[q_1]-S[q_2]
 =\dot\xi_-(t_m)\,\xi_+(t_m)-\xi_-(t_m)\,\dot\xi_+(t_m)
 +O(\ee^{-2\omega\tau}).
 \label{eq:appWronskian}
\end{equation}
The right-hand side is minus the Wronskian of two solutions of the linearized equation, so it does not depend on $t_m$.  With the tails $\xi_-=\alpha_-\ee^{-\omega_v(t-t_1)}$ and $\xi_+=\alpha_+\ee^{\omega_v(t-t_2)}$ of Eq.~\eqref{eq:tails} it equals $-2\omega_v\alpha_-\alpha_+\ee^{-\omega_v\tau}$, which is Eq.~\eqref{eq:tailInteraction}.  For an $\cI\bar\cI$ pair both tails lie on the same side of $x_v$, so $\alpha_-\alpha_+>0$ and the interaction is attractive.  For an $\cI\cI$ or $\bar\cI\bar\cI$ pair they lie on opposite sides and it is repulsive.

In the double well of Eq.~\eqref{eq:dwExamplePotential} the instanton is $q(t)=1/(1+\ee^{-t})$, which solves $\dot q=q(1-q)$ and has action $\int_0^1q(1-q)\,\dd q=\tfrac16=S_I$.  Its tails are $q\simeq\ee^{t}$ as $t\to-\infty$ and $1-q\simeq\ee^{-t}$ as $t\to+\infty$, so $|\alpha_\pm|=1$ and $\omega=1$, and the interaction magnitude of Eq.~\eqref{eq:QZIaction} is $\Lambda=2/\hbar$.  The same calculation, with the same result, is carried out in Refs.~\cite{Bogomolny1980,ZinnJustin1981,ZinnJustinJentschuraI}, and the role of this interaction as the source of a critical point at infinity is discussed in Ref.~\cite{BehtashInfinity}.

\subsection{The double-well bion as a worked example}
\label{app:dwExample}

Every separation of a double-well event sequence is an $\cI\bar\cI$ pair, and both wells have $\omega=1$, so $\varrho_r=0$ in Eq.~\eqref{eq:rhoDef} with $\omega_{\rm ref}=1$.  For the bion, one instanton and one anti-instanton on the circle, the two separations $\tau$ and $T-\tau$ are both attractive, and the separation integral of Eq.~\eqref{eq:fixedTQZI}, stripped of its normalization, is
\begin{equation}
 \mathscr I_2(T)=\int_0^T\dd\tau\;
 \exp\bigl[\Lambda\ee^{-\tau}+\Lambda\ee^{-(T-\tau)}\bigr],
 \qquad \Lambda=\frac2\hbar.
 \label{eq:appBionIntegral}
\end{equation}
The integrand grows toward both ends, where the two events merge, so the integral is defined by the BZJ prescription.  One evaluates it at the repulsive sign $\Lambda\to-\Lambda$, where it converges, and continues the result back.  At the repulsive sign and for $T\gg\log\Lambda$ the two ends decouple, and
\begin{equation}
 \int_0^T\dd\tau\;\ee^{-\Lambda\ee^{-\tau}-\Lambda\ee^{-(T-\tau)}}
 =T+2\int_0^\infty\dd\tau\,\bigl(\ee^{-\Lambda\ee^{-\tau}}-1\bigr)+\cdots
 =T-2\gamma_E-2\log\Lambda+\cdots,
\end{equation}
where we used $\int_0^\infty\dd\tau\,(\ee^{-\Lambda\ee^{-\tau}}-1)=\int_0^\Lambda\dd z\,(\ee^{-z}-1)/z=-\gamma_E-\log\Lambda-E_1(\Lambda)$, with $E_1$ the exponential integral, and the dots denote terms exponentially small in $T$ or in $\Lambda$. Continuing back to the attractive sign replaces $\log\Lambda$ by $\log(-\Lambda)=\log\Lambda+\eta\pi\ii$, where $\eta=\pm1$ labels the logarithm determination, and $\eta=-1$ matches $Q^+$ while $\eta=+1$ matches $Q^-$ in the convention specified in Sec.~\ref{subsec:doubleWell}, so
\begin{equation}
 \mathscr I_2(T)=T-2\gamma_E-2\log\frac2\hbar-2 \eta\pi\ii+\cdots,
\end{equation}
which is Eq.~\eqref{eq:dwTwoEventQZIExample}.  This is the classic result of Bogomolny and Zinn-Justin, in which the instanton--anti-instanton contribution carries the combination $\log(-2/\hbar)+\gamma_E$ and its imaginary part cancels the Borel ambiguity of the perturbative series \cite{Bogomolny1980,ZinnJustin1981,ZinnJustinJentschuraI}.

The Gamma periods of Sec.~\ref{subsec:gammaPeriods} give the same answer.  At the repulsive sign the Laplace transform factorizes,
\begin{equation}
 \int_0^\infty\dd T\,\ee^{-sT}\int_0^T\dd\tau\,f(\tau)f(T-\tau)
 =\Bigl[\int_0^\infty\dd\tau\,\ee^{-s\tau}f(\tau)\Bigr]^2,
 \qquad f(\tau)=\ee^{-\Lambda\ee^{-\tau}},
\end{equation}
which is Eq.~\eqref{eq:LaplaceFactorization}.  Each factor is the period $J^{\rm rep}$ of Eq.~\eqref{eq:JT} with $\omega=1$ and $\mu=s$, up to an exponentially small remainder, and $\Gamma(s)\Lambda^{-s}=1/s-\gamma_E-\log\Lambda+O(s)$.  The double pole $1/s^2$ of the square inverts to $T$ and the simple pole $-2(\gamma_E+\log\Lambda)/s$ to the constant, which reproduces the result above.  At the attractive sign the same steps apply to the rays $\mathsf c_\pm$, whose phases $\ee^{\pm\ii\pi\mu}$ in Eqs.~\eqref{eq:JNplus} and \eqref{eq:JNminus} shift the constant by $\pm\ii\pi$ per separation and so produce the two values of the imaginary term.

\section{Numerical check of the triple-well splitting}
\label{app:twNumerics}

\begin{table}[!b]
\centering
\begin{tabular}{@{}cccc@{}}
\toprule
$\hbar$ & numerical $\hbar\Delta_{\rm out}$ & Eq.~\eqref{eq:twOuterSplitting} & ratio\\
\midrule
$0.030$ & $4.275\times10^{-4}$ & $1.107\times10^{-3}$ & $0.386$\\
$0.020$ & $1.307\times10^{-5}$ & $2.103\times10^{-5}$ & $0.622$\\
$0.015$ & $2.737\times10^{-7}$ & $3.764\times10^{-7}$ & $0.727$\\
$0.0125$ & $1.140\times10^{-8}$ & $1.471\times10^{-8}$ & $0.775$\\
$0.010$ & $9.101\times10^{-11}$ & $1.108\times10^{-10}$ & $0.821$\\
$0.009$ & $6.097\times10^{-12}$ & $7.262\times10^{-12}$ & $0.840$\\
$0.008$ & $2.051\times10^{-13}$ & $2.392\times10^{-13}$ & $0.858$\\
\bottomrule
\end{tabular}
\caption{Splitting of the outer-well ground-state doublet of the triple well \eqref{eq:twExamplePotential}. The numerical values come from the exact Schr\"odinger spectrum, and the prediction is Eq.~\eqref{eq:twOuterSplitting} multiplied by $\hbar$.}
\label{tab:twNumerics}
\end{table}

The coefficient \eqref{eq:twSpectralGammaExample} can be tested against the exact spectrum.  In the harmonic approximation the ground levels of the two outer wells coincide at $\epsilon=\tfrac12$.  Tunneling through the central well splits them into two levels, one even and one odd under $x\to-x$, which we call the {\it outer-well ground-state doublet}.  The harmonic levels of the central well lie at $\epsilon=\tfrac14,\tfrac34,\ldots$, so the doublet is not degenerate with any central level, and its splitting comes from the connected two-barrier sector.  To second order in the barrier weights the two outer levels are then the eigenvalues of an effective two-level Hamiltonian with diagonal entries $\tfrac12+\delta_1$ and $\tfrac12+\delta_3$ and off-diagonal entry $t$.  The shifts $\delta_1$ and $\delta_3$ come from the single barriers and are of order $B_1$ and $B_2$, while the coupling $t$ through the central well is of order $(B_1B_2)^{1/2}$.  The $T^2\ee^{-T/2}$ term of the trace over these two levels is $\tfrac12(\delta_1^2+\delta_3^2)+t^2$, so its part of order $B_1B_2$ is $t^2$, and at the symmetric point the splitting is $\Delta_{\rm out}=2t$.  On the QZI side, Eq.~\eqref{eq:Bromwich} turns the connected coefficient into the contribution $T\Res_{\epsilon=1/2}[\ee^{-T\epsilon}\mathcal C^{\rm spec}_{I,\varsigma}(\epsilon)]$ to the shifted partition function.  Writing $\mathcal C^{\rm spec}_{I,\varsigma}(\epsilon)=g(\epsilon)\,\Gamma(\tfrac12-\epsilon)^2$ and using $\Gamma(\tfrac12-\epsilon)^2=(\epsilon-\tfrac12)^{-2}+2\gamma_E(\epsilon-\tfrac12)^{-1}+O(1)$, the $T^2\ee^{-T/2}$ term of this residue is $-g(\tfrac12)$.  Since $\ee^{-\varsigma\ii\pi}=-1$ and $\Gamma(-\tfrac12)^2=4\pi$, one finds $-g(\tfrac12)=q_{\rm tw}^2/(2\pi\hbar^3)$ for either lateral label, and equating it with $t^2$ gives
\begin{equation}
 \Delta_{\rm out}
 =\sqrt{\frac2\pi}\,\hbar^{-3/2}q_{\rm tw}
 \left[1+O(\hbar)\right].
 \label{eq:twOuterSplitting}
\end{equation}
The corresponding physical splitting is $\hbar\Delta_{\rm out}=\sqrt{2/\pi}\,\hbar^{-1/2}\ee^{-1/(4\hbar)}$.  Table~\ref{tab:twNumerics} compares it with the splitting of the two lowest outer-well levels of the potential \eqref{eq:twExamplePotential}, computed by shooting in the sectors even and odd under $x\to-x$ with an eighth-order Runge--Kutta integrator at relative tolerance $10^{-13}$.  Changing the matching point and the size of the integration region changes the smallest splitting in the table by less than one part in $10^{4}$.  The ratio increases monotonically toward one, and fits linear and quadratic in $\hbar$ over $\hbar\le0.015$ extrapolate to $1.007$ and $0.994$ at $\hbar=0$, with slopes between $-16$ and $-19$ that measure the omitted $O(\hbar)$ correction.  The same procedure applied to the double well gives $\hbar\to0$ intercepts $1.004$ and $0.999$ for the ratio to Eq.~\eqref{eq:dwInstantonSplittingExample} with $n=0$, and slopes between $-5.7$ and $-6.6$, consistent with the known correction factor $1-\tfrac{71}{12}\hbar$ \cite{Turbiner2010}.

\end{document}